\documentclass[acmsmall,screen,nonacm]{acmart}

\usepackage{amsmath,amsthm,amsfonts,bbold,xspace,graphicx,float,mathtools}
\usepackage{braket,subcaption,xcolor,textcomp,pifont}
\usepackage[mathcal]{eucal} %
\usepackage{stmaryrd} %
\usepackage{enumitem} %
\usepackage{hyperref}
\usepackage[nameinlink,capitalize]{cleveref}

\AtEndPreamble{%
\theoremstyle{acmdefinition}
\newtheorem{question}{Question}
\newtheorem{challenge}{Challenge}
}

\crefname{theorem}{Theorem}{Theorems}
\Crefname{theorem}{Theorem}{Theorems}
\crefname{lemma}{Lemma}{Lemmas}
\Crefname{lemma}{Lemma}{Lemmas}
\crefname{corollary}{Corollary}{Corollaries}
\Crefname{corollary}{Corollary}{Corollaries}
\crefname{proposition}{Proposition}{Propositions}
\Crefname{proposition}{Proposition}{Propositions}
\crefname{definition}{Definition}{Definitions}
\Crefname{definition}{Definition}{Definitions}
\crefname{challenge}{Challenge}{Challenges}
\Crefname{challenge}{Challenge}{Challenges}
\crefname{question}{Question}{Questions}
\Crefname{question}{Question}{Questions}
\crefname{example}{Example}{Examples}
\Crefname{example}{Example}{Examples}

\crefname{appendix}{Appendix}{Appendices}
\Crefname{appendix}{Appendix}{Appendices}

\AddToHook{env/lemma/begin}{\crefalias{theorem}{lemma}}
\AddToHook{env/corollary/begin}{\crefalias{theorem}{corollary}}
\AddToHook{env/proposition/begin}{\crefalias{theorem}{proposition}}
\AddToHook{env/definition/begin}{\crefalias{theorem}{definition}}
\AddToHook{env/question/begin}{\crefalias{theorem}{question}}
\AddToHook{env/example/begin}{\crefalias{theorem}{example}}

\usepackage{tabularray}

\usepackage{tikz}
\usetikzlibrary{arrows, shapes.multipart}
\usepackage{algorithm}
\usepackage{algpseudocodex}
\usepackage{wrapfig}
\usepackage{epigraph}

\makeatletter
\renewcommand{\ALG@beginalgorithmic}{\small}
\makeatother

\newfloat{algorithm}{t}{lop}

\DeclareRobustCommand{\algkw}[1]{%
  \textcolor{blue!70!black}{\textsf{\bfseries #1}}%
}

\algrenewcommand\algorithmicif{\algkw{if}}
\algrenewcommand\algorithmicthen{\algkw{then}}
\algrenewcommand\algorithmicelse{\algkw{else}}
\algrenewcommand\algorithmicreturn{\algkw{return}}
\algrenewcommand\algorithmicfunction{\algkw{function}}
\algrenewcommand\algorithmicrequire{\algkw{Init}}

\newcommand{\Z}{\mathbb Z}
\newcommand{\Co}{\mathbb C}

\newcommand{\bbone}{\mathbb 1}
\newcommand{\Id}{\bbone}

\newcommand{\tr}{\operatorname{tr}}
\newcommand{\Span}{\operatorname{span}}
\DeclarePairedDelimiter\Bracks{\llbracket}{\rrbracket}
\DeclarePairedDelimiter\parens{\lparen}{\rparen}
\DeclarePairedDelimiter\abs{\lvert}{\rvert}

\DeclarePairedDelimiter\braces{\lbrace}{\rbrace}
\DeclarePairedDelimiter\bracks{\lbrack}{\rbrack}

\newcommand{\calC}{\mathcal{C}}
\newcommand{\calD}{\mathcal{D}}
\newcommand{\calE}{\mathcal{E}}

\newcommand{\calH}{\mathcal{H}}

\newcommand{\calL}{\mathcal{L}}
\newcommand{\calM}{\mathcal{M}}
\newcommand{\calN}{\mathcal{N}}

\newcommand{\calU}{\mathcal{U}}
\newcommand{\calV}{\mathcal{V}}

\newcommand{\var}{\mathit{var}}
\newcommand{\cvar}{\mathit{cvar}}
\newcommand{\qvar}{\mathit{qvar}}
\newcommand{\hvar}{\mathit{Hvar}}
\newcommand{\lvar}{\mathit{Lvar}}
\newcommand{\lift}{\mathit{lift}}

\newcommand{\dil}{\mathit{dil}}
\newcommand{\cdil}{\mathit{cdil}}
\newcommand{\rev}{\mathit{rev}}
\newcommand{\High}{\ensuremath{\mathsf{High}}} %
\newcommand{\Low}{\ensuremath{\mathsf{Low}}}   %

\begin{document}
 
\title{Secure Information Flow in Quantum Programs} 

\author{Zhicheng Zhang}
 \affiliation{
  \institution{Griffith University}          
    \city{Brisbane}
  \country{Australia}
}
\email{z.zhang2@griffith.edu.au}

\author{Mingsheng Ying}
 \affiliation{
  \institution{University of Technology Sydney}          
    \city{Sydney}
  \country{Australia}
}
\email{Mingsheng.Ying@uts.edu.au}

\author{Lirong Xia}
 \affiliation{        
  \institution{Rutgers University and DIMACS}          
    \city{New Brunswick}
  \country{United States}
}
\email{lirong.xia@rutgers.edu}



\begin{abstract}
    Information-flow security aims to prevent the leakage of confidential information through program execution.
    This problem becomes fundamentally subtler for quantum programs, because their states may involve superposition and entanglement, and their executions may involve intricate interactions between classical and quantum variables.
    This paper develops semantic foundations and verification methods for secure information flow in quantum programs.
    We formulate two notions of information-flow security. The first is a noninterference-based definition that naturally generalizes its classical counterpart. The second is a simulation-based definition that explicitly models an attacker, providing a clear operational interpretation of security.
    We prove that these two notions are equivalent.

    To verify the information-flow security of quantum programs,
    we develop a proof framework based on quantum Hoare logic.
    A key challenge is that the standard self-composition technique,
    which expresses information-flow security in Hoare logic by composing a program with a renamed copy of itself, fails in the quantum setting due to an obstruction analogous to the no-cloning theorem.
    To overcome this challenge, we introduce self-conjugation, a novel technique that leverages the deferred measurement principle and properties of unitary operations. We demonstrate our framework on examples drawn from the quantum one-time pad, quantum error correction, and anonymous entanglement generation.
    
\end{abstract}

\maketitle
 
\section{Introduction}

A central problem in computer security is preventing the improper leakage of confidential information. To achieve end-to-end confidentiality, \textit{information-flow security} addresses this problem by regulating how confidential data may propagate through program execution.
In the standard two-level setting, confidential data is labeled $\High$, and data accessible to an attacker is labeled $\Low$.
\textit{Noninterference}~\cite{GM82}, a basic notion of information-flow security, requires that
any two input states agreeing on $\Low$ variables produce output states that still agree on $\Low$ variables.
Building on this foundation, a wide range of methods for establishing secure information flow in programs have been developed, including program-logical verification~\cite{AR80,JL00,DHS05,BDR11}, security type systems~\cite{VIS96,OP97,SV98,HR98,Myers99,PS02,BN02,SM03}, and other static analysis approaches~\cite{DD77,BDNN99,BDNN01,CHH02,MH99,Zanotti02,GM04}.

Quantum programs require this picture to be revisited.
Prior work~\cite{YFY13,AM20,Mestel22,QAC} indicates that adapting classical security methods to the quantum setting is highly nontrivial.
For information-flow security, the challenges arise from interactions between classical and quantum variables, as well as from entanglement within quantum data.
Indeed, even the fundamental notions of information-flow security require re-examination in quantum programming.

\subsection{Illustrative Example: Quantum One-Time Pad}

We begin with an illustrative example to see what quantum programs look like and how information-flow security arises in this setting.
Consider the \textit{quantum one-time pad}~\cite{AMTdW00,BR03}, a basic quantum encryption protocol.
Like its classical counterpart, it encrypts a secret quantum message using a one-time key. For clarity, 
we focus on the simplest single-qubit case;
the general multi-qubit case is analogous.

\begin{example}[Security of quantum one-time pad]
    \label{exp:qotp}
    The quantum one-time pad for a secret qubit $m$ can be expressed by the following quantum programs:
    \begin{equation*}
        \begin{aligned}
            \mathit{KeyGen}\equiv\ & r_x:=\ket{0};\ \mathit{H}[r_x];\ x:=\mathbf{meas}[r_x];\\
            \qquad & r_z:=\ket{0};\ \mathit{H}[r_z];\ z:=\mathbf{meas}[r_z]\\
            \mathit{Enc}\equiv\ & q:=\ket{0};\ \mathit{SWAP}[m,q];\\
            \qquad & \mathbf{if}\ x=1\ \mathbf{then}\ \mathit{X}[q];\\
            \qquad & \mathbf{if}\ z=1\ \mathbf{then}\ \mathit{Z}[q]
        \end{aligned}
    \end{equation*}
    Here, program $\mathit{KeyGen}$ generates two random bits $(x,z)$ as the one-time key by measuring qubits $r_x$ and $r_z$ after applying the Hadamard gate $H$.
    Program $\mathit{Enc}$ encrypts the secret qubit $m$ by first swapping its state into a working qubit $q$
    and then applying Pauli $X$ and $Z$ gates conditioned on the key $(x,z)$.
    The corresponding decryption program is analogous to $\mathit{Enc}$ and is omitted for simplicity.

    Informally, the security of the quantum one-time pad guarantees that the composed program $\mathit{KeyGen};\mathit{Enc}$ leaks no information about the secret quantum message $m$ to any attacker who can access only the encrypted message $q$ but not the key $(x,z)$.
    This guarantee can be formulated as an information-flow security requirement
    by classifying $m,r_x,r_z,x,z$ as $\High$ variables and $q$ as a $\Low$ variable.
    Intuitively, if $\mathit{KeyGen};\mathit{Enc}$ has secure information flow, then any attacker with access only to $q$ learns nothing about the secret $m$.
\end{example}

\Cref{exp:qotp} is sufficiently simple that its security guarantee may appear evident even without a formal security definition.
However, as in the classical setting, informal reasoning about individual examples does not scale.
Manually checking such guarantees for every new quantum program is infeasible, especially for programs describing complex quantum algorithms or protocols.
We therefore need a formal security notion with a clear operational interpretation, together with a sound verification framework that reduces security claims about new programs to formal proof obligations.
This raises the following two questions.

\begin{question}
    \label{ques:1}
    How do we formally define information-flow security for quantum programs?
\end{question}

\begin{question}
    \label{ques:2}
    How do we formally verify that a quantum program has secure information flow?
\end{question}

\subsection{Challenges from Classical to Quantum}
\label{sub:challenges_from_classical_to_quantum}

The classical noninterference-based definition of information-flow security~\cite{GM82} has been extended to a quantum automata model~\cite{YFY13}. However, this extension does not address \Cref{ques:1}, because it is tailored to an automata model and does not directly account for program structure.
Moreover, this extension does not by itself explain what the definition means operationally.
Historically, foundational security notions often gain clear operational interpretations through equivalent formulations, as in Shannon's equivalent characterizations of perfect secrecy~\cite[Theorem 6]{Shannon49} and the equivalence between Goldwasser-Micali's semantic security and ciphertext indistinguishability~\cite{GM82b,MRS86}.
For classical noninterference, this point is folklore: the observable $\Low$ behavior of a secure program can be simulated from the $\Low$ input alone.
When we move to the quantum setting, the following challenge arises.

\begin{challenge}
    \label{cha:1}
    Find a noninterference-based definition of information-flow security for quantum programs with a \textit{clear operational interpretation}.
\end{challenge}

\Cref{exp:qotp} already indicates some subtleties behind \Cref{cha:1}.
First, classical and quantum data interact in both directions. 
For example, in a measurement $x:=\mathbf{meas}[r_x]$, information flows from quantum to classical data,
whereas in a conditional $\mathbf{if}\ x=1\ \mathbf{then}\ X[q]$, information flows from classical to quantum data.
Thus, a security definition must reason about joint classical-quantum states.
Second, quantum data may involve entanglement.
For example, when the quantum one-time pad is embedded in a larger program (as often happens for security protocols), the $\Low$ variable $q$ may be entangled with other quantum variables, including those accessible to an attacker.
Thus, a proper operational interpretation of the security notion must account for entangled side information.

To address \Cref{ques:2}, one could consider several language-based approaches, including type-based enforcement, static analysis, and program-logical verification.
In this paper, we focus on the program-logical route.
This choice is motivated by the kind of reasoning needed even in \Cref{exp:qotp}:
$\mathbf{if}\ x=1\ \mathbf{then}\ X[q]$ allows the $\High$ variable $x$ to control an operation on the $\Low$ variable $q$, yet the overall program $\mathit{KeyGen}; \mathit{Enc}$ still has secure information flow.
Thus, verification must reason about the semantics of the composed program as a whole.
Classically, a standard program-logical technique is self-composition~\cite{BDR11},
which reduces the verification of noninterference to proof obligations about a composed program in classical Hoare logic~\cite{Hoare69}.

Unfortunately, a direct quantum analogue of self-composition encounters a fundamental obstacle.
Classical self-composition uses an assertion $\overline{l}=\overline{l'}$ to compare $\Low$ variables $\overline{l}$ and their renamed copies $\overline{l'}$.
As we will see in \Cref{sub:self_composition_fail}, 
a direct quantum analogue fails because of an obstruction analogous to the quantum no-cloning theorem~\cite{WZ82,Dieks82}.
This leads to the second challenge.

\begin{challenge}
    \label{cha:2}
    Develop a program-logical verification method for the information-flow security of quantum programs
    despite the \textit{no-cloning-style obstruction to direct self-composition}.
\end{challenge}

\subsection{Main Contributions}

The aim of this paper is to study foundational notions of information-flow security for quantum programs and develop a Hoare-style framework for their verification.
We address \Cref{cha:1,cha:2}, and thereby \Cref{ques:1,ques:2}, through two main contributions.%
\begin{itemize}
    \item
        We formally define two notions of information-flow security for quantum programs:
        a noninterference-based notion \textbf{NI} and a simulation-based notion \textbf{SIM}.
        The former naturally generalizes classical noninterference and serves as the main specification for verification.
        The latter explicitly models an attacker and provides a clear operational interpretation of security.
        We prove that the two notions are equivalent, thereby connecting \textbf{NI} to the operational meaning provided by \textbf{SIM}.
    \item
        We develop \textit{self-conjugation}, a novel language-based technique that reduces verification of the information-flow security of a quantum program $P$ to proof obligations in quantum Hoare logic for a self-conjugated program derived from $P$ by following its syntax.
        In contrast to classical self-composition, self-conjugation composes a program with its reverse rather than with a renamed copy.
        By exploiting the deferred measurement principle and properties of unitary operations, self-conjugation circumvents the obstruction to direct self-composition.
        We demonstrate our verification framework on illustrative examples, including the quantum one-time pad in \Cref{exp:veri-qotp} and more complex examples drawn from quantum error correction and anonymous entanglement generation.
\end{itemize}

\subsection{Organization}

The rest of the paper is organized as follows. \Cref{sec:quantum_programs} provides the necessary background on a loop-free quantum programming language. \Cref{sec:defining_secure_information_flow} develops two information-flow security notions for quantum programs and proves their equivalence. Building on these results, \Cref{sec:verification} introduces self-conjugation and develops a formal verification framework. \Cref{sec:illustrative_examples} demonstrates the framework through illustrative examples. Finally, \Cref{sec:related_work} discusses related work and \Cref{sec:conclusion} concludes the paper.

\section{Background on Quantum Programs}
\label{sec:quantum_programs}

In this section, we provide the background on quantum programs needed for our study of information-flow security.
We consider a simple quantum programming language supporting classical and quantum variables,
mid-circuit measurements, and both classical and quantum control flow.
With these features, the language can describe dynamic quantum circuits~\cite{NJ21,CTILMCG21}, the kind of quantum programs executable on current quantum hardware.
As usual, loops with a fixed bound can be unrolled into the language.

\subsection{Variables and States}

A quantum program may contain both classical and quantum variables.
Classical variables are standard: each classical variable $x$ has a set $T_x$ as its type (i.e., state space),
such as $\Z_2$ (Boolean) or $\Z_d$ (bounded integer).
Each quantum variable $q$ has a Hilbert space $\calH_q$ as its type,
such as $\Co^2$ (qubit) or $\Co^d$ (qudit).
We implicitly assume that the set of variable names is countably infinite and the type of every variable has finite cardinality or dimension.

For convenience, we use $\overline{x}=x_1,\ldots,x_n$ (and $\overline{q}=q_1,\ldots,q_n$) to denote a list of variables, sometimes simply called variables. By a slight abuse of notation,
we may also use $\overline{x}=x_1,\ldots,x_n$ to denote the set $\braces*{x_1,\ldots,x_n}$ of variables. 
A list of quantum variables $\overline{q}=q_1,\ldots,q_n$ has Hilbert space
$\calH_{\overline{q}}=\bigotimes_{j=1}^{n}\calH_{q_j}$.
For quantum variable $q$ and classical variable $x$,
if $\calH_q=\Co^{T_x}$, then we say that $q$ has a lifted type from $x$; for example, a qubit variable has a lifted type from a bit variable. This convention extends naturally to lists of variables.

We adopt standard conventions from quantum computing.
For a Hilbert space $\calH=\Co^{d}$, its computational-basis states are denoted by $\ket{0},\ket{1},\ldots,\ket{d-1}$.
A pure state of quantum variables $\overline{q}$ is represented by a unit vector $\ket{\psi}$ in $\calH_{\overline{q}}$; for example, a qubit variable may be in a superposition state $\ket{\psi}=\frac{1}{\sqrt{2}}\parens*{\ket{0}-i\ket{1}}$.
As usual, the covector corresponding to $\ket{\psi}$ is denoted by $\bra{\psi}$.
More generally, a mixed state of $\overline{q}$ is represented by a density operator $\rho$, a positive semidefinite operator on $\calH_{\overline{q}}$.
For example, $\rho=\frac{1}{3}\ket{0}\!\bra{0}+\frac{2}{3}\ket{1}\!\bra{1}$, or equivalently, $\rho=\parens*{\begin{smallmatrix}1/3 & 0\\ 0 & 2/3\end{smallmatrix}}$.
Unless explicitly stated otherwise, a mixed state in this paper may be normalized (with $\tr(\rho)=1$) or subnormalized (with $0\leq \tr(\rho)\leq 1$).

\subsection{Syntax}
\label{sub:syntax}

We first introduce the syntax of quantum circuits without measurements,
which involve only quantum variables.

\begin{definition}[Syntax of quantum circuits]
    \label{def:syntax_for_circuits}
    Quantum circuits are generated by the following syntax:
    \begin{equation}
        \label{eq:syn_circuit}
        C::= \ \mathbf{skip}\mid C_1; C_2\mid U[\overline{q}]\mid \mathbf{qif}[q](\ket{0}\rightarrow C_0)\square(\ket{1}\rightarrow C_1).
    \end{equation}
\end{definition}

Here, $C,C_0,C_1,C_2$ denote quantum circuits,
and $U$ denotes an elementary quantum gate.
The $\mathbf{skip}$ statement and sequential composition $C_1;C_2$ are standard.
We further explain \Cref{def:syntax_for_circuits} below. The corresponding semantics will be presented in \Cref{sub:semantics}.

\begin{itemize}
    \item 
        The quantum gate $U[\overline{q}]$ applies an elementary quantum gate $U$ to the quantum variables $\overline{q}$.
        Here, we require $U$ to be a unitary operator acting on the Hilbert space $\calH_{\overline{q}}$.
        For example, in \Cref{exp:qotp}, $H,X,Z$ are single-qubit gates and $\mathit{SWAP}$ is a two-qubit gate.
    \item
        The quantum if-statement $\mathbf{qif}\bracks*{q}\parens*{\ket{0}\rightarrow C_0}\square\parens*{\ket{1}\rightarrow C_1}$
		executes $C_j$, conditioned on the qubit variable $q$:
		when $q$ is in state $\ket{0}$, $C_0$ is executed;
		when $q$ is in state $\ket{1}$, $C_1$ is executed.
        Since $q$ can be in a superposition state $\alpha\ket{0}+\beta\ket{1}$, the two branches $C_0$ and $C_1$ may also be executed in superposition.
        Here, we require $q\notin \qvar(C_0)\cup\qvar(C_1)$,
        where $\qvar(C)$ denotes the quantum variables occurring in quantum circuit $C$ and will be formally defined in \Cref{def:quantum_variables}.

        Unlike a classical if-statement, whose control flow only runs through one of the two classical branches (see \Cref{def:syntax_for_programs} below),
        the quantum if-statement here manages quantum control flow~\cite{AG05,SVV18,BBGV20,VLRH23,YYF12,YVC24} that can run through both quantum branches in superposition.
        A formal distinction is given in \Cref{sub:semantics}.
\end{itemize}

Next, we introduce the syntax of quantum programs by extending quantum circuits with classical statements and quantum measurements. 

\begin{definition}[Syntax of quantum programs]
    \label{def:syntax_for_programs}
    Quantum programs are generated by the following syntax:
    \begin{equation}
            P::=\ \mathbf{skip}\mid P_1; P_2 \mid C\mid x:=e %
            \mid \mathbf{if}\ b\ \mathbf{then}\ P_1\ \mathbf{else}\ P_2
            \mid q:=\ket{e} \mid x:= \mathbf{meas}[q].
    \end{equation}
\end{definition}

Here, $P,P_1,P_2$ denote quantum programs and $C$ denotes a quantum circuit as defined in \Cref{def:syntax_for_circuits}.
The $\mathbf{skip}$ and sequential composition $P_1;P_2$ are standard. 
We further explain \Cref{def:syntax_for_programs} below.
The corresponding semantics will be presented in \Cref{sub:semantics}.
\begin{itemize}
    \item
        Every quantum circuit $C$ (see \Cref{def:syntax_for_circuits}) is also a quantum program.
    \item 
        The assignment $x:=e$ assigns the value of expression $e$ to the classical variable $x$.
        As usual, the types of $e$ and $x$ must match.
        We omit the formal typing rules for simplicity. 
    \item 
        The if-statement $\mathbf{if}\ b\ \mathbf{then}\ P_1\ \mathbf{else}\ P_2$ evaluates the Boolean expression $b$
        and executes program $P_1$ or $P_2$, according to whether $b$ evaluates to $\mathit{true}$ or $\mathit{false}$. As usual, we write $\mathbf{if}\ b\ \mathbf{then}\ P_1$ whenever $P_2\equiv \mathbf{skip}$.
    \item 
        The reset $q:=\ket{e}$ sets a quantum variable $q$ to the computational-basis state $\ket{e}$ determined by the value of expression $e$. Here, $q$ should have a lifted type from $e$.
    \item 
        The measurement $x:=\mathbf{meas}[q]$ performs the full computational-basis measurement $\braces*{\ket{m}\!\bra{m}}_{m}$ on quantum variable $q$,
        and assigns the measurement outcome to the classical variable $x$.
        Here, $q$ should have a lifted type from $x$.
        For example, in $x:=\mathbf{meas}[r_x]$ from \Cref{exp:qotp}, $x$ is a classical bit and $r_x$ is a qubit.
        Restricting the language to computational-basis measurements incurs no loss of generality, since arbitrary quantum measurements can be implemented using computational-basis measurements together with ancillary quantum variables.
\end{itemize}

For convenience, let us further define the following syntactic sugar.
Suppose $\overline{x}=x_1,\ldots,x_n$, $\overline{q}=q_1,\ldots,q_n$,
and $\overline{e}=e_1,\ldots,e_n$ are lists of classical variables, quantum variables, and expressions, respectively.
\begin{itemize}
    \item 
        $\overline{x}:=\overline{e}$ abbreviates $x_1:=e_1;\ldots;x_n:=e_n$.
    \item 
        $\overline{q}:=\ket{\overline{e}}$ abbreviates $q_1:=\ket{e_1};\ldots; q_n:=\ket{e_n}$.
    \item
        $\overline{x}:=\mathbf{meas}[\overline{q}]$ abbreviates
        $x_1:=\mathbf{meas}[q_1];\ldots; x_n:=\mathbf{meas}[q_n]$.
\end{itemize}

We now define the variables occurring in a quantum program.
Let $\var(\overline{e})$ denote the set of classical variables occurring in a list of expressions $\overline{e}$.

\begin{definition}[Classical and quantum variables]
    \label{def:quantum_variables}
    The set of classical variables in a program $P$, denoted by $\cvar(P)$, is recursively defined as:
    \begin{itemize}
        \item 
            $\cvar(P)=\emptyset$ if $P\equiv \mathbf{skip}\mid C$;
        \item 
            $\cvar(P_1;P_2)=\cvar(P_1)\cup \cvar(P_2)$;
        \item 
            $\cvar(x:=e)=\braces*{x}\cup \var(e)$;
        \item
            $\cvar(\mathbf{if}\ b\ \mathbf{then}\ P_1\ \mathbf{else}\ P_2)=\var(b)\cup \cvar(P_1)\cup \cvar(P_2)$;
        \item
            $\cvar(q:=\ket{e})=\var(e)$; and
        \item 
            $\cvar(x:=\mathbf{meas}[q])=\braces*{x}$.
    \end{itemize}
    Similarly, the set of quantum variables occurring in a program $P$, denoted by $\qvar(P)$, is recursively defined as:
    \begin{itemize}
        \item 
            $\qvar(P)=\emptyset$ if $P\equiv \mathbf{skip}\mid x:=e$;
        \item 
            $\qvar(P_1;P_2)=\qvar(P_1)\cup \qvar(P_2)$;
        \item
            $\qvar(U[\overline{q}])=\overline{q}$;
        \item 
            $\qvar(P)=\braces*{q}$ if $P\equiv q:=\ket{e}\mid x:=\mathbf{meas}[q]$;
        \item
            $\qvar(\mathbf{if}\ b\ \mathbf{then}\ P_1\ \mathbf{else}\ P_2)=\qvar(P_1)\cup \qvar(P_2)$; and
        \item
            $\qvar(\mathbf{qif}[q](\ket{0}\rightarrow C_0)\square(\ket{1}\rightarrow C_1))=\braces*{q}\cup\qvar(C_0)\cup \qvar(C_1)$.
    \end{itemize}
\end{definition}
In this paper, we implicitly assume that in any quantum program $P$, the names of all classical and quantum variables are distinct. We use $\var(P)=\cvar(P)\uplus \qvar(P)$ to denote the set of all variables in $P$.

\subsection{Semantics}
\label{sub:semantics}

Now we introduce the semantics of quantum circuits and quantum programs.
We first consider the quantum circuits in \Cref{def:syntax_for_circuits}.
A configuration of quantum circuits has the form $\parens*{C,\ket{\psi}}$,
where $C$ is the remaining circuit to be executed or the termination symbol $\downarrow$, and $\ket{\psi}$ is a pure state of quantum variables.
We use the convention $\downarrow; C\equiv C$.
Since the execution of a quantum circuit corresponds to a unitary operator, its semantics can be defined entirely in terms of pure states.
Let $\qvar(\ket{\psi})$ denote the set of quantum variables involved in $\ket{\psi}$.
A configuration $\parens*{C,\ket{\psi}}$ is valid only if $\qvar(C)\subseteq \qvar(\ket{\psi})$.
Let $\calC_{\textup{circ}}$ be the set of valid configurations of quantum circuits.

The operational semantics of quantum circuits is defined as follows.

\begin{definition}[Operational semantics of quantum circuits]
    \label{def:operational_semantics_circ}
    The operational semantics of quantum circuits is given by the
    transition relation $\rightarrow\subseteq \calC_{\textup{circ}}\times \calC_{\textup{circ}}$
    defined by the rules in \Cref{fig:operational-semantics-circuit}.
\end{definition} 

\begin{figure}[h]
    \centering 
    \begin{equation*}
        \begin{split}
            &\textbf{\textup{(SK)}}\quad\parens*{\mathbf{skip},\ket{\psi}}\rightarrow\parens*{\downarrow,\ket{\psi}}
            \qquad \textbf{\textup{(SC)}}\quad\frac{\parens*{C_1,\ket{\psi}}\rightarrow\parens*{C_1',\ket{\psi'}}}{\parens*{C_1;C_2,\ket{\psi}}\rightarrow\parens*{C_1';C_2,\ket{\psi'}}}\\
            &\textbf{\textup{(GA)}}\quad\parens*{U[\overline{q}],\ket{\psi}}\rightarrow\parens*{\downarrow, (U_{\overline{q}}\otimes I)\ket{\psi}}\\ 
            &\textbf{\textup{(QIF)}}\quad \frac{\ket{\psi}=\alpha\ket{0}_q\ket{\theta_0}+\beta\ket{1}_q\ket{\theta_1}\qquad (C_j,\ket{\theta_j})\rightarrow^*(\downarrow,\ket{\theta_j'})\quad (j=0,1)}{(\mathbf{qif}[q](\ket{0}\rightarrow C_0)\square(\ket{1}\rightarrow C_1),\ket{\psi})\rightarrow (\downarrow,\alpha\ket{0}_q\ket{\theta_0'}+\beta\ket{1}_q\ket{\theta_1'})}
        \end{split}
    \end{equation*}
    \caption{Transition rules for the operational semantics of quantum circuits.}
    \label{fig:operational-semantics-circuit}
\end{figure}

In \Cref{fig:operational-semantics-circuit}, $\rightarrow^k$ denotes the composition of $k$ copies of $\rightarrow$,
and $\rightarrow^*=\bigcup_{k=0}^{\infty}\rightarrow^k$.
Most rules are standard and self-explanatory.
We comment only on the (QIF) rule. %
Here, $j=0$ and $j=1$ correspond to the two quantum branches created by the $\mathbf{qif}$ statement.
The two branches are executed in superposition and may take different numbers of steps, so the premise of (QIF) uses $\rightarrow^*$~\cite{QRM,YZ24}.
The relation $\rightarrow$ is still well-defined by induction on the structure of quantum circuits.
The semantics of the $\mathbf{qif}$ statement is equivalent to a quantum multiplexor with $q$ as the control qubit.
The (QIF) rule also explains why the condition $q\notin \qvar(C_0)\cup \qvar(C_1)$ is required in \Cref{def:syntax_for_circuits}.
For simplicity, we assume throughout the paper that this condition holds for every $\mathbf{qif}$ statement.

Next, we consider the semantics of quantum programs.
A configuration of quantum programs has the form $\parens*{P,\sigma, \rho}$,
where $P$ is the remaining program to be executed or the termination symbol $\downarrow$, 
$\sigma$ is a state of classical variables, 
and $\rho$ is a mixed state of quantum variables.
We use the convention $\downarrow; P\equiv P$.
As in the case of quantum circuits, a configuration $\parens*{P,\sigma, \rho}$ is valid only if $\cvar(P)\subseteq \cvar(\sigma)$ and $\qvar(P)\subseteq \qvar(\rho)$,
where $\cvar(\sigma)$ and $\qvar(\rho)$ denote the set of classical and quantum variables involved in states $\sigma$ and $\rho$, respectively.
Let $\calC_{\textup{prog}}$ be the set of valid configurations of quantum programs.

The operational semantics of quantum programs is defined as follows.

\begin{definition}[Operational semantics of quantum programs]
    \label{def:operational_semantics}
    The operational semantics of quantum programs is given by the
    transition relation $\rightarrow\subseteq \calC_{\textup{prog}}\times \calC_{\textup{prog}}$
    defined by the rules in \Cref{fig:operational-semantics}.
\end{definition} 

\begin{figure}[h]
    \centering 
    \begin{equation*}
        \begin{split}
            &\textbf{\textup{(SK)}}\quad \parens*{\mathbf{skip},\sigma,\rho}\rightarrow\parens*{\downarrow,\sigma,\rho}
            \qquad\textbf{\textup{(QC)}}\quad \frac{\parens*{C,\ket{\psi_j}}\rightarrow^* \parens{\downarrow,\ket{\psi'_j}}\quad (j=1,\ldots,k)}{\parens{C,\sigma,\sum_{j=1}^{k}p_j\ket{\psi_j}\!\bra{\psi_j}}\rightarrow \parens{\downarrow, \sigma,\sum_{j=1}^{k} p_j\ket{\psi_j'}\!\bra{\psi_j'}}}\\
            &\textbf{\textup{(SC)}}\quad \frac{\parens*{P_1,\sigma,\rho}\rightarrow\parens*{P_1',\sigma',\rho'}}{\parens*{P_1;P_2,\sigma,\rho}\rightarrow\parens*{P_1';P_2,\sigma',\rho'}}
            \qquad\textbf{\textup{(AS)}}\quad \parens*{x:=e,\sigma, \rho}\rightarrow \parens*{\downarrow,\sigma[x:=\sigma(e)],\rho}\\
            &\textbf{\textup{(IF)}}\quad \frac{\sigma\models b}{\parens*{\mathbf{if}\ b\ \mathbf{then}\ P_1\ \mathbf{else}\ P_2,\sigma,\rho}\rightarrow \parens*{P_1,\sigma,\rho}}\qquad \frac{\sigma\models \neg b}{\parens*{\mathbf{if}\ b\ \mathbf{then}\ P_1\ \mathbf{else}\ P_2,\sigma,\rho}\rightarrow \parens*{P_2,\sigma,\rho}}\\
            &\textbf{\textup{(RS)}}\quad \parens*{q:=\ket{e},\sigma,\rho}\rightarrow\parens*{\downarrow,\sigma,\tr_{q}(\rho)\otimes\ket{\sigma(e)}\!\bra{\sigma(e)}_{q}}\\
            &\textbf{\textup{(MS)}}\quad \parens*{x:=\mathbf{meas}[q], \sigma, \rho}\rightarrow \parens*{\downarrow, \sigma[x:=m], \parens{\ket{m}\!\bra{m}_{q}\otimes\Id}\rho\parens{\ket{m}\!\bra{m}_{q}\otimes\Id}^\dagger}\\
            &\qquad\qquad\textup{ for every outcome $m$ from the measurement $\braces*{\ket{m}\!\bra{m}}_{m}$}
        \end{split}
    \end{equation*}
    \caption{Transition rules for the operational semantics of quantum programs.}
    \label{fig:operational-semantics}
\end{figure}

In \Cref{fig:operational-semantics}, most rules are self-explanatory.
We comment only on the (QC) and (RS) rules. In the premise of (QC), $\rightarrow^*$ is as defined in~\Cref{def:operational_semantics_circ}.
In the conclusion, $\rho=\sum_{j=1}^k p_j \ket{\psi_j}\!\bra{\psi_j}$ is a mixed state.
It is easy to verify that (QC) remains consistent under alternative decompositions of $\rho$, such as $\rho=\sum_{j=1}^{l} r_j\ket{\phi_j}\!\bra{\phi_j}$.
When applying (QC), the semantics moves from pure states to mixed states,
thereby discarding the global phase information stored in pure states $\ket{\psi_j}$ (and $\ket{\psi'_j}$).
This subtlety is also reflected in the verification framework developed later in \Cref{sec:verification}.
In (RS), $\tr_q(\rho)=\sum_{j}\bra{j}_{q}\rho\ket{j}_q$ stands for the reduced density operator obtained by tracing out the quantum variable $q$, where $\ket{j}$ ranges over the computational-basis states.

We now turn to the denotational semantics of quantum programs.
Unlike the standard approach that treats classical and quantum variables separately, 
we adopt a definition that uses a density operator to represent the joint state of both classical and quantum variables.
This representation can simplify the formulation of the security notions introduced in later sections.
More precisely, a joint state of classical variables $\overline{c}$ and quantum variables $\overline{q}$ is a mixed state of the form $s=\sum_{j} \ket{\sigma_j}\!\bra{\sigma_j}\otimes \rho_j$ with $\tr(s)=1$,
where each $\sigma_j$ is a state of $\overline{c}$ and $\rho_j$ is a mixed state of $\overline{q}$.
We define $\cvar(s)=\overline{c}$ and $\qvar(s)=\overline{q}$,
and let $\var(s)=\cvar(s)\cup \qvar(s)$.
When the context is clear, 
we simply use ``state'' to refer to ``joint state''. 

Finally, we define the denotational semantics of quantum programs.
For simplicity, we derive it from the operational semantics.

\begin{definition}[Denotational semantics]
    \label{def:denotational_semantics}
    The denotational semantics of a quantum program $P$ is defined as a linear map $\Bracks*{P}$
    such that for any input state $s=\ket{\sigma}\!\bra{\sigma}\otimes \rho$ with $\cvar(s)\supseteq \cvar(P)$ and  $\qvar(s)\supseteq \qvar(P)$,
    \begin{equation}
        \Bracks*{P}(s)=\sum\braces*{\abs*{\ket{\sigma'}\!\bra{\sigma'}\otimes\rho^{\prime}:\parens*{P,\sigma,\rho}\rightarrow^*\parens*{\downarrow,\sigma',\rho'}}},
    \end{equation} 
    where $\{|\cdot|\}$ denotes a multiset, and occurrences of the same state $\ket{\sigma'}\!\bra{\sigma'}\otimes\rho^{\prime}$ produced from distinct execution paths $\rightarrow^*$ are counted separately.
\end{definition}

The denotational semantics $\Bracks*{P}$ in \Cref{def:denotational_semantics} is uniquely determined, since its definition on states of the form $\ket{\sigma}\!\bra{\sigma}\otimes\rho$ extends linearly to general $\sum_j \ket{\sigma_j}\!\bra{\sigma_j}\otimes \rho_j$.
We call $s$ an input state to program $P$ whenever $\var(P)\subseteq \var(s)$.
Equivalently, $\Bracks*{P}$ may be viewed as a hybrid classical-quantum channel that can act on any input state $s$ to $P$.
If $P\equiv C$ is a quantum circuit, then $\Bracks*{P}$ is a unitary channel.

For illustration, consider the program $\mathit{KeyGen};\mathit{Enc}$ in \Cref{exp:qotp}.
Suppose we start with classical state $\sigma=\braces*{x\gets 0, z\gets 0}$ and quantum state
$\ket{\psi}\!\bra{\psi}_{m}\otimes\ket{\overline{0}}\!\bra{\overline{0}}_{q,r_x,r_z}$,
where $\ket{\psi}=\frac{1}{\sqrt{3}}\ket{0}+\frac{\sqrt{2}}{\sqrt{3}}\ket{1}$ and $\ket{\overline{0}}$ denotes $\ket{0,\ldots,0}$.
Alternatively, the initial state is a joint state $s=\ket{\psi}\!\bra{\psi}_m\otimes \ket{\overline{0}}\!\bra{\overline{0}}_{q,x,z,r_x,r_z}$.
Using \Cref{def:operational_semantics,def:denotational_semantics}, we obtain
\begin{align*}
    \Bracks*{\mathit{KeyGen}}(s)&=\frac{1}{4}\sum\nolimits_{a,b\in \braces{0,1}}\ket{\psi}\!\bra{\psi}_m\otimes \ket{a,a,b,b}\!\bra{a,a,b,b}_{x,r_x,z,r_z} \otimes\ket{0}\!\bra{0}_q,\\
    \Bracks*{\mathit{KeyGen};\mathit{Enc}}(s)&=\frac{1}{4}\sum\nolimits_{a,b\in \braces{0,1}}\ket{0}\!\bra{0}_m\otimes \ket{a,a,b,b}\!\bra{a,a,b,b}_{x,r_x,z,r_z} \otimes \ket{\psi^{a,b}}\!\bra{\psi^{a,b}}_q,
\end{align*}
where $\ket{\psi^{a,b}}=\frac{1}{\sqrt{3}} \ket{a}+(-1)^{b}\frac{\sqrt{2}}{\sqrt{3}}\ket{a\oplus 1}$ and $\oplus$ denotes the XOR operator.

\section{Defining Secure Information Flow in Quantum Programs}
\label{sec:defining_secure_information_flow}

In this section, we address \Cref{ques:1} by formulating two equivalent notions of information-flow security for quantum programs.
We fix a two-level security labeling that assigns each variable a security level, $\High$ or $\Low$.
We focus on confidentiality:
the security guarantee is to prevent confidential information from flowing from $\High$ variables to $\Low$ variables.
For a quantum program $P$, let $\hvar(P)$ and $\lvar(P)$ denote the sets of $\High$ and $\Low$ variables in $P$, respectively.
For example, in \Cref{exp:qotp},
$\hvar(\mathit{KeyGen})=\braces*{x,z,r_x,r_z}$, $\lvar(\mathit{KeyGen})=\emptyset$,
$\hvar(\mathit{Enc})=\braces*{m,x,z}$, and $\lvar(\mathit{Enc})=\braces*{q}$.
Note that $\hvar(P)$ and $\lvar(P)$ may contain both classical and quantum variables. 

\subsection{Information-Flow Security}
\label{sub:information_flow_security}

\subsubsection{Noninterference-Based Security}
\label{sub:NI}

A standard notion of information-flow security is noninterference~\cite{GM82}.
Intuitively, a program $P$ satisfies noninterference if any two input states that are indistinguishable to a $\Low$ attacker produce output states that remain indistinguishable to that attacker.
In other words, changes to the $\High$ part of the input state cannot be detected through an execution of $P$, ensuring that no confidential information flows from $\High$ to $\Low$ variables.

Prior work~\cite{YFY13} studied this idea in a quantum automata model,
closely paralleling the classical counterpart~\cite{GM82}.
Here, we need to instantiate it for quantum programs instead.
Let us first define \textit{low equivalence} between states,
building on the syntax and semantics introduced in \Cref{sec:quantum_programs}.

\begin{definition}[Low equivalence]
    \label{def:low-equivalence}
    Consider two states $s_1$ and $s_2$ with $\var(s_1)=\var(s_2)$.
    Let $\overline{h}$ and $\overline{l}$ be the lists of $\High$ and $\Low$ variables, respectively, in $\var(s_1)=\var(s_2)$.
    Then, $s_1$ and $s_2$ are $\Low$ equivalent, denoted by $s_1 =_{\Low} s_2$, iff
    $\tr_{\overline{h}}(s_1)=\tr_{\overline{h}}(s_2)$.
\end{definition}

Here, $\tr_{\overline{h}}(s)=\sum_{j}\bra{j}_{\overline{h}}s\ket{j}_{\overline{h}}$ traces out the $\High$ part of $s$, yielding exactly the reduced density operator for a $\Low$ attacker who can only access $\Low$ variables $\overline{l}$.
Thus, low equivalence captures indistinguishability to any such $\Low$ attacker.

For illustration, consider \Cref{exp:qotp}. Let $y$ be a fresh $\Low$ quantum variable and
$\ket{\beta}_{m,y}=\frac{1}{\sqrt{2}}(\ket{00}_{m,y}+\ket{11}_{m,y})$. Define
\begin{equation}
    \label{eq:eg-for-low-equiv}
    s_1=\frac{1}{2}(\ket{00}\!\bra{00}_{m,y}+\ket{11}\!\bra{11}_{m,y})\otimes \ket{\overline{0}}\!\bra{\overline{0}}_{x,z,r_x,r_z,q}\quad\textup{and}\quad
    s_2=\ket{\beta}\!\bra{\beta}_{m,y}\otimes \ket{\overline{0}}\!\bra{\overline{0}}_{x,z,r_x,r_z,q}.
\end{equation}
Then, $\overline{h}=m,x,z,r_x,r_z$ and $\overline{l}=q,y$.
Here, $s_2$ contains entanglement between $\High$ and $\Low$ parts, whereas $s_1$ only has classical correlation.
Still, we have $s_1=_{\Low}s_2$, because $\tr_{\overline{h}}(s_1)=\tr_{\overline{h}}(s_2)=\frac{\Id_y}{2}\otimes \ket{0}\!\bra{0}_q$. 

\begin{figure}[h]
    \centering
    \includegraphics[width=0.56\textwidth]{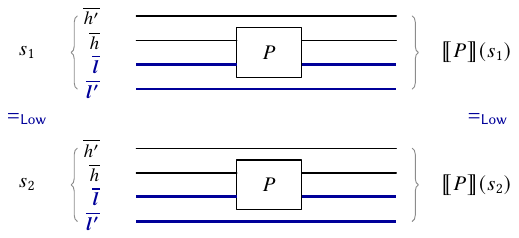}
    \caption{Illustration of \textbf{NI}. Here, $\overline{h}=\hvar(P)$ and $\overline{l}=\lvar(P)$. Variables $\overline{h'},\overline{l'}$ are $\High$ and $\Low$ variables outside $\var(P)$.
    All $\Low$ variables are colored blue.
    The input states $s_1,s_2$ are states over $\overline{h},\overline{l},\overline{h'},\overline{l'}$ and may contain entanglement across these variables.}
    \label{fig:NI_illustrate}
\end{figure}

Next, we define the noninterference-based information-flow security, abbreviated as \textbf{NI}, for quantum programs.
Its intuition is illustrated in \Cref{fig:NI_illustrate}.

\begin{definition}[Noninterference-based information-flow security (\textbf{NI})]
    \label{def:noninterference}
    A quantum program $P$ satisfies \textbf{NI} iff
    for any two input states $s_1$ and $s_2$ to $P$:
    \begin{equation*}
        s_1 =_{\Low} s_2 \Rightarrow \Bracks*{P}(s_1)=_{\Low} \Bracks*{P}(s_2).
    \end{equation*}
\end{definition}

The condition for \textbf{NI} in~\Cref{def:noninterference} formalizes the intuition of noninterference described above.
Continuing the illustration, consider the outputs of $\mathit{KeyGen};\mathit{Enc}$ on the input states $s_1,s_2$ from~\Cref{eq:eg-for-low-equiv}:
\begin{align*}
    s_1'=\Bracks*{\mathit{KeyGen};\mathit{Enc}}(s_1)&=\frac{1}{8}\sum\nolimits_{a,b,c\in \braces{0,1}}\ket{0}\!\bra{0}_m\otimes \ket{aabb}\!\bra{aabb}_{x,r_x,z,r_z} \otimes \ket{c\oplus a}\!\bra{c\oplus a}_{q}\otimes \ket{c}\!\bra{c}_{y},\\
    s_2'=\Bracks*{\mathit{KeyGen};\mathit{Enc}}(s_2)&=\frac{1}{4}\sum\nolimits_{a,b\in \braces{0,1}}\ket{0}\!\bra{0}_m\otimes \ket{aabb}\!\bra{aabb}_{x,r_x,z,r_z} \otimes \ket{\phi^{a,b}}\!\bra{\phi^{a,b}}_{q,y},
\end{align*}
where $\ket{\phi^{a,b}}_{q,y}=\frac{1}{\sqrt{2}}(\ket{a,0}_{q,y}+(-1)^{b}\ket{a\oplus1, 1}_{q,y})$.
It is straightforward to verify that $s_1'=_{\Low}s_2'$ since
$\tr_{\overline{h}}(s_1')=\tr_{\overline{h}}(s_2')=\frac{\Id_{q,y}}{4}$.
This calculation illustrates only the preservation of low equivalence required by \textbf{NI}. The formal verification that $\mathit{KeyGen};\mathit{Enc}$ satisfies \textbf{NI} is given in \Cref{exp:veri-qotp}.

\subsubsection{Simulation-Based Security}

While \textbf{NI} in \Cref{def:noninterference} is clean and intuitive,
it is formulated in terms of indistinguishability rather than an explicit attacker model.
To address \Cref{cha:1}, we introduce an alternative simulation-based definition that explicitly models an attacker
and admits a direct operational interpretation.
We then show that it is equivalent to \Cref{def:noninterference}.
This definition is in the same spirit as Shannon's perfect secrecy~\cite{Shannon49} and Goldwasser-Micali semantic security~\cite{GM82b}.

Before presenting the security definition, we introduce a notion of leakage probability
that quantifies how well a $\Low$ attacker can recover confidential information encoded in the $\High$ part of the input state.

\begin{definition}[Leakage probability]
    \label{def:leaking-pr}
    Consider a quantum program $P$ and classical variables $\overline{x},\overline{y}$ of the same types.
    The leakage probability of $P$ from $\overline{x}$ to $\overline{y}$ is defined as
    \begin{equation}
        \label{eq:leak-pr}
        \Pr_{\textup{leak}}\bracks*{P,\overline{x}\rightarrow \overline{y}}=\inf_{s}
            f\parens*{\tr_{\overline{rx}}(s),\tr_{\overline{ry}}\circ \Bracks*{P}(s)},
    \end{equation}
    where $f(\rho_1,\rho_2)=(\tr\sqrt{\sqrt{\rho_1}\rho_2\sqrt{\rho_1}})^2$ is the (squared) fidelity function, 
    $s$ ranges over all (normalized) states with $\var(P)\cup \overline{x}\cup \overline{y}\subseteq \var(s)$,
    $\overline{rx}=\var(s)-\overline{x}$, and
    $\overline{ry}=\var(s)-\overline{y}$.
\end{definition}

\begin{figure}[h]
    \centering
    \includegraphics[width=0.57\textwidth]{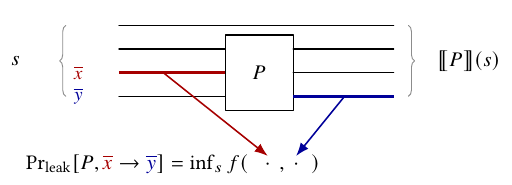}
    \caption{Illustration of leakage probability $\Pr_{\textup{leak}}$. 
    Here, the input state of $\overline{x}$ and output state of $\overline{y}$,
    corresponding to $\tr_{\overline{rx}}(s)$ and $\tr_{\overline{ry}}\circ\Bracks*{P}(s)$ in \Cref{eq:leak-pr},
    are colored red and blue, respectively.}
    \label{fig:leak_illustrate}
\end{figure}

Intuitively, $\Pr_{\textup{leak}}\bracks*{P,\overline{x}\rightarrow \overline{y}}$ characterizes the worst-case success probability of recovering the input state of $\overline{x}$
from the output state of $\overline{y}$ after executing $P$.
\Cref{fig:leak_illustrate} illustrates this intuition.
A formal justification is presented in \Cref{sub:justification}.
We further explain \Cref{def:leaking-pr} below.
\begin{itemize}
    \item 
    In \Cref{eq:leak-pr}, $\tr_{\overline{rx}}(s)$ is the input state of $\overline{x}$, and $\tr_{\overline{ry}}\circ \Bracks*{P}(s)$ is the output state of $\overline{y}$ after executing $P$.
    \item
    The (squared) fidelity $f$ quantifies the statistical overlap between the output state of $\overline{y}$ and the input state of $\overline{x}$.
    Taking the infimum over all input states $s$ yields the worst-case recovery probability.
    \item
    \Cref{def:leaking-pr} remains well-defined even when $\overline{x},\overline{y}\not\subseteq \cvar(P)$.
\end{itemize}

Now we are ready to introduce the simulation-based information-flow security, abbreviated as \textbf{SIM}, for quantum programs.
We call a program $P$ a $\Low$ program if $\hvar(P)=\emptyset$.

\begin{definition}[Simulation-based information-flow security (\textbf{SIM})]
    \label{def:classical-inf-sec}
    A quantum program $P$ satisfies \textbf{SIM} iff
    for any $\High$ classical variables $\overline{x}$ %
    and any $\Low$ classical variables $\overline{y}$ of the same type,
    any program $\mathit{E}$, %
    and any $\Low$ program $\mathit{D}$, %
    there exists a $\Low$ program $\mathit{D}'$ %
    such that
    \begin{equation}
        \label{eq:leak-cond}
        \Pr_{\textup{leak}}\bracks*{\mathit{E}; P; \mathit{D}, \overline{x}\rightarrow \overline{y}} \leq \Pr_{\textup{leak}}\bracks*{\mathit{E}; \mathit{D}', \overline{x}\rightarrow \overline{y}}.
    \end{equation}
\end{definition}

\begin{figure}[h]
    \centering
    \includegraphics[width=0.95\textwidth]{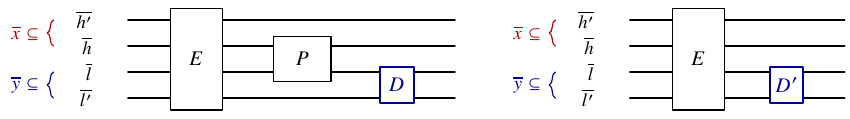}
    \caption{Illustration of \textbf{SIM}. 
    Here, $\overline{h}=\hvar(P)$ and $\overline{l}=\lvar(P)$.
    $\overline{h'},\overline{l'}$ are $\High$ and $\Low$ variables outside $\var(P)$,
    and $\overline{x},\overline{y}$ are as in \Cref{def:leaking-pr}.
    The LHS shows the program $E;P;D$, where the $\Low$ attacker runs program $D$ after executing $P$.
    The RHS shows $E;D'$, where the $\Low$ attacker runs program $D'$ without executing $P$.
    }
    \label{fig:SIM_illustrate}
\end{figure}

Intuitively, the condition for \textbf{SIM} in \Cref{def:classical-inf-sec} states that, regardless of how classical information is encoded in the input state, a $\Low$ attacker cannot recover it with a higher worst-case probability after executing the program $P$ than before.
\Cref{fig:SIM_illustrate} illustrates this intuition.
We further explain \Cref{def:classical-inf-sec} as follows.
\begin{itemize}
    \item 
    The program $\mathit{E}$ models how confidential information initially stored in $\High$ variable $\overline{x}$ is encoded in the state before executing $P$.
    Although the confidential information is classical, 
    $\mathit{E}$ can encode it into the joint state of both classical and quantum variables.
    Further discussion about quantum confidential information can be found in \Cref{sub:secure_quantum_information_flow}.
    \item
    The $\Low$ programs $\mathit{D}$ and $\mathit{D}'$ model the possible behaviors of a $\Low$ attacker with access to the $\Low$ part of the state after and before executing $P$, respectively.
    \Cref{eq:leak-cond} says that, even without executing $P$, $\mathit{D}'$ can recover the confidential information into $\Low$ variables $\overline{y}$ at least as well as $\mathit{D}$.
\end{itemize}

For illustration, consider again the program $\mathit{KeyGen};\mathit{Enc}$ from \Cref{exp:qotp}.
We focus on illustrating the role of the program $\mathit{E}$ in \Cref{def:classical-inf-sec}.
Let $u$ and $v$ be $\High$ and $\Low$ bit variables, playing the roles of $\overline{x}$ and $\overline{y}$, respectively.
For example, $E$ may encode the confidential input of $u$ into the $\High$ qubit $m$.
\begin{equation*}
    E\equiv m:=\ket{0};\ \mathbf{if}\ u=1\ \mathbf{then}\ X[m],
\end{equation*}
Alternatively, $E$ may store $u$ directly in a fresh $\Low$ bit $w$,
meaning that the confidential information is already available in $\Low$ variables before executing $\mathit{KeyGen};\mathit{Enc}$:
\begin{equation*}
    E\equiv w:=u,
\end{equation*}
Finally, $E$ may encode $u$ into entanglement between $m$ and a fresh $\Low$ qubit $o$.
\begin{equation*}
    E\equiv m,o:=\ket{\overline{0}};\ H[m];\ \mathit{CNOT}[m,o];\ \mathbf{if}\ u=1\ \mathbf{then}\ Z[m].
\end{equation*}
Informally, if $u=b$, then the encoded state is $\frac{1}{\sqrt{2}}(\ket{00}_{m,o}+(-1)^b \ket{11}_{m,o})$.
The above examples illustrate several ways in which confidential information may be encoded, 
such as in classical variables, quantum variables, or entanglement between $\High$ and $\Low$ variables.
Similarly, $\mathit{D}$ and $\mathit{D}'$ are arbitrary $\Low$ programs modeling an attacker and may perform any permitted operations on $\Low$ variables. 

\Cref{sub:secure_quantum_information_flow} further discusses the case when the source of confidential information is quantum rather than classical.

\subsubsection{Equivalence of \textbf{NI} and \textbf{SIM}}

We now prove that \textbf{NI} in \Cref{def:noninterference} and \textbf{SIM} in \Cref{def:classical-inf-sec} are equivalent notions of information-flow security.
This equivalence is in the same spirit as Shannon's equivalent characterizations of secrecy~\cite[Theorem 6]{Shannon49}
and the equivalence between ciphertext indistinguishability (also known as polynomial security in early formulations) and semantic security~\cite{GM82b,MRS86} in cryptography.
This equivalence gives \textbf{NI} an operational interpretation from \textbf{SIM}.

\begin{theorem}[\textbf{NI} $\Leftrightarrow$ \textbf{SIM}]
    \label{thm:non-inter-eq-classical-sec}
    \textbf{NI} in \Cref{def:noninterference}
    and \textbf{SIM} in \Cref{def:classical-inf-sec} are equivalent.
\end{theorem}

\begin{proof}
    \quad
    \begin{enumerate}
        \item 
            \textbf{SIM} $\Rightarrow$ \textbf{NI}.
            For clarity, we first consider the special case when every required state preparation, unitary gate, and measurement can be expressed in our programming language.
            Consider any quantum program $P$ %
            that satisfies \textbf{SIM}.
            Suppose, for contradiction, that $P$ does not satisfy \textbf{NI}.
            Then, there exist two input states $s_1$ and $s_2$ to program $P$ such that
            \begin{equation}
            \label{eq:assume-not-NI}
            s_1=_{\Low} s_2 \quad\textup{and}\quad \Bracks*{P}(s_1)\neq_{\Low} \Bracks*{P}(s_2).
            \end{equation}

            Let us use $\overline{c}$ and $\overline{q}$ to denote classical variables in $\cvar(s_1)=\cvar(s_2)$ and quantum variables in $\qvar(s_1)=\qvar(s_2)$, respectively.
            Suppose $s_1$ and $s_2$ can be generated by some quantum programs $Q_1$ and $Q_2$ 
            from the input state $\ket{\overline{0}}\!\bra{\overline{0}}_{\overline{c}}\otimes \ket{\overline{0}}\!\bra{\overline{0}}_{\overline{q}}$, respectively;
            i.e., $s_b=\Bracks*{Q_b}(\ket{\overline{0}}\!\bra{\overline{0}}_{\overline{c}}\otimes \ket{\overline{0}}\!\bra{\overline{0}}_{\overline{q}})$ for $b=1,2$.

            Consider the following program:
            \begin{equation}
                \label{eq:E-SIM-NI}
                \mathit{E}\equiv \overline{c}:=\overline{0};\ \overline{q}:=\ket{\overline{0}};\ \mathbf{if}\ x=0\ \mathbf{then}\ Q_1\ \mathbf{else}\ Q_2,
            \end{equation}
            where $x\notin \overline{c}$ is a $\High$ classical bit variable.
            Intuitively, the program $\mathit{E}$ encodes the confidential bit stored in $x$ into its output state while discarding the input states of other variables in $\overline{c}$ and $\overline{q}$.

            Let $\overline{h}$ and $\overline{l}$ be the lists of $\High$ and $\Low$ variables in $\overline{c},\overline{q}$, respectively.
            Consider
            \begin{equation*}
                s_1'=\tr_{\overline{h}}\circ\Bracks*{Q_1;P}(\ket{\overline{0}}\!\bra{\overline{0}}_{\overline{c}}\otimes \ket{\overline{0}}\!\bra{\overline{0}}_{\overline{q}}) 
                \quad\textup{and}\quad s_2'=\tr_{\overline{h}}\circ\Bracks*{Q_2;P}(\ket{\overline{0}}\!\bra{\overline{0}}_{\overline{c}}\otimes \ket{\overline{0}}\!\bra{\overline{0}}_{\overline{q}}).
            \end{equation*}
            By~\Cref{eq:assume-not-NI}, $s_1'\neq s_2'$. We now construct a $\Low$ program $\mathit{D}$ that can distinguish these two states with worst-case success probability strictly greater than $\frac{1}{2}$, using the pretty good measurement~\cite{Belavkin75,Holevo79,HW94}.
            
            More precisely, let $\overline{lc}$ and $\overline{lq}$ be classical and quantum variables in $\overline{l}$.
            Let $\overline{lc'}$ be fresh quantum variables that have lifted types from classical variables $\overline{lc}$.
            For $b\in \braces*{1,2}$,
            suppose $s_b'=\sum_{j} \ket{\sigma_j^b}\!\bra{\sigma_j^b}_{\overline{lc}}\otimes \parens{\rho_j^b}_{\overline{lq}}$
            and define $\widehat{s_b}=\sum_{j} \ket{\sigma_j^b}\!\bra{\sigma_j^b}_{\overline{lc'}}\otimes \parens{\rho_j^b}_{\overline{lq}}$ of quantum variables $\overline{l'}=\overline{lc'},\overline{lq}$.
            Now suppose the pretty good measurement (see~\Cref{sec:quantum_state_discrimination}) for $\widehat{s_1},\widehat{s_2}$ has measurement operators $M=\braces*{M_1,M_2}$ acting on $\overline{l'}$.
            To implement measurement $M$, by Stinespring's dilation theorem, one can find a unitary $U$
            acting on variables $\overline{l'},r$ with $r\notin \overline{l'}$ being a $\Low$ qubit variable such that
            $U(\ket{\psi}_{\overline{l'}}\ket{0}_r)=(M_1\ket{\psi}_{\overline{l'}})\ket{0}_r+(M_2\ket{\psi}_{\overline{l'}})\ket{1}_r$.
            Then, we can set the $\Low$ program $\mathit{D}$ to be
            \begin{equation}
                \label{eq:D-SIM-NI}
                \mathit{D}\equiv \overline{lc'}:=\ket{\overline{lc}};\ r:=\ket{0};\ U[\overline{lc'},\overline{lq},r];\ y:= \mathbf{meas}[r],
            \end{equation}
            where $y$ is a $\Low$ classical bit variable.
            Consequently, by properties of the pretty good measurement (see \Cref{sec:quantum_state_discrimination}), there exists some $\delta>0$ such that
            \begin{equation}
                \label{eq:E-D-SIM-NI}
                \Pr_{\textup{leak}}\bracks*{\mathit{E};P;\mathit{D}, x\rightarrow y}\geq \frac{1}{2}+\delta.
            \end{equation}

            On the other hand, since $s_1=_{\Low}s_2$,
            they are indistinguishable to any $\Low$ program $\mathit{D}'$.
            In particular, for any $\Low$ program $\mathit{D}'$, the output state $\tr_{\overline{ry}}\circ\Bracks*{\mathit{E};\mathit{D}'}(s)$ is independent of whether the confidential input $x=0$ or $x=1$ in the state $s$.
            Thus,
            \begin{equation}
                \label{eq:leak-below}
                \Pr_{\textup{leak}}[\mathit{E};\mathit{D}',x\rightarrow y]\leq \frac{1}{2},
            \end{equation} 
            which combined with \Cref{eq:E-D-SIM-NI} leads to a contradiction with \Cref{eq:leak-cond}, from our assumption that $P$ satisfies \textbf{SIM}.

            In the general case, the preparation of $s_1,s_2$ and the unitary $U$ cannot be exactly expressed in our language, but they can be approximately implemented arbitrarily well by quantum programs using fresh variables (see Solovay-Kitaev theorem in~\cite{Kitaev97}). 
            The approximation precision can be chosen such that $s_1'\neq s_2'$ and \Cref{eq:E-D-SIM-NI} still hold (with a different $\delta$),
            while the RHS of \Cref{eq:leak-below} becomes $\frac{1}{2}+\epsilon$ for some $\epsilon<\delta$.
            The details are tedious and thus omitted.
        \item 
            \textbf{NI} $\Rightarrow$ \textbf{SIM}.
            Consider any quantum program $P$ that satisfies \textbf{NI}.
            We want to verify that $P$ also satisfies \textbf{SIM}.
            Fix any $\High$ classical variables $\overline{x}$ and $\Low$ classical variables $\overline{y}$ of the same type,
            any program $\mathit{E}$,
            and any $\Low$ program $\mathit{D}$.
            Our goal is to find a $\Low$ program $\mathit{D}'$ such that
            \Cref{eq:leak-cond} holds.

            Suppose $\hvar(P)=\overline{h}$, and
            let $\overline{hc}$ and $\overline{hq}$ be lists of classical and quantum variables in $\overline{h}$.
            Let us consider another program
            \begin{equation*}
                Q\equiv\overline{hc}:=\overline{0};\ \overline{hq}:=\ket{\overline{0}};\ P.%
            \end{equation*}
            Note that for any input state $s$ to program $P$,
            \begin{align}
                \label{eq:Q-reset-h-1}
                \Bracks*{Q}(s)&=_{\Low}\Bracks*{P}\circ\Bracks*{\overline{hc}:=\overline{0};\overline{hq}:=\ket{\overline{0}}}(s)\\
                \label{eq:Q-reset-h-2}
                &=_{\Low}\Bracks*{P}\parens*{\tr_{\overline{h}}(s)\otimes \ket{\overline{0}}\!\bra{\overline{0}}_{\overline{h}}}\\
                \label{eq:Q-reset-h-3}
                &=_{\Low}\Bracks*{P}(s),
            \end{align}
            where \Cref{eq:Q-reset-h-1,eq:Q-reset-h-2} come from the semantics in \Cref{sec:quantum_programs}.
            \Cref{eq:Q-reset-h-3} holds because $P$ satisfies \textbf{NI} and
                $\tr_{\overline{h}}(s)\otimes \ket{\overline{0}}\!\bra{\overline{0}}_{\overline{h}}
                =_{\Low} s$.
            
            Let $\overline{v}$ be the list of variables in $\var(P)\cup \var(\mathit{E})\cup \var(\mathit{D})\cup \overline{x}\cup\overline{y}$.
            Let $R\equiv Q\parens*{\overline{h}\leftarrow \overline{g}}$ be a new program obtained in $Q$ by replacing $\High$ variables $\overline{h}$ by $\Low$ variables $\overline{g}$ with the same type and $\overline{g}\cap \overline{v}=\emptyset$.
            Then, for any input state $s$ with $\var(P)\cup \overline{g}\subseteq \var(s)$,
            \begin{align}
                \label{eq:RP-id-1}
                    \tr_{\overline{g}}\circ\Bracks*{R}(s)&=_{\Low}\tr_{\overline{g}}\circ\Bracks*{P(\overline{h}\leftarrow \overline{g})}(\tr_{\overline{g}}(s)\otimes \ket{\overline{0}}\!\bra{\overline{0}}_{\overline{g}})\\
                \label{eq:RP-id-2}
                    &=_{\Low}\tr_{\overline{g}}\circ\Bracks*{P(\overline{h}\leftarrow \overline{g})}(\tr_{\overline{h},\overline{g}}(s)\otimes \ket{\overline{0}}\!\bra{\overline{0}}_{\overline{g}}\otimes \ket{\overline{0}}\!\bra{\overline{0}}_{\overline{h}})\\
                \label{eq:RP-id-3}
                    &=_{\Low}\tr_{\overline{g}}\circ\Bracks*{P}(\tr_{\overline{h},\overline{g}}(s)\otimes \ket{\overline{0}}\!\bra{\overline{0}}_{\overline{h}}\otimes \ket{\overline{0}}\!\bra{\overline{0}}_{\overline{g}})\\
                \label{eq:RP-id-4}
                    &=_{\Low} \tr_{\overline{g}}\circ \Bracks*{P}(\tr_{\overline{h}}(s)\otimes \ket{\overline{0}}\!\bra{\overline{0}}_{\overline{h}})\\
                \label{eq:RP-id-5}
                    &=_{\Low}\tr_{\overline{g}}\circ \Bracks*{P}(s),
            \end{align}
            where \Cref{eq:RP-id-1} is obtained similarly to \Cref{eq:Q-reset-h-1,eq:Q-reset-h-2}.
            \Cref{eq:RP-id-2} uses the fact that $\overline{h}$ are $\High$ variables and $\Bracks*{P(\overline{h}\leftarrow \overline{g})}$ acts trivially on $\overline{h}$.
            \Cref{eq:RP-id-3} follows by renaming variables and that $\Bracks*{P}$ acts trivially on $\overline{g}$,
            which immediately implies \Cref{eq:RP-id-4}.
            The derivation of \Cref{eq:RP-id-5} is similar to \Cref{eq:Q-reset-h-3}.

            Now we set the $\Low$ program $\mathit{D}'$ to be $\mathit{D}'\equiv R; \mathit{D}$. %
            To see that $\mathit{D}'$ satisfies \Cref{eq:leak-cond},
            we calculate, for any input state $s$ with $\overline{v}\cup \overline{g}\subseteq \var(s)$:
            \begin{align}
                \label{eq:commute-b-tr-1}
                    \tr_{\overline{g}}\circ\Bracks*{\mathit{E};\mathit{D}'}(s)
                    &=_{\Low}\tr_{\overline{g}}\circ \Bracks*{\mathit{D}}\circ\Bracks*{R}\circ\Bracks*{\mathit{E}}(s) \\
                \label{eq:commute-b-tr-2}
                    &=_{\Low}\tr_{\overline{g}}\circ\Bracks*{\mathit{D}}\circ\tr_{\overline{g}}\circ\Bracks*{R}\circ\Bracks*{\mathit{E}}(s)\\
                \label{eq:commute-b-tr-3}
                    &=_{\Low}\tr_{\overline{g}}\circ\Bracks*{\mathit{D}}\circ\tr_{\overline{g}}\circ\Bracks*{P}\circ\Bracks*{\mathit{E}}(s)\\
                \label{eq:commute-b-tr-4}
                    &=_{\Low}\tr_{\overline{g}}\circ\Bracks*{\mathit{D}}\circ\Bracks*{P}\circ\Bracks*{\mathit{E}}(s)\\
                \label{eq:commute-b-tr-5}
                    &=_{\Low}\tr_{\overline{g}}\circ\Bracks*{\mathit{E}; P; \mathit{D}}\parens*{s}.
            \end{align}
            In the above, \Cref{eq:commute-b-tr-1,eq:commute-b-tr-5} come from the semantics in \Cref{sec:quantum_programs}.
            In \Cref{eq:commute-b-tr-2,eq:commute-b-tr-4}, we use that $\Bracks*{\mathit{D}}$ acts trivially on $\overline{g}$.
            \Cref{eq:commute-b-tr-3} follows from \Cref{eq:RP-id-5} and the fact that $\mathit{D}$ is a $\Low$ program.
            
            Since $\overline{g}\cap \overline{y}=\emptyset$ and $\overline{y}$ consists of $\Low$ variables, from \Cref{eq:commute-b-tr-5} we obtain,
            for any state $s$ with $\overline{v} \cup \overline{g}\subseteq \var(s)$:
            \begin{equation}
                \label{eq:dec-simulate}
                \tr_{\overline{ry}}\circ \Bracks*{\mathit{E};\mathit{D}'}(s)
                =\tr_{\overline{ry}}\circ\Bracks*{\mathit{E}; P; \mathit{D}}\parens*{s}.
            \end{equation}

            For any state $s'$ with $\overline{v} \subseteq \var(s')$,
            let $\overline{g'}$ be the list of variables appearing in $\overline{g}$ but not in $\var(s')$.
            Note that $s=s'\otimes \ket{0}\!\bra{0}_{\overline{g'}}$ satisfies $\overline{v}\cup \overline{g}\subseteq \var(s)$ and
            \begin{equation}
                \label{eq:enc-dec-fid-1}
                f\parens*{\tr_{\overline{ry}}\circ\Bracks*{\mathit{E};P;\mathit{D}}(s'), \tr_{\overline{rx}}(s')}
                =f\parens*{\tr_{\overline{ry}}\circ\Bracks*{\mathit{E};P;\mathit{D}}(s), \tr_{\overline{rx}}(s)},
            \end{equation}
            where \Cref{eq:enc-dec-fid-1} uses $\overline{g}\cap \overline{v}=\emptyset$ and the invariance of fidelity under tensoring with the same state.
            Combining \Cref{eq:enc-dec-fid-1,def:leaking-pr}, we obtain
            \begin{equation}
                \label{eq:enc-dec-left}
                \Pr_{\textup{leak}}\bracks*{E; P; D,\overline{x}\rightarrow \overline{y}}
                =\inf_{s:\overline{v}\cup\overline{g}\subseteq \var(s)} f \parens*{\tr_{\overline{ry}}\circ \Bracks*{E;P;D}(s),\tr_{\overline{rx}}(s)}.
            \end{equation}
            Similarly, we have
            \begin{equation}
                \label{eq:enc-dec-right}
                \Pr_{\textup{leak}}\bracks*{E; D',\overline{x}\rightarrow \overline{y}}
                =\inf_{s:\overline{v}\cup\overline{g}\subseteq \var(s)} f \parens*{\tr_{\overline{ry}}\circ \Bracks*{E;D'}(s),\tr_{\overline{rx}}(s)}.
            \end{equation}
            Finally, combining \Cref{eq:dec-simulate,eq:enc-dec-left,eq:enc-dec-right}
            immediately yields \Cref{eq:leak-cond}.
            Hence $P$ satisfies \textbf{SIM}.
    \end{enumerate}
\end{proof}

\subsection{Properties of Information-Flow Security in Quantum Programs}

In this subsection, we derive several useful properties of information-flow security from the \textbf{NI} characterization in \Cref{def:noninterference}. These properties can facilitate direct checking of information-flow security for quantum programs, as will be seen in \Cref{sec:illustrative_examples} and \Cref{sec:more_examples}.
The proofs of the following lemmas are deferred to \Cref{sub:proofs_properties}.

The first property states that, to determine whether a program $P$ satisfies \textbf{NI}, it suffices to consider all input states $s$ with $\var(s)=\var(P)$.

\begin{lemma}[Restricted inputs suffice]
    \label{lmm:res-suffice}
    A quantum program $P$ satisfies \textbf{NI} iff for any two input states with $\var(s_1)=\var(s_2)=\var(P)$,
    \begin{equation}
        \label{eq:low-eq-var-P}
        s_1 =_{\Low} s_2 \Rightarrow \Bracks*{P}(s_1)=_{\Low} \Bracks*{P}(s_2).
    \end{equation}
\end{lemma}

The second property is compositionality.

\begin{lemma}[Compositionality]
    \label{thm:compose}
    If two programs $P_1$ and $P_2$ both satisfy \textbf{NI}, 
    then their sequential composition $P_1;P_2$ also satisfies \textbf{NI}.
\end{lemma}

\section{Verifying Secure Information Flow by Self-Conjugation}
\label{sec:verification}

Even with the help of \Cref{lmm:res-suffice,thm:compose} from the previous section, directly checking the information-flow security of quantum programs remains difficult. 
In this section, we develop a verification framework that reduces the security condition \textbf{NI} in \Cref{def:noninterference} to formal proof obligations in Hoare-style logic.
The key technique we introduce is self-conjugation,
which overcomes the no-cloning-style obstruction to direct self-composition~\cite{BDR11} in the quantum setting, thereby addressing~\Cref{cha:2}.

\subsection{Why Does Direct Self-Composition Fail?}
\label{sub:self_composition_fail}

In classical programming, a standard technique for formalizing and verifying secure information flow is \textit{self-composition}~\cite{BDR11}. 
The idea is to reduce the noninterference condition (see \Cref{sub:NI}) of a classical program $P$ to a Hoare triple for the composed program $P;P'$, where $P'$ is a renamed copy of $P$.
More precisely, suppose $\hvar(P)=\overline{h}$ and $\lvar(P)=\overline{l}$.
Then the noninterference of $P$ can be established by proving the Hoare-style judgment
\begin{equation}
    \label{eq:classical-self-composition}
    \braces*{\overline{l}=\overline{l'}}\
    P;\ P'
    \ \braces*{\overline{l}=\overline{l'}},
\end{equation}
where $P'\equiv P(\overline{h},\overline{l}\leftarrow \overline{h'},\overline{l'})$ is obtained from $P$ by renaming
$\overline{h},\overline{l}$ to fresh variables $\overline{h'},\overline{l'}$ of the same types.

Intuitively, \Cref{eq:classical-self-composition} says that whenever we start from two $\Low$ equivalent input states (specified by the precondition $\braces{\overline{l}=\overline{l'}}$),
after two independent executions of $P$,
the output states remain $\Low$ equivalent (specified by the postcondition $\braces{\overline{l}=\overline{l'}}$).
Thus, \Cref{eq:classical-self-composition} captures the noninterference condition.

When moving to quantum programs, a direct analogue of \Cref{eq:classical-self-composition} would require a quantum assertion $\Lambda$ comparing two quantum variables.
One would naturally expect such an assertion to have the following properties: (1) if the two quantum variables are in the same quantum state, then $\Lambda$ is satisfied; and (2) if they are in distinct quantum states that can be perfectly distinguished, then $\Lambda$ is not satisfied.
However, no such quantum assertion, which in general is represented by a quantum observable~\cite{DP06}, exists. The obstruction is analogous to the quantum no-cloning theorem~\cite{WZ82,Dieks82} and related no-go theorems for universal comparison of quantum states~\cite{BCJ03,JAC04}, as formalized by the following lemma.

\begin{lemma}
    \label{lmm:obstruction}
    Let $q,q'$ be two quantum variables of the same type $\calH$ with $\dim(\calH)\geq 2$. Then, there exists no quantum observable $\Lambda$ on $\calH_q\otimes \calH_{q'}$ with $0\leq \Lambda\leq \Id$ such that
    \begin{equation}
        \label{eq:quantum-equality-accept}
        \tr\parens*{\Lambda(\rho\otimes\rho)}=1
    \end{equation}
    for every pure state $\rho$ on $\calH$, and
    \begin{equation}
        \label{eq:quantum-equality-reject}
        \tr\parens*{\Lambda(\rho\otimes\sigma)}=0
    \end{equation}
    for every pair of orthogonal pure states $\rho,\sigma$ with $\tr(\rho\sigma)=0$.
\end{lemma}

Here, \Cref{eq:quantum-equality-accept,eq:quantum-equality-reject} formalize properties (1) and (2) above, respectively.
The proof is similar to those of prior no-go theorems~\cite{WZ82,Dieks82,BCJ03,JAC04}, relying on the linearity of quantum operators. We include it in \Cref{sub:proof_no_quantum_equal} for completeness.

\Cref{lmm:obstruction} indicates that
direct self-composition encounters a no-cloning-style obstruction in the quantum setting.
To illustrate how \Cref{lmm:obstruction} rules out natural candidates for a quantum assertion comparing $q$ and $q'$, let $\ket{\pm}=\frac{1}{\sqrt{2}}(\ket{0}\pm\ket{1})$ and consider the following two examples:
\begin{itemize}
    \item
        \textit{Comparison in the computational basis}: $\Lambda_1=\ket{0,0}\!\bra{0,0}+\ket{1,1}\!\bra{1,1}$ violates \Cref{eq:quantum-equality-accept} for $\rho=\ket{+}\!\bra{+}$, since ${\tr(\Lambda_1(\rho\otimes\rho))}=\frac{1}{2}$.
        Much like copying with a CNOT gate, $\Lambda_1$ satisfies \Cref{eq:quantum-equality-accept,eq:quantum-equality-reject} for computational-basis states but not for their superpositions.
    \item
        \textit{Comparison via the swap test}~\cite{BCWdW01}: $\Lambda_2=\frac{1}{2}(\Id+\mathit{SWAP})$ violates \Cref{eq:quantum-equality-reject} for $\rho=\ket{0}\!\bra{0}$ and $\sigma=\ket{1}\!\bra{1}$, since ${\tr(\Lambda_2(\rho\otimes\sigma))}=\frac{1}{2}$.
\end{itemize}
With $\Lambda_1$ or $\Lambda_2$ in place of $\braces{\overline{l}=\overline{l'}}$, the judgment in \Cref{eq:classical-self-composition} no longer captures the noninterference condition.
To see this, consider a $\Low$ qubit $q$ and a $\High$ bit $x$.
The program $P_1\equiv\mathbf{if}\ x=1\ \mathbf{then}\ Z[q]$ violates \textbf{NI}: for $q$ initially in state $\ket{+}$, the output state of $q$ is $\ket{+}$ or $\ket{-}$, depending on $x$.
Yet the judgment $\braces*{\Lambda_1}\ P_1;\ P_1'\ \braces*{\Lambda_1}$ holds, as $\Lambda_1$ commutes with $Z$ gates.
Conversely, the program $P_2\equiv x:=\mathbf{meas}[q]$ satisfies \textbf{NI}, but the judgment $\braces*{\Lambda_2}\ P_2;\ P_2'\ \braces*{\Lambda_2}$ does not hold.
Indeed, $\Lambda_2$ is satisfied when $q$ and $q'$ are both initially in state $\ket{+}$, but not by their output states, which are both $\frac{1}{2}\Id$.

To formalize and verify secure information flow in quantum programs, we therefore develop a new technique, called self-conjugation, that exploits distinctively quantum properties.
Rather than comparing the outputs of two executions, it composes a program with its \textit{reverse}.
We will introduce it in \Cref{sub:verification_by_self_conjugation}.

\subsection{Background on Quantum Hoare Logic}
\label{sub:qhl}

In this section, we briefly introduce a Hoare-style logic for the quantum circuit language (see \Cref{def:syntax_for_circuits}).
Here, the circuit language suffices for our verification framework, even though the source programs use the more general language of \Cref{def:syntax_for_programs}. As will be explained in \Cref{sub:verification_by_self_conjugation}, we can reduce the \textbf{NI} condition (see \Cref{def:noninterference}) of a program $P$ to a Hoare-style correctness condition for a derived quantum circuit $P'$.
In this setting, we further adapt ideas from a projective variant~\cite{ZYY19} of quantum Hoare logic~\cite{Ying11}.

We start with quantum Hoare triples, which specify the correctness of quantum circuits.
For a set $\Gamma$ of quantum states over quantum variables $\overline{q}$,
we write $\qvar(\Gamma)=\overline{q}$.

\begin{definition}[Syntax of quantum Hoare triples]
    \label{def:qht}
    A quantum Hoare triple for a quantum circuit $C$ has the form
    \begin{equation}
        \label{eq:quantum-Hoare-triple}
        \braces*{\Gamma}\ C\ \braces*{\Gamma'},
    \end{equation}
    where the quantum assertions $\Gamma$ and $\Gamma'$ are sets of quantum states with $\qvar(C)\subseteq \qvar(\Gamma)=\qvar(\Gamma')$.
\end{definition}

As in the classical setting, the quantum Hoare triple in~\Cref{eq:quantum-Hoare-triple} means that whenever we start with an input pure state $\ket{\psi}$ to $C$ that satisfies the precondition $\Gamma$, the output quantum state $\ket{\psi'}$ obtained after executing $C$ satisfies the postcondition $\Gamma'$.
We formalize this as follows.

\begin{definition}[Semantics of quantum Hoare triples]
    \label{def:qht-semantics}
    A quantum Hoare triple $\braces*{\Gamma}\ C\ \braces*{\Gamma'}$ is true, denoted as 
    \begin{equation*}
        \models \braces*{\Gamma}\ C\ \braces*{\Gamma'},
    \end{equation*}
    iff for any input state $\ket{\psi}$ to $C$ that satisfies $\ket{\psi} \models \Gamma$
    and $(C,\ket{\psi})\rightarrow^*(\downarrow,\ket{\psi'})$ (see \Cref{def:operational_semantics_circ}), we have $\ket{\psi'}\models \Gamma'$.
    Here, the satisfaction $\ket{\psi}\models \Gamma$ is interpreted as:
    \begin{equation*}
        \ket{\psi}\models \Gamma \Leftrightarrow 
        \begin{cases}
            \ket{\psi}=\ket{\phi}, & \Gamma=\braces*{\ket{\phi}},\\
            \ket{\psi}\in \Span(\Gamma), & o.w.
        \end{cases}
    \end{equation*}
\end{definition}

In \Cref{def:qht-semantics},
the interpretation of $\ket{\psi}\models \Gamma$ deliberately distinguishes the cases $\abs*{\Gamma}=1$ and $\abs*{\Gamma}>1$.
When $\abs*{\Gamma}=1$, the assertion $\Gamma$ specifies a single state vector, which carries global phase information; for example, $\ket{\psi}\neq e^{i\theta}\ket{\psi}$ whenever $\theta\neq 2k\pi$. When $\abs*{\Gamma}>1$, $\Gamma$ instead specifies a projective subspace $\Span(\Gamma)$, which does not carry global phase information.
This distinction is necessary because the semantics of the $\mathbf{qif}$ statement is sensitive to relative phases between quantum states (see the (QIF) rule in \Cref{fig:operational-semantics-circuit} and discussion there).
Projections alone (as used in~\cite{ZYY19}) cannot capture such phase information and are therefore insufficient for reasoning about $\mathbf{qif}$ statements.

It is also worth pointing out that \Cref{def:qht-semantics} does not distinguish between partial correctness and total correctness, because every quantum circuit in the language (\Cref{def:syntax_for_circuits}) terminates on every input state. 

\begin{figure}
    \begin{align*}
        &\textbf{\textup{(SK-P)}}\quad \braces*{\Gamma}\ \mathbf{skip}\ \braces*{\Gamma}\qquad
        \textbf{\textup{(SC-P)}}\quad \frac{\braces*{\Gamma}\ C_1\ \braces*{\Gamma'}\quad \braces*{\Gamma'}\ C_2\ \braces*{\Gamma''}}
        {\braces*{\Gamma}\ C_1;C_2\ \braces*{\Gamma''}}\qquad
        \textbf{\textup{(GA-P)}}\quad \braces*{\Gamma}\ U[\overline{q}]\ \braces*{U_{\overline{q}}(\Gamma)}\\
        &\textbf{\textup{(QIF-P)}}\quad \frac{\braces*{\ket{\psi_j}}\ C_j\ \braces{\ket{\psi_j'}}\quad (j=0,1)\qquad \abs*{\alpha}^2+\abs*{\beta}^2=1}
        {\braces*{\alpha\ket{0}_q\ket{\psi_0}+\beta\ket{1}_q\ket{\psi_1}}\ \mathbf{qif}[q](\ket{0}\rightarrow C_0)\square(\ket{1}\rightarrow C_1)\ \braces*{\alpha\ket{0}_q\ket{\psi_0'}+\beta\ket{1}_q\ket{\psi_1'}}}
           \\
        &\textbf{\textup{(Consequence)}}\quad \frac{\Lambda\models\Gamma\quad \braces*{\Gamma}\ C\ \braces*{\Gamma'}\quad \Gamma'\models \Lambda'}{\braces*{\Lambda}\ C\ \braces*{\Lambda'}}\qquad
        \textbf{\textup{(Conjunction)}}\quad \frac{\braces*{\Gamma_j}\ C\ \braces{\Gamma_j'}\quad (j=0,1)}
        {\braces*{\Gamma_0\sqcap \Gamma_1}\ C\ \braces*{\Gamma_0'\sqcap \Gamma_1'}}\\
        &\textbf{\textup{(Disjunction)}}\quad \frac{\braces*{\Gamma_j}\ C\ \braces{\Gamma_j'}\quad (j=0,1)}
        {\braces*{\Gamma_0\sqcup \Gamma_1}\ C\ \braces*{\Gamma_0'\sqcup \Gamma_1'}}\qquad
        \textbf{\textup{(Frame)}}\quad \frac{\braces*{\Gamma}\ C\ \braces*{\Gamma'}\quad \qvar(\Lambda)\cap \qvar(C)=\emptyset}
        {\braces*{\Gamma\otimes \Lambda}\ C\ \braces*{\Gamma'\otimes \Lambda}}
    \end{align*}
    \caption{Proof rules for the quantum circuit language defined in \Cref{def:syntax_for_circuits}.}
    \label{fig:proof_system}
\end{figure}

We now introduce a simple proof system for verifying the correctness of a quantum circuit. The system consists of the inference rules shown in~\Cref{fig:proof_system}.
We use the following notation
\begin{equation}
    \label{eq:wedge-vee}
    \begin{split}
    &U(\Gamma)=\braces*{U\ket{\psi}:\ket{\psi}\in \Gamma},\qquad
    \Gamma\otimes \Lambda = \braces*{\ket{\psi}\ket{\phi}:\ket{\psi}\in \Gamma,\ket{\phi}\in \Lambda},\\
    &\Gamma_0\sqcap\Gamma_1=\Span(\Gamma_0)\cap\Span(\Gamma_1),\qquad
    \Gamma_0\sqcup \Gamma_1=\overline{\operatorname{span}(\Gamma_0\cup \Gamma_1)},\\
    \end{split}
\end{equation}
where $\operatorname{span}(T)$ denotes the subspace spanned by the set $T$,
$\overline{T}$ denotes the closure of $T$,
and we identify a subspace with the set of normalized pure states in it by a slight abuse of notation.
Moreover, the entailment $\Lambda\models \Gamma$ in (Consequence) is understood as:
\begin{enumerate}
    \item 
        If $\Gamma=\braces*{\ket{\psi}}$, then $\Lambda=\Gamma$; and
    \item
        Otherwise, $\Span(\Lambda)\subseteq\Span(\Gamma)$.
\end{enumerate}

We make a few remarks.
The rules (SK-P), (SC-P), (GA-P) in \Cref{fig:proof_system} are self-explanatory.
The structural rules (Frame), (Conjunction), (Disjunction) are adapted from~\cite{ZYY19}.
In the (QIF-P) rule, each assertion contains a single pure state and therefore carries phase information, as required by the semantics of the $\mathbf{qif}$ statement (see (QIF) in \Cref{fig:operational-semantics-circuit}).
For the same reason, case (1) in the definition of $\Lambda\models \Gamma$ above is sensitive to the phase information (see also the definition of $\ket{\psi}\models \Gamma$ in \Cref{def:qht-semantics}), which is important when (QIF-P) and (Consequence) are used together.

As in the standard presentations of classical Hoare logic~\cite{Cook78,ADO09} and quantum Hoare logic~\cite{Ying11,ZYY19}, the entailments $\Lambda\models\Gamma$ are discharged as side conditions of (Consequence).
By cases (1) and (2) above, every entailment in our examples (see \Cref{sec:illustrative_examples,sec:more_examples}) is a finite linear-algebra computation.

The soundness of the proof rules in \Cref{fig:proof_system} follows routinely from the semantics.
A complete proof is deferred to \Cref{sub:soundness}.

\subsection{Verification by Self-Conjugation}
\label{sub:verification_by_self_conjugation}

In this section, we show how to verify secure information flow in quantum programs using quantum Hoare logic. 
In particular, we reduce the noninterference-based security condition \textbf{NI} of a quantum program $P$
to a Hoare triple $\braces*{\Gamma_1} P'\braces*{\Gamma_2}$, where
$P'$ is a quantum circuit derived from $P$.

We begin by defining the dilation of a quantum program, inspired by the Stinespring dilation of a quantum channel.
A list of classical variables $\overline{c}$ can be lifted to a list of quantum variables of the corresponding lifted types.
When the context is clear, we also use $\overline{c}$ to denote these lifted quantum variables;
and when the distinction needs to be emphasized, we write $\lift(\overline{c})$.

\begin{definition}[Dilation]
    \label{def:dilation}
    A quantum circuit $Q$ (see \Cref{def:syntax_for_circuits}) is called a dilation of another program $P$ iff $\qvar(Q)=\lift(\cvar(P))\uplus \qvar(P)\uplus \overline{ev}$ for some environmental quantum variables $\overline{ev}$,
    and for every input state $s$ to $P$ with $\var(s)\cap \overline{ev}=\emptyset$:
    \begin{equation*}
        \tr_{\overline{ev}}\circ\Bracks*{Q}\parens*{s\otimes \ket{\overline{0}}\!\bra{\overline{0}}_{\overline{ev}}}=\Bracks*{P}\parens*{s},
    \end{equation*}
    where we implicitly identify the lifted variables $\lift(\cvar(P))$ on the LHS with classical variables $\cvar(P)$ on the RHS.
\end{definition}

\begin{figure}[h]
    \centering
    \includegraphics[width=0.62\textwidth]{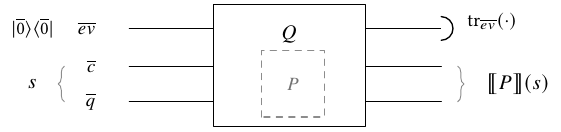}
    \caption{Illustration of dilation. 
    Here, $Q$ is a dilation of $P$. $\overline{c}=\lift(\cvar(P))$, $\overline{q}=\qvar(P)$, and $\overline{ev}$ are environmental quantum variables. Whenever the input state of $\overline{ev}$ is $\ket{\overline{0}}\!\bra{\overline{0}}$, the output state of $\overline{c},\overline{q}$ is $\Bracks*{P}(s)$.
    }
    \label{fig:dilation_illustrate}
\end{figure}

\begin{algorithm}
    \caption{Finding a dilation of a quantum program.}
    \label{alg:dilation}
    \begin{algorithmic}[1]
        \Require $\overline{ev}=\emptyset$
        \Function{$\dil$}{$P$}
        \If {$P\equiv \mathbf{skip}$} \Return $\mathbf{skip}$.
        \ElsIf {$P\equiv C$} \Return $C$.
        \ElsIf {$P\equiv P_1;P_2$} \Return $\dil(P_1);\dil(P_2)$.
        \ElsIf {$P\equiv x:=e$}
            \State Lift $x$ to a quantum variable
            \State Add to $\overline{ev}$
            a fresh quantum variable $x'$ with the same type as $x$
            \State %
            Suppose $e'=e[x\leftarrow x']$, $\overline{u}$ is a list of variables in $\var(e')$ and $C_{e'}$ is a reversible circuit such that $C_{e'}[x,\overline{u}]$ computes the value of $e'$ into $x$
            \State \Return $\mathit{SWAP}[x,x']; C_{e'}[x,\overline{u}]$
        \ElsIf {$(P\equiv \mathbf{if}\ y\ \mathbf{then}\ P_1\ \mathbf{else}\ P_2)\wedge (\textup{$y$ is a bit variable})\wedge (y\notin \var(P_1)\cup \var(P_2))$}
            \State Lift $y$ to a qubit variable
            \State \Return $\mathbf{qif}[y](\ket{0}\rightarrow \dil(P_2))\square(\ket{1}\rightarrow \dil(P_1))$
        \ElsIf {$P\equiv \mathbf{if}\ b\ \mathbf{then}\ P_1\ \mathbf{else}\ P_2$}
            \State Add to $\overline{ev}$ a fresh bit variable $y$
            \State \Return $\dil(y:=b; \mathbf{if}\ y\ \mathbf{then}\ P_1\ \mathbf{else}\ P_2)$
        \ElsIf {$P\equiv q:=\ket{e}$}
            \State Add to $\overline{ev}$ a fresh quantum variable $q'$ with the same type as $q$ 
            \State Suppose $\overline{u}$ is a list of variables in $\var(e)$ and $C_{e}$ is a reversible circuit such that $C_{e}[q',\overline{u}]$ computes the value of $e$ into $q'$
            \State \Return $C_{e}[q',\overline{u}];\mathit{SWAP}[q,q']$
        \ElsIf {$P\equiv x:=\mathbf{meas}[q]$} 
            \State Lift $x$ to a quantum variable
            \State Add to $\overline{ev}$ fresh quantum variables $x'$ and $q'$ with the same types as $x$ and $q$, respectively
                \State \Return $\mathit{SWAP}[x,x']; \mathit{CX}[q,q'];\mathit{CX}[q,x]$
        \EndIf
        \EndFunction
    \end{algorithmic}
\end{algorithm}

\begin{algorithm}
    \caption{Finding the reverse of a quantum circuit.}
    \label{alg:reverse}
    \begin{algorithmic}[1]
        \Function{$\rev$}{$C$}
        \If {$C\equiv \mathbf{skip}$} \Return $\mathbf{skip}$
        \ElsIf {$C\equiv C_1;C_2$} \Return $\rev(C_2);\rev(C_1)$
        \ElsIf {$C\equiv U[\overline{q}]$} \Return $U^\dagger [\overline{q}]$
        \ElsIf {$C\equiv \mathbf{qif}[q](\ket{0}\rightarrow C_0)\square(\ket{1}\rightarrow C_1)$}
        \Return $\mathbf{qif}[q](\ket{0}\rightarrow \rev(C_0))\square(\ket{1}\rightarrow \rev(C_1))$
        \EndIf
        \EndFunction
    \end{algorithmic}
\end{algorithm}
 
Although a quantum program $P$ may admit multiple dilations,
\Cref{alg:dilation} gives a recursive construction of one such dilation,
following the syntax in \Cref{def:syntax_for_programs}.
Here, the construction of the environmental variables $\overline{ev}$ (which are traced out at the end) and corresponding gates can be viewed as an application of the \textit{deferred measurement principle}~\cite{AKN98} that follows the program structure.
We further explain \Cref{alg:dilation} below.
\begin{itemize}
    \item 
    In Line 4, when parsing $P\equiv P_1;P_2$,
    we implicitly split at the leftmost top-level semicolon ``$;$''.
    \item
    In Lines 9, 19, and 23, unitary gate $\mathit{SWAP}$ performs $\ket{a}\ket{b}\mapsto \ket{b}\ket{a}$ and $\mathit{CX}$ performs $\ket{a}\ket{b}\mapsto \ket{a}\ket{b\oplus a}$, where $\oplus$ denotes bitwise XOR.
    \item
    In Line 8, $e'=e[x\leftarrow x']$ denotes the expression obtained from $e$ by replacing $x$ with $x'$.
    \item 
    In Lines 8 and 18, constructing the reversible circuit $C_{e}$ (or $C_{e'}$) is standard~\cite{Landauer61,Bennett73}.
    Suppose $\overline{u}$ is a list of variables occurring in an expression $e$.
    Then, %
    $C_e$ performs
    $\ket{a}\ket{\sigma(u)}\mapsto \ket{a\oplus \sigma(e)}\ket{\sigma(u)}$,
    where $\sigma$ is a classical state.
    \item
    In the assignment $x:=e$, erasing the old value of $x$ is in general not reversible.
    Thus, $\mathit{SWAP}[x,x']$ in Line 9 first moves this value into the fresh environmental variable $x'$ and leaves $x$ in the initial state $\ket{0}$ of $x'$.
    Then $C_{e'}$ reads the old value of $x$ from $x'$ (see Line 8) and computes the value of $e$ into $x$.
    Tracing out $x'$ recovers the semantics of assignment (see \Cref{lmm:dil-is-dilation}).
    Similarly, Lines 19 and 23 move the old states of $q$ and $x$, respectively, into fresh environmental variables.
    \item
    For an if-statement whose guard is already a bit $y$ with
    $y\notin \var(P_1)\cup\var(P_2)$, \Cref{alg:dilation} lifts $y$ directly
    and returns a $\mathbf{qif}$ statement guarded by $y$.
    For a general Boolean guard $b$, \Cref{alg:dilation} first stores $b$ in a fresh
    bit $y$ and then applies the preceding case. In either case, the resulting $\mathbf{qif}$ statement satisfies $y\notin \qvar(\dil(P_1))\cup \qvar(\dil(P_2))$ and is therefore well-defined. Upon termination of~\Cref{alg:dilation}, $\overline{ev}$ contains only quantum variables.
    \item
    Every fresh variable introduced by \Cref{alg:dilation} is disjoint from $\var(P)$ and the current set $\overline{ev}$ of environmental variables.
\end{itemize}

The following lemma follows by induction on the structure of quantum programs (see \Cref{def:syntax_for_programs}).
The proof is tedious and thus deferred to \Cref{sub:proof_dilation}.

\begin{lemma}
    \label{lmm:dil-is-dilation}
    For every quantum program $P$, the program $\dil(P)$ produced by \Cref{alg:dilation} is a dilation of $P$ with respect to \Cref{def:dilation}.
\end{lemma}

Further, we define the \textit{canonical dilation} $\cdil(P)$ of a program $P$.

\begin{definition}[Canonical dilation]
    \label{def:cdil}
    The canonical dilation of a quantum program $P$ is the quantum circuit:
    \begin{equation}
        \label{eq:cdil}
        \cdil(P)\equiv \mathit{CX}[\overline{c},\overline{c^\circ}];\ \dil(P),
    \end{equation}
    where $\overline{c}=\lift(\cvar(P))$ are quantum variables lifted from $\cvar(P)$ in the construction of $\dil(P)$ by \Cref{alg:dilation},
    and $\overline{c^\circ}$ are fresh quantum variables of the same types as $\overline{c}$.
    The environmental variables of $\cdil(P)$ are $\overline{ev},\overline{c^\circ}$, where $\overline{ev}$ denotes the environmental variables introduced by $\dil(P)$.
\end{definition}

\begin{figure}[h]
    \centering
    \includegraphics[width=0.62\textwidth]{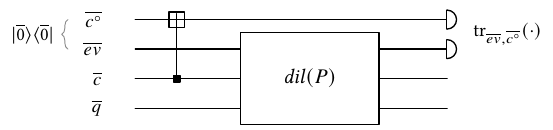}
    \caption{Illustration of canonical dilation.
    Here, the square-shaped controlled gate corresponds to $\mathit{CX}[\overline{c},\overline{c^\circ}]$.
    }
    \label{fig:cdilation_illustrate}
\end{figure}

\Cref{fig:cdilation_illustrate} illustrates the canonical dilation.
Intuitively, $\mathit{CX}[\overline{c},\overline{c^\circ}]$ and the final tracing out of $\overline{c^\circ}$ serve to ``dephase'' the initial state of the quantum variables $\overline{c}$, because they are lifted from classical variables $\cvar(P)$.
In this way, the state of $\overline{c}$ is made effectively classical before executing $\dil(P)$.

Next, we define the reverse of a quantum circuit $C$, exploiting the fact that its denotational semantics $\Bracks*{C}$ is a \textit{unitary} channel and therefore has an inverse $\Bracks*{C}^\dagger$.

\begin{definition}[Reverse]
    \label{def:reverse}
    A quantum circuit $D$ is called a reverse of $C$ iff $\qvar(D)=\qvar(C)$ and $\Bracks*{D}=\Bracks*{C}^{\dagger}$.
\end{definition}

\Cref{alg:reverse} gives a recursive construction of a reverse of any quantum circuit $C$.
As before, when parsing $C\equiv C_1;C_2$,
we implicitly split at the leftmost top-level semicolon ``$;$''.
The following lemma follows by induction on the structure of quantum circuits (see \Cref{def:syntax_for_circuits}).
We omit the proof for simplicity.

\begin{lemma}
    For every quantum circuit $C$, the quantum circuit $\rev(C)$ produced by \Cref{alg:reverse} is a reverse of $C$ with respect to \Cref{def:reverse}.
\end{lemma}

Now we are ready to reduce the security condition \textbf{NI} to a formal proof obligation in quantum Hoare logic.
For convenience, let us use $\ket{\overline{\Phi}}_{\overline{q},\overline{q'}}$
to denote the tensor product of maximally entangled states
$\bigotimes_{j=1}^n \ket{\Phi}_{q_j,q_j'}$ for quantum variables $\overline{q}=q_1,\ldots,q_n$ and $\overline{q'}=q_1',\ldots,q_n'$, where $q_j,q_j'$ are of the same types.
For a pure state $\ket{\psi}$, let $\ket{\psi}^\bot=\braces*{\ket{\phi}:\braket{\phi|\psi}=0}$.
When the context is clear, we omit the braces $\braces{}$ around an assertion containing only a single quantum state in preconditions and postconditions.

\begin{theorem}
    \label{thm:test-NI}
    Consider a quantum program $P$ with $\High$ variables $\hvar(P)=\overline{h}$ and $\Low$ variables $\lvar(P)=\overline{l}$.
    Define a quantum circuit
    \begin{equation}
        \label{eq:test}
        \mathit{Test}\equiv \cdil(P); \mathit{SWAP}[\overline{l},\overline{l'}]; \rev\parens*{\cdil(P)},
    \end{equation}
    where $\cvar(P)$ are implicitly lifted to quantum variables, $\overline{l'}$ are fresh quantum variables with the same types as $\overline{l}$, and $\cdil(P)$ is defined in \Cref{def:cdil}.
    Suppose $\cdil(P)$ has fresh environmental variables $\overline{ev}$.
    Let $\overline{h'}$ be fresh quantum variables with the same types as $\overline{h}$.
    Then, $P$ satisfies \textbf{NI} iff 
    the following judgment holds:
    \begin{equation}
        \label{eq:test-triple}
            \models \braces*{\ket{\overline{0}}_{\overline{ev}}\otimes \ket{\overline{\Phi}}_{\overline{h},\overline{h'}}\otimes \calH_{\overline{l},\overline{l'}}}
            \ \mathit{Test} 
            \ \braces*{
                \parens*{\ket{\overline{0}}_{\overline{ev}}\otimes
                \ket{\overline{\Phi}}_{\overline{h},\overline{h'}}\otimes
                \calH_{\overline{l},\overline{l'}}}
                \sqcup\parens*{\ket{\overline{0}}_{\overline{ev}}^\bot\otimes
                \calH_{\overline{h},\overline{h'},\overline{l},\overline{l'}}}},
    \end{equation}
    where $\ket{\overline{\Phi}}$ and $\ket{\overline{0}}^\bot$ are defined above,
    $\sqcup$ and $\otimes$ are defined in \Cref{eq:wedge-vee}, and Hilbert spaces $\calH$ are identified with quantum assertions by a slight abuse of notation.
\end{theorem}

\begin{figure}[h]
    \centering
    \includegraphics[width=0.95\textwidth]{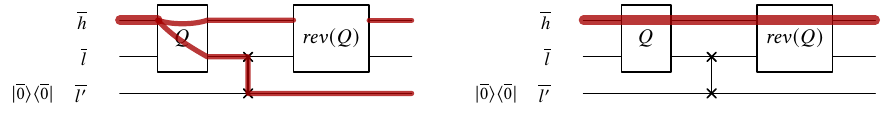}
    \caption{Intuition for self-conjugation in the simple case where $P$ is a circuit $Q$.
    Here, program $\mathit{Test}$ is visualized, with $\overline{h}=\hvar(Q)$ and $\overline{l}=\lvar(Q)$.
    The red flow indicates confidential information. The LHS illustrates an insecure $Q$, where confidential information stored in $\overline{h}$ leaks to $\overline{l}$ and is swapped to $\overline{l'}$, so $\rev(Q)$ cannot fully recover the original state of $\overline{h}$. The RHS illustrates a secure $Q$, where $\rev(Q)$ can fully recover the confidential information because it has not leaked.
    }
    \label{fig:selfconj_illustrate}
\end{figure}

In a sentence, self-conjugation composes $P$ with its reverse, swapping the $\Low$ variables out in between (see \Cref{eq:test}): whatever has flowed from the $\High$ into the $\Low$ variables cannot flow back, so the initial state of the $\High$ variables is restored exactly when nothing has flowed.
\Cref{fig:selfconj_illustrate} illustrates this intuition.
By \Cref{lmm:obstruction}, no quantum assertion can play the role of $\braces{\overline{l}=\overline{l'}}$ in \Cref{eq:classical-self-composition}.
In contrast, the restoration of the $\High$ variables can be specified by quantum assertions, as in \Cref{eq:test-triple}.

It is important to note that verification does not require executing the program $\mathit{Test}$ on a quantum computer. It suffices to use the proof rules in \Cref{sub:qhl} to derive the target Hoare triple in~\Cref{eq:test-triple} on a classical computer.

\begin{proof}[Proof sketch of \Cref{thm:test-NI}]
We only show a brief sketch of the proof idea. The full proof is rather involved and deferred to \Cref{sub:proof_self_conjugation}.
For illustration, we first consider the special case where $P$ is already a quantum circuit $Q$. Then, $\cdil(Q)=Q$ and hence $\mathit{Test}\equiv Q; \mathit{SWAP}[\overline{l},\overline{l'}]; \rev(Q)$, as illustrated in \Cref{fig:selfconj_illustrate}. Intuitively, if $Q$ violates \textbf{NI}, then confidential information initially stored in $\overline{h}$ can flow into $\overline{l}$. After $\mathit{SWAP}[\overline{l},\overline{l'}]$, this information is moved into $\overline{l'}$, which lies outside $\var(Q)$. Consequently, applying $\rev(Q)$ cannot recover the initial state of $\overline{h}$. Conversely, if $Q$ satisfies \textbf{NI}, then $\mathit{SWAP}[\overline{l},\overline{l'}]$ does not remove confidential information needed by $\rev(Q)$ to restore $\overline{h}$.

For a general program $P$, the situation is more complex. The canonical dilation $\cdil(P)$ (see \Cref{def:cdil}) now plays the role of the circuit $Q$.
Here, $\dil(P)$ makes $P$ reversible, lifting the classical variables $\cvar(P)$ to quantum variables $\overline{c}$.
The ``dephasing'' $\mathit{CX}[\overline{c},\overline{c^\circ}]$ in $\cdil(P)$ makes the initial state of $\overline{c}$ effectively classical, so that the dilated program behaves the same as $P$ (see the explanation after \Cref{def:cdil}).
Without it, only the direction \Cref{eq:test-triple}~$\Rightarrow$~\textbf{NI} would hold.
Unlike in the special case above, we only care about executions of $\cdil(P)$ whose environmental variables $\overline{ev}$ are initially in state $\ket{\overline{0}}$ because of \Cref{def:dilation}.
Similarly, we only care about executions of $\rev(\cdil(P))$ for which $\overline{ev}$ ends in state $\ket{\overline{0}}$.

The preceding intuition suggests the following condition: whenever an execution of $\mathit{Test}$ starts and ends with $\overline{ev}$ in state $\ket{\overline{0}}$, the initial state of $\overline{h}$ should be restored. By introducing reference quantum variables $\overline{h'}$, restoration of $\overline{h}$ can be expressed as preservation of the maximally entangled state $\ket{\overline{\Phi}}_{\overline{h},\overline{h'}}$.
This leads to the term $\ket{\overline{0}}_{\overline{ev}}\otimes \ket{\overline{\Phi}}_{\overline{h},\overline{h'}}$ in both the precondition and postcondition of \Cref{eq:test-triple}.
The second part of the postcondition accounts for executions in which $\overline{ev}$ ends in a state orthogonal to $\ket{\overline{0}}$, for which no recovery condition is imposed on $\overline{h}$. 

\end{proof}

\section{Illustrative Examples}
\label{sec:illustrative_examples}

In this section, we present several illustrative examples showing how the framework developed in the preceding sections can be used to specify and verify secure information flow in quantum programs.
The examples are intentionally compact enough to display the resulting proof obligations explicitly,
while still illustrating natural security properties arising in quantum cryptography, quantum error correction, and multiparty quantum communication.
They also show how the verification follows the program structure, with local reasoning via the (Frame) rule and compositional reasoning via \Cref{thm:compose}.

The first example illustrates how we can verify the security of the quantum one-time pad from \Cref{exp:qotp} using the self-conjugation technique developed in \Cref{sec:verification}.

\begin{example}[Security verification of quantum one-time pad]
    \label{exp:veri-qotp}
    Consider the program
    \begin{equation*}
        P\equiv \mathit{KeyGen};\mathit{Enc}
    \end{equation*}
    from~\Cref{exp:qotp}.
    Since $x$ and $z$ are bit variables, we identify the guards ``$x=1$'' and ``$z=1$'' with $x$ and $z$, respectively.
    We first construct the corresponding program $\mathit{Test}$ (see \Cref{eq:test}) for $P$.
    Applying \Cref{alg:dilation,def:cdil} yields:
    \begin{equation*}
        \begin{aligned}
            \cdil(P)\equiv\ & \mathit{CX}[x,x^\circ]; \ \mathit{CX}[z,z^\circ];\\
            &\mathit{SWAP}[r_x,r_x'];\ \mathit{H}[r_x];\ \mathit{SWAP}[x,x'];\ \mathit{CX}[r_x,r_x''];\ \mathit{CX}[r_x,x];\\
            \qquad & \mathit{SWAP}[r_z,r_z'];\ \mathit{H}[r_z];\ \mathit{SWAP}[z,z'];\ \mathit{CX}[r_z,r_z''];\ \mathit{CX}[r_z,z];\\
            \qquad & \mathit{SWAP}[q,q'];\ \mathit{SWAP}[m,q];\\
            \qquad & \mathbf{qif}[x](\ket{0}\rightarrow \mathbf{skip})\square(\ket{1}\rightarrow \mathit{X}[q]);\\
            \qquad & \mathbf{qif}[z](\ket{0}\rightarrow \mathbf{skip})\square(\ket{1}\rightarrow \mathit{Z}[q]).
        \end{aligned}
    \end{equation*}
    where $r_x',r_x'',x',r_z',r_z'',z',q',x^\circ,z^\circ$ are fresh environmental variables.

    Further applying \Cref{alg:reverse} to the above circuit gives:
    \begin{equation*}
        \begin{aligned}
            \rev\parens*{\cdil(P)}\equiv\ & \mathbf{qif}[z](\ket{0}\rightarrow \mathbf{skip})\square(\ket{1}\rightarrow \mathit{Z}[q]);\\
            \qquad & \mathbf{qif}[x](\ket{0}\rightarrow \mathbf{skip})\square(\ket{1}\rightarrow \mathit{X}[q]);\\
            \qquad & \mathit{SWAP}[m,q];\ \mathit{SWAP}[q,q'];\\
            \qquad & \mathit{CX}[r_z,z];\ \mathit{CX}[r_z,r_z''];\ \mathit{SWAP}[z,z'];\ \mathit{H}[r_z];\ \mathit{SWAP}[r_z,r_z'];\\
            \qquad & \mathit{CX}[r_x,x];\ \mathit{CX}[r_x,r_x''];\ \mathit{SWAP}[x,x'];\ \mathit{H}[r_x];\ \mathit{SWAP}[r_x,r_x'];\\
            \qquad & \mathit{CX}[z,z^\circ];\ \mathit{CX}[x,x^\circ].
        \end{aligned}
    \end{equation*}

    According to \Cref{eq:test}, the corresponding $\mathit{Test}$ circuit for program $P$ is
    \begin{equation}
        \label{eq:qotp-test}
        \mathit{Test}\equiv \cdil(P);\mathit{SWAP}[q,q^\star];\rev\parens*{\cdil(P)}.
    \end{equation}
    where for clarity we use $q^\star$ to denote the fresh variable with the same type as $q$.

    To show that $P$ satisfies \textbf{NI}, our goal is to establish the judgment in \Cref{eq:test-triple}.
    This can be achieved by using the proof rules introduced in \Cref{sub:qhl}, as shown in \Cref{fig:qotp_proof}.

    \begin{figure}
        \centering
        \includegraphics[width=\textwidth]{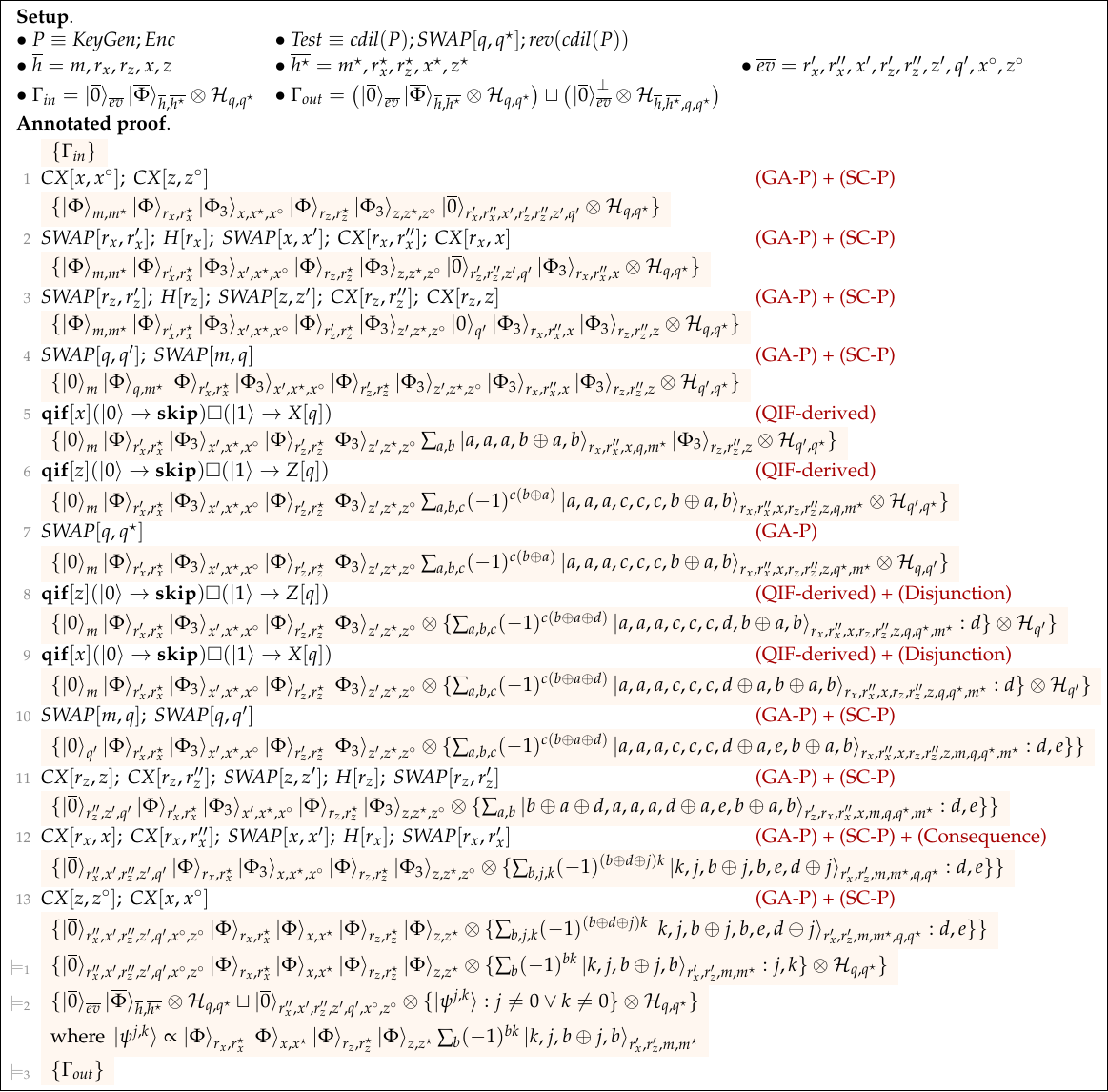}
        \caption{Annotated proof of information-flow security for the quantum one-time pad.}
        \label{fig:qotp_proof}
    \end{figure}

    Let us explain the proof in \Cref{fig:qotp_proof} in more detail.
    The proof is annotated with intermediate quantum assertions (see \Cref{sub:qhl}); or equivalently, decomposed into a sequence of intermediate Hoare triples using the (SC-P) rule. Whenever needed, the (Frame) rule is applied implicitly to handle variables untouched by the current program segment, thereby enabling \textit{local reasoning}.
    On the right, we also indicate the proof rules used to derive the Hoare triple of each program segment.
    We further explain the main steps below.
    \begin{itemize}
        \item 
            \textbf{Assertions}: 
            \begin{enumerate}
                \item
                    All indices $a,b,c,d,e,j,k$ range over $\braces*{0,1}$. $\oplus$ denotes binary XOR.
                \item
                    By a slight abuse of notation, we always use $\ket{\Phi}_{q_1,q_2}$ to represent the maximally entangled state $\frac{1}{\sqrt{d}}\sum_{j=0}^{d-1}\ket{j,j}_{q_1,q_2}$ for quantum variables $q_1,q_2$ with $\dim(\calH_{q_1})=\dim(\calH_{q_2})=d$, regardless of the value of $d$.
                    Similarly, we denote $\ket{\Phi_3}_{q_1,q_2,q_3}=\frac{1}{\sqrt{d}}\sum_{j=0}^{d-1} \ket{j,j,j}_{q_1,q_2,q_3}$.
                \item
                    When the context is clear, normalizing factors are omitted from assertions. For example, after Step 5, $\sum_{a,b} \ket{a,a,a,b\oplus a,b}$ actually represents $\frac{1}{2}\sum_{a,b}\ket{a,a,a,b\oplus a,b}$.
            \end{enumerate}             
        \item
            \textbf{Proof rules}:
            In Steps 5,6,8,9, the (QIF-derived) denotes the following proof pattern: (1) apply (QIF-P),
            with the premises for the two $\mathbf{qif}$ branches derived using (SK-P) or (GA-P); (2) apply (Frame) to handle quantum variables untouched; and (3) apply (Consequence) to strengthen the precondition or weaken the postcondition.

            For illustration, let us consider Step 5. Applying (SK-P), (GA-P), and (Frame) yields 
            \begin{equation*}
                \begin{split}
                    &\vdash \braces{\ket{0,0}_{r_x,r_x''}\ket{\Phi}_{q,m^\star}}\ \mathbf{skip}\ \braces{\ket{0,0}_{r_x,r_x''}\ket{\Phi}_{q,m^\star}}\\
                    &\vdash \braces{\ket{1,1}_{r_x,r_x''}\ket{\Phi}_{q,m^\star}}\ X[q]\ \braces{\ket{1,1}_{r_x,r_x''}\sum\nolimits_b \ket{b\oplus 1,b}_{q,m^\star}}.
                \end{split}
            \end{equation*}
            Then, we apply (QIF-P) to obtain
            \begin{equation}
                \label{eq:qotp-step5-inter}
                \begin{split}
                    \vdash&\braces*{\ket{0}_{x}(\ket{0,0}_{r_x,r_x''}\ket{\Phi}_{q,m^\star})+\ket{1}_{x}(\ket{1,1}_{r_x,r_x''}\ket{\Phi}_{q,m^\star})}\\
                    &\qquad \mathbf{qif}[x](\ket{0}\rightarrow \mathbf{skip})\square(\ket{1}\rightarrow \mathit{X}[q])\\
                    &\qquad\qquad\braces*{\ket{0}_{x}(\ket{0,0}_{r_x,r_x''}\ket{\Phi}_{q,m^\star})+\ket{1}_{x}(\ket{1,1}_{r_x,r_x''}\sum\nolimits_b \ket{b\oplus 1,b}_{q,m^\star})}.
                \end{split}
            \end{equation} 
            In \Cref{eq:qotp-step5-inter}, the precondition is $\braces*{\sum_a\ket{a,a,a}_{r_x,r_x'',x}\ket{\Phi}_{q,m^\star}}$.
            By applying (Consequence), the postcondition can be rewritten as $\braces*{\sum_{a,b}\ket{a,a,a,b\oplus a,b}_{r_x,r_x'',x,q,m^\star}}$.
            Finally, applying (Frame) yields the full Hoare triple in Step 5. 
        \item
            \textbf{Further details}: 
            \begin{enumerate}
                \item
                In Step 8, we first apply the (QIF-derived) proof pattern to obtain, for $d\in \braces{0,1}$:
                \begin{equation*}
                    \begin{split}
                        \vdash&\braces*{\sum\nolimits_{a,b,c}(-1)^{c(b\oplus a)}\ket{a,a,a,c,c,c,d,b\oplus a, b}_{r_x,r_x'',x,r_z,r_z'',z,q,q^\star,m^\star}}\\
                        &\qquad\mathbf{qif}[z](\ket{0}\rightarrow \mathbf{skip})\square(\ket{1}\rightarrow Z[q])\\
                        &\qquad\qquad \braces*{\sum\nolimits_{a,b,c}(-1)^{c(b\oplus a\oplus d)}\ket{a,a,a,c,c,c,d,b\oplus a, b}_{r_x,r_x'',x,r_z,r_z'',z,q,q^\star,m^\star}}
                    \end{split}
                \end{equation*}
                Applying (Disjunction) and (Consequence) further leads to 
                \begin{equation*}
                    \begin{split}
                        \vdash&\braces*{\sum\nolimits_{a,b,c}(-1)^{c(b\oplus a)}\ket{a,a,a,c,c,c,b\oplus a, b}_{r_x,r_x'',x,r_z,r_z'',z,q^\star,m^\star}\otimes \calH_q};\\
                        &\qquad\mathbf{qif}[z](\ket{0}\rightarrow \mathbf{skip})\square(\ket{1}\rightarrow Z[q])\\
                        &\qquad\qquad\braces*{\sum\nolimits_{a,b,c}(-1)^{c(b\oplus a\oplus d)}\ket{a,a,a,c,c,c,d,b\oplus a, b}_{r_x,r_x'',x,r_z,r_z'',z,q,q^\star,m^\star}:d}
                    \end{split}
                \end{equation*}
                Finally, using (Frame) yields the full Hoare triple in Step 8. 
                \item
                    Step 9 is analogous to Step 8. 
                \item 
                    In Step 12, we take $j=b\oplus a\oplus d$ and sum over $j$ rather than over $a$.
                \item
                    The three $\models$ after Step 13 exploit (Frame) and (Consequence).
                    The first $\models_1$ uses %
                    \begin{equation*}
                        \begin{split}
                            &\braces*{\sum\nolimits_{b,j,k}(-1)^{(b\oplus d\oplus j)k} \ket{k,j,b\oplus j,b,e,d\oplus j}_{r_x',r_z',m,m^\star,q,q^\star}:d,e}\\
                            \models & \braces*{\sum\nolimits_b (-1)^{bk}\ket{k,j,b\oplus j,b}_{r_x',r_z',m,m^\star}:j,k}\otimes \calH_{q,q^\star}.
                        \end{split}
                    \end{equation*}
                    The second $\models_2$ groups the states in the assertion according to the component $\ket{k,j}_{r_x',r_z'}$, yielding the disjunction $\sqcup$ of two assertions corresponding to $j=0\wedge k=0$ and $j\neq 0\vee k\neq 0$.
                    The third $\models_3$ uses $\braces*{\ket{\overline{0}}_{r_x'',x',r_z'',z',q',x^\circ,z^\circ}\ket{\psi^{j,k}}}\models\ket{\overline{0}}_{\overline{ev}}^\bot\otimes \calH_{\overline{h},\overline{h^\star}}$ for $j\neq 0\vee k\neq 0$.
            \end{enumerate}
    \end{itemize}
\end{example}

The second example concerns the security of logical data in quantum error correction (QEC)~\cite{Gottesman97}.
In fault-tolerant quantum computing, QEC protects logical qubits by encoding them into physical qubits, extracting error syndromes, decoding the syndromes classically, and applying recovery operations to correct physical errors. Recent work on side-channel leakage and hardware security shows that information exposed during QEC and fault-tolerant execution can be security-sensitive. For example, surface-code execution traces can reveal circuit dataflow~\cite{TKW+25}, while error syndromes can be used as hardware fingerprints~\cite{MCM+25}. 
The following example illustrates a basic confidentiality requirement for QEC:
syndrome data accessible to a decoder should reveal information about physical errors, but not confidential logical data.
For readability, we present only the specification of the requirement here
and defer its verification to \Cref{sec:more_examples}.

\begin{example}[Security of logical data in quantum error correction]
    \label{exp:qec}
    For clarity, consider the simplest three-qubit bit-flip code
    with a fixed single-qubit bit-flip error on the second qubit.
    More general error-correcting codes and error patterns can be treated analogously.
    The following program consists of four stages: encoding, error injection, syndrome extraction, and recovery:
    \begin{equation*}
        P\equiv
        \mathit{Enc};\mathit{Err};\mathit{Syn};\mathit{Rec},
    \end{equation*}
    where
    \begin{equation*}
        \begin{aligned}
            \mathit{Enc}\equiv\ & q_2:=\ket{0};\ q_3:=\ket{0};\
            \mathit{CNOT}[q_1,q_2];\ \mathit{CNOT}[q_1,q_3]\\
            \mathit{Err}\equiv\ & \mathit{X}[q_2]\\
            \mathit{Syn}\equiv\ & r_1:=\ket{0};\ r_2:=\ket{0};\\
            &\mathit{CNOT}[q_1,r_1];\ \mathit{CNOT}[q_2,r_1];
             \mathit{CNOT}[q_2,r_2];\ \mathit{CNOT}[q_3,r_2];\\
            &m_1:=\mathbf{meas}[r_1];\ m_2:=\mathbf{meas}[r_2]\\
            \mathit{Rec}\equiv\ &
            \mathbf{if}\ m_1=1\ \mathbf{then}\\
                &\qquad\mathbf{if}\ m_2=1\ \mathbf{then}\ \mathit{X}[q_2]\ \mathbf{else}\ \mathit{X}[q_1]\\
            & \mathbf{else}\
                \mathbf{if}\ m_2=1\ \mathbf{then}\ \mathit{X}[q_3].
        \end{aligned}
    \end{equation*}
    Here, qubits $q_1,q_2,q_3$ form the encoded block, with $q_1$ initially holding the logical qubit.
    The qubits $r_1,r_2$ are syndrome ancillas, and classical bits $m_1,m_2$ store the syndrome measurement outcomes.
    The gate $\mathit{X}[q_2]$ injects a bit-flip error on the second qubit $q_2$. 

    We classify $q_1,q_2,q_3,r_1,r_2$ as $\High$ variables
    and $m_1,m_2$ as $\Low$ variables.
    Under this labeling, the information-flow security of program $P$ guarantees that the syndrome measurement outcomes $m_1,m_2$ do not leak confidential logical quantum data initially stored in $q_1$.
\end{example}

The third example concerns the security of sender and receiver identities in anonymous entanglement generation.
In multiparty quantum communication, anonymous entanglement generation establishes a Bell pair between two selected parties while hiding their identities as the sender and receiver~\cite{CW05,BBF+07}.
Such protocols use shared multipartite entanglement and classical broadcast messages to coordinate local corrections.
The following example illustrates a basic anonymity requirement for anonymous entanglement generation:
the broadcast messages should not reveal the identities of the sender and receiver.
For readability, the verification is deferred to \Cref{sec:more_examples}.

\begin{example}[Security of anonymous entanglement generation]
    \label{exp:aeg}
    For clarity, 
    we consider a simplified four-party protocol in the honest-party setting and focus on anonymity with respect to the broadcast messages.
    Let $\mathit{id}$ be a classical variable of type
    \begin{equation}
        \label{eq:id-type}
        \braces*{(s,r):s,r\in\braces*{1,2,3,4},\ s\ne r},
    \end{equation}
    and let $s(\mathit{id})$ and $r(\mathit{id})$ denote the first and the second components of $\mathit{id}$, respectively.
    The following program consists of two stages: transmission and recovery:
    \begin{equation*}
        P\equiv \mathit{Tran};\mathit{Rec},
    \end{equation*}
    where
    \begin{equation*}
        \begin{aligned}
            \mathit{Tran}\equiv\ &
            u_1,u_2,u_3,u_4:=
            \braces*{(1/2,0),(1/2,1)},
            \braces*{(1/2,0),(1/2,1)},
            \braces*{(1/2,0),(1/2,1)},
            \braces*{(1/2,0),(1/2,1)};\\
            &q_1,q_2,q_3,q_4:=\ket{0},\ket{0},\ket{0},\ket{0};\\
            &\mathit{H}[q_1];\
            \mathit{CNOT}[q_1,q_2];\
            \mathit{CNOT}[q_1,q_3];\
            \mathit{CNOT}[q_1,q_4];\\
            &\mathbf{if}\ s(\mathit{id})=1\vee r(\mathit{id})=1\ \mathbf{then}\ v_1:=u_1\
              \mathbf{else}\ \mathit{H}[q_1];\ v_1:=\mathbf{meas}[q_1];\\
            &\mathbf{if}\ s(\mathit{id})=2\vee r(\mathit{id})=2\ \mathbf{then}\ v_2:=u_2\
              \mathbf{else}\ \mathit{H}[q_2];\ v_2:=\mathbf{meas}[q_2];\\
            &\mathbf{if}\ s(\mathit{id})=3\vee r(\mathit{id})=3\ \mathbf{then}\ v_3:=u_3\
              \mathbf{else}\ \mathit{H}[q_3];\ v_3:=\mathbf{meas}[q_3];\\
            &\mathbf{if}\ s(\mathit{id})=4\vee r(\mathit{id})=4\ \mathbf{then}\ v_4:=u_4\
              \mathbf{else}\ \mathit{H}[q_4];\ v_4:=\mathbf{meas}[q_4]
        \end{aligned}
    \end{equation*}
    and
    \begin{equation*}
        \begin{aligned}
            \mathit{Rec}\equiv\ &
            \mathbf{if}\ s(\mathit{id})=1\wedge u_1=1\ \mathbf{then}\ \mathit{Z}[q_1];\\
            &\mathbf{if}\ s(\mathit{id})=2\wedge u_2=1\ \mathbf{then}\ \mathit{Z}[q_2];\\
            &\mathbf{if}\ s(\mathit{id})=3\wedge u_3=1\ \mathbf{then}\ \mathit{Z}[q_3];\\
            &\mathbf{if}\ s(\mathit{id})=4\wedge u_4=1\ \mathbf{then}\ \mathit{Z}[q_4];\\
            &\mathbf{if}\ r(\mathit{id})=1\wedge v_2\oplus v_3\oplus v_4=1\ \mathbf{then}\ \mathit{Z}[q_1];\\
            &\mathbf{if}\ r(\mathit{id})=2\wedge v_1\oplus v_3\oplus v_4=1\ \mathbf{then}\ \mathit{Z}[q_2];\\
            &\mathbf{if}\ r(\mathit{id})=3\wedge v_1\oplus v_2\oplus v_4=1\ \mathbf{then}\ \mathit{Z}[q_3];\\
            &\mathbf{if}\ r(\mathit{id})=4\wedge v_1\oplus v_2\oplus v_3=1\ \mathbf{then}\ \mathit{Z}[q_4].
        \end{aligned}
    \end{equation*}
    Here, the qubits $q_j$ are used to prepare the four-party Greenberger–Horne–Zeilinger (GHZ) state,
    $u_j$ are random bits generated privately by the parties,
    and $v_j$ are the broadcast bits.
    The probabilistic assignment $u_j:=\braces*{(1/2,0),(1/2,1)}$ is syntactic sugar for $w_j:=\ket{0};\ H[w_j]; u_j:=\mathbf{meas}[w_j]$, where $w_j$ is a fresh qubit variable introduced implicitly in $P$.

    For $j=1$ to $4$, we classify $\mathit{id},u_j,w_j,q_j$ as $\High$ variables
    and $v_j$ as $\Low$ variables.
    Under this labeling, the information-flow security of program $P$ guarantees that the broadcast bits
    $v_j$ reveal no information about the sender and receiver identities encoded in $\mathit{id}$.
\end{example}

\section{Related Work}
\label{sec:related_work}

\paragraph{Quantum information-flow security}
The closest prior work~\cite{YFY13} studied information-flow security and access control in quantum automata models, closely paralleling the original noninterference framework of Goguen and Meseguer~\cite{GM82}.
Recent work~\cite{AM20} studied quantum quantitative information flow in a specific setting where a collection of quantum states encodes a secret classical value and attackers may choose the measurements to perform.
Quantum entanglement has also been shown to increase the capacity of classical covert channels~\cite{Mestel22}.
In addition, recent work~\cite{QAC} identified threats to the security guarantees of classical access control models when classical operating systems are integrated with quantum computing.

In contrast, this paper focuses on end-to-end confidentiality for quantum programs with both classical and quantum variables, providing useful abstractions for quantum applications that manipulate confidential data, as illustrated by the examples in \Cref{sec:illustrative_examples}.
In addition to the noninterference-based security notion \textbf{NI}, which naturally extends the classical idea~\cite{GM82} and accounts for entangled side information,
we introduce an equivalent simulation-based notion \textbf{SIM} that gives security a clear operational interpretation.
Building on these semantic foundations, we further develop a formal verification framework.

\paragraph{Relational and Hoare-style verification}
In classical programming, the standard self-composition technique reduces verification of noninterference to a Hoare triple for a composed program in traditional Hoare logic~\cite{JL00,DHS05,BDR11,BCK11}.
Alternatively, noninterference can be viewed as a relational property of two program executions and verified using relational Hoare logic~\cite{Benton04}, a general framework for reasoning about relations between programs.
More recently, quantum relational Hoare logics have also been studied~\cite{Unruh19,Unruh20,BHY+20,LU21,YJY24,BGWZ25}, with applications including the verification of quantum cryptographic protocols.

Our goal in this paper is not to develop a relational proof system.
Instead, we begin with the classical self-composition approach to verifying secure information flow in programs.
The failure of its direct quantum analogue (see \Cref{sub:challenges_from_classical_to_quantum,sub:self_composition_fail}) motivates our new self-conjugation technique.
By exploiting the deferred measurement principle and properties of unitary operations in a nontrivial way, self-conjugation reduces verification of \textbf{NI} to proof obligations in a variant of the well-studied quantum Hoare logic~\cite{Ying11}.

\section{Conclusion}
\label{sec:conclusion}

This paper develops semantic foundations and verification methods for secure information flow in quantum programs.
We introduced two notions of information-flow security: the noninterference-based notion \textbf{NI} and the simulation-based notion \textbf{SIM}.
By proving their equivalence, we address \Cref{ques:1}: the resulting security notion is both operationally meaningful and suitable for specification.
Building on these semantic foundations, we developed a Hoare-style verification framework that addresses \Cref{ques:2}.
At its core is self-conjugation, a novel technique that circumvents the obstruction to direct self-composition~\cite{BDR11} in the quantum setting.
We demonstrated the framework through a series of illustrative examples.

We conclude with several directions for future research.
First, an immediate next step is to mechanize the verification framework developed in this paper.
This could build on existing frameworks and tools for quantum computation and quantum program verification, such as Quantomatic~\cite{KZ15}, ZX-calculus~\cite{Bob11,KW19}, QWire~\cite{Paykin17}, QBricks~\cite{Char21}, VOQC~\cite{Rand21}, Gleipnir~\cite{Tao21}, QHLProver~\cite{Liu19}, CoqQ~\cite{Zhou23}, QEDIA~\cite{Ren26}, and AutoQ~\cite{Chen25,Chen25b}.
A key prerequisite is support for projective reasoning about programs with quantum control flow (see \Cref{sub:qhl}), extending existing implementations of quantum Hoare logics.
The verification conditions arising from such reasoning can be discharged by linear algebra.
Second, it would be interesting to extend the framework to incorporate richer programming features such as loops and recursion~\cite{Selinger04,DTPW24,YC22,YZ24,QRM}.
Since \textbf{NI} and \textbf{SIM} are defined semantically, we expect the results of \Cref{sec:defining_secure_information_flow} to carry over to such programs.
However, a key challenge is to generalize self-conjugation to support these features.
For example, the dilation of a loop may require an unbounded number of environmental variables (see \Cref{alg:dilation}).
Third, the challenges identified in this paper suggest revisiting other classical methods~\cite{PS02,GM10,BCK16,CSG16,DFT20,DBJ21,GBTB21,ISV24,DM24} for establishing secure information flow in the quantum setting.
Adapting these methods could build on the security foundations developed here and yield new techniques complementary to self-conjugation.

\section*{Disclosure of Use of AI in Research}
In the later stages of this work, we used ChatGPT (GPT-5.4 and GPT-5.5) to assist with preparing the figures and testing the correctness of our constructions.

\section*{Data Availability Statement}
This work is theoretical and has no associated artifacts or datasets.

\begin{acks}
This work was supported in part by the Australian Research Council under Grant DP250102952.
Part of this work was carried out while Zhicheng Zhang and Mingsheng Ying were visiting DIMACS, Rutgers University, United States.
Zhicheng Zhang thanks Robert Koenig, Marco Tomamichel, and Mark Wilde for helpful discussions.
\end{acks}

\bibliographystyle{ACM-Reference-Format}
\bibliography{mybib}


\begin{thebibliography}{100}


\ifx \showCODEN    \undefined \def \showCODEN     #1{\unskip}     \fi
\ifx \showISBNx    \undefined \def \showISBNx     #1{\unskip}     \fi
\ifx \showISBNxiii \undefined \def \showISBNxiii  #1{\unskip}     \fi
\ifx \showISSN     \undefined \def \showISSN      #1{\unskip}     \fi
\ifx \showLCCN     \undefined \def \showLCCN      #1{\unskip}     \fi
\ifx \shownote     \undefined \def \shownote      #1{#1}          \fi
\ifx \showarticletitle \undefined \def \showarticletitle #1{#1}   \fi
\ifx \showURL      \undefined \def \showURL       {\relax}        \fi
\providecommand\bibfield[2]{#2}
\providecommand\bibinfo[2]{#2}
\providecommand\natexlab[1]{#1}
\providecommand\showeprint[2][]{arXiv:#2}

\bibitem[Aharonov et~al\mbox{.}(1998)]%
        {AKN98}
\bibfield{author}{\bibinfo{person}{Dorit Aharonov}, \bibinfo{person}{Alexei Kitaev}, {and} \bibinfo{person}{Noam Nisan}.} \bibinfo{year}{1998}\natexlab{}.
\newblock \showarticletitle{Quantum circuits with mixed states}. In \bibinfo{booktitle}{\emph{Proceedings of the Thirtieth Annual ACM Symposium on Theory of Computing}}. \bibinfo{pages}{20--30}.
\newblock


\bibitem[Altenkirch and Grattage(2005)]%
        {AG05}
\bibfield{author}{\bibinfo{person}{Thorsten Altenkirch} {and} \bibinfo{person}{Jonathan Grattage}.} \bibinfo{year}{2005}\natexlab{}.
\newblock \showarticletitle{A functional quantum programming language}. In \bibinfo{booktitle}{\emph{20th Annual IEEE Symposium on Logic in Computer Science (LICS'05)}}. \bibinfo{pages}{249--258}.
\newblock
\href{https://doi.org/10.1109/LICS.2005.1}{doi:\nolinkurl{10.1109/LICS.2005.1}}


\bibitem[Ambainis et~al\mbox{.}(2000)]%
        {AMTdW00}
\bibfield{author}{\bibinfo{person}{Andris Ambainis}, \bibinfo{person}{Michele Mosca}, \bibinfo{person}{Alain Tapp}, {and} \bibinfo{person}{Ronald {de Wolf}}.} \bibinfo{year}{2000}\natexlab{}.
\newblock \showarticletitle{Private quantum channels}. In \bibinfo{booktitle}{\emph{Proceedings 41st Annual Symposium on Foundations of Computer Science}}. \bibinfo{pages}{547--553}.
\newblock


\bibitem[Am{\'e}rico and Malacaria(2020)]%
        {AM20}
\bibfield{author}{\bibinfo{person}{Arthur Am{\'e}rico} {and} \bibinfo{person}{Pasquale Malacaria}.} \bibinfo{year}{2020}\natexlab{}.
\newblock \showarticletitle{{QQIF}: Quantum quantitative information flow}. In \bibinfo{booktitle}{\emph{2020 IEEE European Symposium on Security and Privacy Workshops (EuroS\&PW)}}. \bibinfo{pages}{261--270}.
\newblock


\bibitem[Andrews and Reitman(1980)]%
        {AR80}
\bibfield{author}{\bibinfo{person}{Gregory~R. Andrews} {and} \bibinfo{person}{Richard~P. Reitman}.} \bibinfo{year}{1980}\natexlab{}.
\newblock \showarticletitle{An axiomatic approach to information flow in programs}.
\newblock \bibinfo{journal}{\emph{ACM Transactions on Programming Languages and Systems (TOPLAS)}} \bibinfo{volume}{2}, \bibinfo{number}{1} (\bibinfo{year}{1980}), \bibinfo{pages}{56--76}.
\newblock


\bibitem[Apt et~al\mbox{.}(2009)]%
        {ADO09}
\bibfield{author}{\bibinfo{person}{Krzysztof~R. Apt}, \bibinfo{person}{Frank~S. de Boer}, {and} \bibinfo{person}{Ernst-R{\"u}diger Olderog}.} \bibinfo{year}{2009}\natexlab{}.
\newblock \bibinfo{booktitle}{\emph{Verification of sequential and concurrent programs}}.
\newblock \bibinfo{publisher}{Springer}.
\newblock


\bibitem[Banerjee and Naumann(2002)]%
        {BN02}
\bibfield{author}{\bibinfo{person}{Anindya Banerjee} {and} \bibinfo{person}{David~A. Naumann}.} \bibinfo{year}{2002}\natexlab{}.
\newblock \showarticletitle{Secure Information Flow and Pointer Confinement in a Java-like Language}. In \bibinfo{booktitle}{\emph{Proceedings of the 15th IEEE Workshop on Computer Security Foundations}}. \bibinfo{pages}{253--267}.
\newblock


\bibitem[Barnett et~al\mbox{.}(2003)]%
        {BCJ03}
\bibfield{author}{\bibinfo{person}{Stephen~M. Barnett}, \bibinfo{person}{Anthony Chefles}, {and} \bibinfo{person}{Igor Jex}.} \bibinfo{year}{2003}\natexlab{}.
\newblock \showarticletitle{Comparison of two unknown pure quantum states}.
\newblock \bibinfo{journal}{\emph{Physics Letters A}} \bibinfo{volume}{307}, \bibinfo{number}{4} (\bibinfo{year}{2003}), \bibinfo{pages}{189--195}.
\newblock


\bibitem[Barthe et~al\mbox{.}(2011a)]%
        {BCK11}
\bibfield{author}{\bibinfo{person}{Gilles Barthe}, \bibinfo{person}{Juan~Manuel Crespo}, {and} \bibinfo{person}{C{\'e}sar Kunz}.} \bibinfo{year}{2011}\natexlab{a}.
\newblock \showarticletitle{Relational verification using product programs}. In \bibinfo{booktitle}{\emph{International Symposium on Formal Methods}}. \bibinfo{pages}{200--214}.
\newblock


\bibitem[Barthe et~al\mbox{.}(2016)]%
        {BCK16}
\bibfield{author}{\bibinfo{person}{Gilles Barthe}, \bibinfo{person}{Juan~Manuel Crespo}, {and} \bibinfo{person}{C{\'e}sar Kunz}.} \bibinfo{year}{2016}\natexlab{}.
\newblock \showarticletitle{Product programs and relational program logics}.
\newblock \bibinfo{journal}{\emph{Journal of Logical and Algebraic Methods in Programming}} \bibinfo{volume}{85}, \bibinfo{number}{5} (\bibinfo{year}{2016}), \bibinfo{pages}{847--859}.
\newblock


\bibitem[Barthe et~al\mbox{.}(2011b)]%
        {BDR11}
\bibfield{author}{\bibinfo{person}{Gilles Barthe}, \bibinfo{person}{Pedro~R. {D'Argenio}}, {and} \bibinfo{person}{Tamara Rezk}.} \bibinfo{year}{2011}\natexlab{b}.
\newblock \showarticletitle{Secure information flow by self-composition}.
\newblock \bibinfo{journal}{\emph{Mathematical Structures in Computer Science}} \bibinfo{volume}{21}, \bibinfo{number}{6} (\bibinfo{year}{2011}), \bibinfo{pages}{1207--1252}.
\newblock


\bibitem[Barthe et~al\mbox{.}(2025)]%
        {BGWZ25}
\bibfield{author}{\bibinfo{person}{Gilles Barthe}, \bibinfo{person}{Minbo Gao}, \bibinfo{person}{Theo Wang}, {and} \bibinfo{person}{Li Zhou}.} \bibinfo{year}{2025}\natexlab{}.
\newblock \showarticletitle{Complete quantum relational {Hoare} logics from optimal transport duality}. In \bibinfo{booktitle}{\emph{40th Annual ACM/IEEE Symposium on Logic in Computer Science (LICS)}}. \bibinfo{pages}{884--925}.
\newblock


\bibitem[Barthe et~al\mbox{.}(2020)]%
        {BHY+20}
\bibfield{author}{\bibinfo{person}{Gilles Barthe}, \bibinfo{person}{Justin Hsu}, \bibinfo{person}{Mingsheng Ying}, \bibinfo{person}{Nengkun Yu}, {and} \bibinfo{person}{Li Zhou}.} \bibinfo{year}{2020}\natexlab{}.
\newblock \showarticletitle{Relational proofs for quantum programs}.
\newblock \bibinfo{journal}{\emph{Proceedings of the ACM on Programming Languages}} \bibinfo{volume}{4}, \bibinfo{number}{POPL}, Article \bibinfo{articleno}{21} (\bibinfo{year}{2020}), \bibinfo{numpages}{29}~pages.
\newblock
\href{https://doi.org/10.1145/3371089}{doi:\nolinkurl{10.1145/3371089}}


\bibitem[Belavkin(1975)]%
        {Belavkin75}
\bibfield{author}{\bibinfo{person}{Viacheslav~P. Belavkin}.} \bibinfo{year}{1975}\natexlab{}.
\newblock \showarticletitle{Optimal multiple quantum statistical hypothesis testing}.
\newblock \bibinfo{journal}{\emph{Stochastics: an international journal of probability and stochastic processes}} \bibinfo{volume}{1}, \bibinfo{number}{1-4} (\bibinfo{year}{1975}), \bibinfo{pages}{315--345}.
\newblock


\bibitem[{B}ennett(1973)]%
        {Bennett73}
\bibfield{author}{\bibinfo{person}{Charles~H. {B}ennett}.} \bibinfo{year}{1973}\natexlab{}.
\newblock \showarticletitle{Logical reversibility of computation}.
\newblock \bibinfo{journal}{\emph{IBM Journal of Research and Development}} \bibinfo{volume}{17}, \bibinfo{number}{6} (\bibinfo{year}{1973}), \bibinfo{pages}{525--532}.
\newblock
\href{https://doi.org/10.1147/rd.176.0525}{doi:\nolinkurl{10.1147/rd.176.0525}}


\bibitem[Benton(2004)]%
        {Benton04}
\bibfield{author}{\bibinfo{person}{Nick Benton}.} \bibinfo{year}{2004}\natexlab{}.
\newblock \showarticletitle{Simple relational correctness proofs for static analyses and program transformations}.
\newblock \bibinfo{journal}{\emph{ACM SIGPLAN Notices}} \bibinfo{volume}{39}, \bibinfo{number}{1} (\bibinfo{year}{2004}), \bibinfo{pages}{14--25}.
\newblock


\bibitem[Bichsel et~al\mbox{.}(2020)]%
        {BBGV20}
\bibfield{author}{\bibinfo{person}{Benjamin Bichsel}, \bibinfo{person}{Maximilian Baader}, \bibinfo{person}{Timon Gehr}, {and} \bibinfo{person}{Martin Vechev}.} \bibinfo{year}{2020}\natexlab{}.
\newblock \showarticletitle{Silq: A high-level quantum language with safe uncomputation and intuitive semantics}. In \bibinfo{booktitle}{\emph{Proceedings of the 41st ACM SIGPLAN Conference on Programming Language Design and Implementation}}. \bibinfo{pages}{286--300}.
\newblock
\href{https://doi.org/10.1145/3385412.3386007}{doi:\nolinkurl{10.1145/3385412.3386007}}


\bibitem[Bodei et~al\mbox{.}(1999)]%
        {BDNN99}
\bibfield{author}{\bibinfo{person}{Chiara Bodei}, \bibinfo{person}{Pierpaolo Degano}, \bibinfo{person}{Flemming Nielson}, {and} \bibinfo{person}{Hanne~Riis Nielson}.} \bibinfo{year}{1999}\natexlab{}.
\newblock \showarticletitle{Static analysis of processes for no read-up and no write-down}. In \bibinfo{booktitle}{\emph{International Conference on Foundations of Software Science and Computation Structure}}. \bibinfo{pages}{120--134}.
\newblock


\bibitem[Bodei et~al\mbox{.}(2001)]%
        {BDNN01}
\bibfield{author}{\bibinfo{person}{Chiara Bodei}, \bibinfo{person}{Pierpaolo Degano}, \bibinfo{person}{Flemming Nielson}, {and} \bibinfo{person}{Hanne~Riis Nielson}.} \bibinfo{year}{2001}\natexlab{}.
\newblock \showarticletitle{Static analysis for the $\pi$-calculus with applications to security}.
\newblock \bibinfo{journal}{\emph{Information and Computation}} \bibinfo{volume}{168}, \bibinfo{number}{1} (\bibinfo{year}{2001}), \bibinfo{pages}{68--92}.
\newblock


\bibitem[Boykin and Roychowdhury(2003)]%
        {BR03}
\bibfield{author}{\bibinfo{person}{P.~Oscar Boykin} {and} \bibinfo{person}{Vwani Roychowdhury}.} \bibinfo{year}{2003}\natexlab{}.
\newblock \showarticletitle{Optimal encryption of quantum bits}.
\newblock \bibinfo{journal}{\emph{Physical Review A}} \bibinfo{volume}{67}, \bibinfo{number}{4} (\bibinfo{year}{2003}), \bibinfo{pages}{042317}.
\newblock
\href{https://doi.org/10.1103/physreva.67.042317}{doi:\nolinkurl{10.1103/physreva.67.042317}}


\bibitem[Brassard et~al\mbox{.}(2007)]%
        {BBF+07}
\bibfield{author}{\bibinfo{person}{Gilles Brassard}, \bibinfo{person}{Anne Broadbent}, \bibinfo{person}{Joseph Fitzsimons}, \bibinfo{person}{S{\'e}bastien Gambs}, {and} \bibinfo{person}{Alain Tapp}.} \bibinfo{year}{2007}\natexlab{}.
\newblock \showarticletitle{Anonymous quantum communication}. In \bibinfo{booktitle}{\emph{International Conference on the Theory and Application of Cryptology and Information Security}}. \bibinfo{pages}{460--473}.
\newblock


\bibitem[Buhrman et~al\mbox{.}(2001)]%
        {BCWdW01}
\bibfield{author}{\bibinfo{person}{Harry Buhrman}, \bibinfo{person}{Richard Cleve}, \bibinfo{person}{John Watrous}, {and} \bibinfo{person}{Ronald de Wolf}.} \bibinfo{year}{2001}\natexlab{}.
\newblock \showarticletitle{Quantum fingerprinting}.
\newblock \bibinfo{journal}{\emph{Physical Review Letters}} \bibinfo{volume}{87}, \bibinfo{number}{16} (\bibinfo{year}{2001}), \bibinfo{pages}{167902}.
\newblock
\href{https://doi.org/10.1103/PhysRevLett.87.167902}{doi:\nolinkurl{10.1103/PhysRevLett.87.167902}}


\bibitem[Chareton et~al\mbox{.}(2021)]%
        {Char21}
\bibfield{author}{\bibinfo{person}{Christophe Chareton}, \bibinfo{person}{S{\'e}bastien Bardin}, \bibinfo{person}{Fran\c{c}ois Bobot}, \bibinfo{person}{Valentin Perrelle}, {and} \bibinfo{person}{Beno{\^\i}t Valiron}.} \bibinfo{year}{2021}\natexlab{}.
\newblock \showarticletitle{An Automated Deductive Verification Framework for Circuit-Building Quantum Programs}. In \bibinfo{booktitle}{\emph{Programming Languages and Systems ({ESOP})}}, Vol.~\bibinfo{volume}{12648}. \bibinfo{pages}{148--177}.
\newblock
\href{https://doi.org/10.1007/978-3-030-72019-3_6}{doi:\nolinkurl{10.1007/978-3-030-72019-3_6}}


\bibitem[Chen et~al\mbox{.}(2025a)]%
        {Chen25b}
\bibfield{author}{\bibinfo{person}{Yu{-}Fang Chen}, \bibinfo{person}{Kai{-}Min Chung}, \bibinfo{person}{Min{-}Hsiu Hsieh}, \bibinfo{person}{Wei{-}Jia Huang}, \bibinfo{person}{Ond{\v r}ej Leng{\'a}l}, \bibinfo{person}{Jyun{-}Ao Lin}, {and} \bibinfo{person}{Wei{-}Lun Tsai}.} \bibinfo{year}{2025}\natexlab{a}.
\newblock \showarticletitle{{AutoQ} 2.0: From Verification of Quantum Circuits to Verification of Quantum Programs}. In \bibinfo{booktitle}{\emph{Tools and Algorithms for the Construction and Analysis of Systems ({TACAS})}}, Vol.~\bibinfo{volume}{15698}. \bibinfo{pages}{87--108}.
\newblock
\href{https://doi.org/10.1007/978-3-031-90660-2_5}{doi:\nolinkurl{10.1007/978-3-031-90660-2_5}}


\bibitem[Chen et~al\mbox{.}(2025b)]%
        {Chen25}
\bibfield{author}{\bibinfo{person}{Yu{-}Fang Chen}, \bibinfo{person}{Kai{-}Min Chung}, \bibinfo{person}{Ond{\v r}ej Leng{\'a}l}, \bibinfo{person}{Jyun{-}Ao Lin}, \bibinfo{person}{Wei{-}Lun Tsai}, {and} \bibinfo{person}{Di{-}De Yen}.} \bibinfo{year}{2025}\natexlab{b}.
\newblock \showarticletitle{An Automata-Based Framework for Verification and Bug Hunting in Quantum Circuits}.
\newblock \bibinfo{journal}{\emph{Commun. ACM}} \bibinfo{volume}{68}, \bibinfo{number}{6} (\bibinfo{year}{2025}), \bibinfo{pages}{85--93}.
\newblock
\href{https://doi.org/10.1145/3725728}{doi:\nolinkurl{10.1145/3725728}}


\bibitem[Christandl and Wehner(2005)]%
        {CW05}
\bibfield{author}{\bibinfo{person}{Matthias Christandl} {and} \bibinfo{person}{Stephanie Wehner}.} \bibinfo{year}{2005}\natexlab{}.
\newblock \showarticletitle{Quantum anonymous transmissions}. In \bibinfo{booktitle}{\emph{International Conference on the Theory and Application of Cryptology and Information Security}}. \bibinfo{pages}{217--235}.
\newblock


\bibitem[Clark et~al\mbox{.}(2002)]%
        {CHH02}
\bibfield{author}{\bibinfo{person}{David Clark}, \bibinfo{person}{Chris Hankin}, {and} \bibinfo{person}{Sebastian Hunt}.} \bibinfo{year}{2002}\natexlab{}.
\newblock \showarticletitle{Information flow for Algol-like languages}.
\newblock \bibinfo{journal}{\emph{Computer Languages, Systems \& Structures}} \bibinfo{volume}{28}, \bibinfo{number}{1} (\bibinfo{year}{2002}), \bibinfo{pages}{3--28}.
\newblock


\bibitem[Coecke and Duncan(2011)]%
        {Bob11}
\bibfield{author}{\bibinfo{person}{Bob Coecke} {and} \bibinfo{person}{Ross Duncan}.} \bibinfo{year}{2011}\natexlab{}.
\newblock \showarticletitle{Interacting Quantum Observables: Categorical Algebra and Diagrammatics}.
\newblock \bibinfo{journal}{\emph{New Journal of Physics}} \bibinfo{volume}{13}, \bibinfo{number}{4} (\bibinfo{year}{2011}), \bibinfo{pages}{043016}.
\newblock
\href{https://doi.org/10.1088/1367-2630/13/4/043016}{doi:\nolinkurl{10.1088/1367-2630/13/4/043016}}


\bibitem[Cook(1978)]%
        {Cook78}
\bibfield{author}{\bibinfo{person}{Stephen~A. Cook}.} \bibinfo{year}{1978}\natexlab{}.
\newblock \showarticletitle{Soundness and completeness of an axiom system for program verification}.
\newblock \bibinfo{journal}{\emph{SIAM J. Comput.}} \bibinfo{volume}{7}, \bibinfo{number}{1} (\bibinfo{year}{1978}), \bibinfo{pages}{70--90}.
\newblock
\href{https://doi.org/10.1137/0207005}{doi:\nolinkurl{10.1137/0207005}}


\bibitem[{C}{\'o}rcoles et~al\mbox{.}(2021)]%
        {CTILMCG21}
\bibfield{author}{\bibinfo{person}{A.~D. {C}{\'o}rcoles}, \bibinfo{person}{Maika {T}akita}, \bibinfo{person}{Ken {I}noue}, \bibinfo{person}{Scott {L}ekuch}, \bibinfo{person}{Zlatko~K. {M}inev}, \bibinfo{person}{Jerry~M. {C}how}, {and} \bibinfo{person}{Jay~M. {G}ambetta}.} \bibinfo{year}{2021}\natexlab{}.
\newblock \showarticletitle{Exploiting dynamic quantum circuits in a quantum algorithm with superconducting qubits}.
\newblock \bibinfo{journal}{\emph{Physical Review Letters}} \bibinfo{volume}{127}, \bibinfo{number}{10} (\bibinfo{year}{2021}), \bibinfo{pages}{100501}.
\newblock


\bibitem[Costanzo et~al\mbox{.}(2016)]%
        {CSG16}
\bibfield{author}{\bibinfo{person}{David Costanzo}, \bibinfo{person}{Zhong Shao}, {and} \bibinfo{person}{Ronghui Gu}.} \bibinfo{year}{2016}\natexlab{}.
\newblock \showarticletitle{End-to-end verification of information-flow security for {C} and assembly programs}.
\newblock \bibinfo{journal}{\emph{ACM SIGPLAN Notices}} \bibinfo{volume}{51}, \bibinfo{number}{6} (\bibinfo{year}{2016}), \bibinfo{pages}{648--664}.
\newblock


\bibitem[Dardinier and M{\"u}ller(2024)]%
        {DM24}
\bibfield{author}{\bibinfo{person}{Thibault Dardinier} {and} \bibinfo{person}{Peter M{\"u}ller}.} \bibinfo{year}{2024}\natexlab{}.
\newblock \showarticletitle{Hyper {Hoare} logic: ({Dis-})proving program hyperproperties}.
\newblock \bibinfo{journal}{\emph{Proceedings of the ACM on Programming Languages}} \bibinfo{volume}{8}, \bibinfo{number}{PLDI} (\bibinfo{year}{2024}), \bibinfo{pages}{1485--1509}.
\newblock


\bibitem[Darvas et~al\mbox{.}(2005)]%
        {DHS05}
\bibfield{author}{\bibinfo{person}{\'{A}d\'{a}m Darvas}, \bibinfo{person}{Reiner H\"{a}hnle}, {and} \bibinfo{person}{David Sands}.} \bibinfo{year}{2005}\natexlab{}.
\newblock \showarticletitle{A theorem proving approach to analysis of secure information flow}. In \bibinfo{booktitle}{\emph{Proceedings of the Second International Conference on Security in Pervasive Computing}} \emph{(\bibinfo{series}{SPC'05})}. \bibinfo{pages}{193--209}.
\newblock
\href{https://doi.org/10.1007/978-3-540-32004-3_20}{doi:\nolinkurl{10.1007/978-3-540-32004-3_20}}


\bibitem[Deng et~al\mbox{.}(2024)]%
        {DTPW24}
\bibfield{author}{\bibinfo{person}{Haowei Deng}, \bibinfo{person}{Runzhou Tao}, \bibinfo{person}{Yuxiang Peng}, {and} \bibinfo{person}{Xiaodi Wu}.} \bibinfo{year}{2024}\natexlab{}.
\newblock \showarticletitle{A case for synthesis of recursive quantum unitary programs}.
\newblock \bibinfo{journal}{\emph{Proceedings of the ACM on Programming Languages}} \bibinfo{volume}{8}, \bibinfo{number}{POPL} (\bibinfo{year}{2024}), \bibinfo{pages}{1759--1788}.
\newblock
\href{https://doi.org/10.1145/3632901}{doi:\nolinkurl{10.1145/3632901}}


\bibitem[Denning and Denning(1977)]%
        {DD77}
\bibfield{author}{\bibinfo{person}{Dorothy~E. Denning} {and} \bibinfo{person}{Peter~J. Denning}.} \bibinfo{year}{1977}\natexlab{}.
\newblock \showarticletitle{Certification of programs for secure information flow}.
\newblock \bibinfo{journal}{\emph{Commun. ACM}} \bibinfo{volume}{20}, \bibinfo{number}{7} (\bibinfo{year}{1977}), \bibinfo{pages}{504--513}.
\newblock


\bibitem[Derakhshan et~al\mbox{.}(2021)]%
        {DBJ21}
\bibfield{author}{\bibinfo{person}{Farzaneh Derakhshan}, \bibinfo{person}{Stephanie Balzer}, {and} \bibinfo{person}{Limin Jia}.} \bibinfo{year}{2021}\natexlab{}.
\newblock \showarticletitle{Session logical relations for noninterference}. In \bibinfo{booktitle}{\emph{2021 36th Annual ACM/IEEE Symposium on Logic in Computer Science (LICS)}}. IEEE, \bibinfo{pages}{1--14}.
\newblock


\bibitem[{D'Hondt} and Panangaden(2006)]%
        {DP06}
\bibfield{author}{\bibinfo{person}{Ellie {D'Hondt}} {and} \bibinfo{person}{Prakash Panangaden}.} \bibinfo{year}{2006}\natexlab{}.
\newblock \showarticletitle{Quantum weakest preconditions}.
\newblock \bibinfo{journal}{\emph{Mathematical Structures in Computer Science}} \bibinfo{volume}{16}, \bibinfo{number}{3} (\bibinfo{year}{2006}), \bibinfo{pages}{429--451}.
\newblock


\bibitem[Dieks(1982)]%
        {Dieks82}
\bibfield{author}{\bibinfo{person}{Dennis Dieks}.} \bibinfo{year}{1982}\natexlab{}.
\newblock \showarticletitle{Communication by {EPR} devices}.
\newblock \bibinfo{journal}{\emph{Physics Letters A}} \bibinfo{volume}{92}, \bibinfo{number}{6} (\bibinfo{year}{1982}), \bibinfo{pages}{271--272}.
\newblock


\bibitem[Dimitrova et~al\mbox{.}(2020)]%
        {DFT20}
\bibfield{author}{\bibinfo{person}{Rayna Dimitrova}, \bibinfo{person}{Bernd Finkbeiner}, {and} \bibinfo{person}{Hazem Torfah}.} \bibinfo{year}{2020}\natexlab{}.
\newblock \showarticletitle{Probabilistic hyperproperties of {Markov} decision processes}. In \bibinfo{booktitle}{\emph{International Symposium on Automated Technology for Verification and Analysis}}. \bibinfo{pages}{484--500}.
\newblock


\bibitem[Giacobazzi and Mastroeni(2004)]%
        {GM04}
\bibfield{author}{\bibinfo{person}{Roberto Giacobazzi} {and} \bibinfo{person}{Isabella Mastroeni}.} \bibinfo{year}{2004}\natexlab{}.
\newblock \showarticletitle{Abstract non-interference: Parameterizing non-interference by abstract interpretation}.
\newblock \bibinfo{journal}{\emph{ACM SIGPLAN Notices}} \bibinfo{volume}{39}, \bibinfo{number}{1} (\bibinfo{year}{2004}), \bibinfo{pages}{186--197}.
\newblock


\bibitem[Giacobazzi and Mastroeni(2010)]%
        {GM10}
\bibfield{author}{\bibinfo{person}{Roberto Giacobazzi} {and} \bibinfo{person}{Isabella Mastroeni}.} \bibinfo{year}{2010}\natexlab{}.
\newblock \showarticletitle{A proof system for abstract non-interference}.
\newblock \bibinfo{journal}{\emph{Journal of Logic and Computation}} \bibinfo{volume}{20}, \bibinfo{number}{2} (\bibinfo{year}{2010}), \bibinfo{pages}{449--479}.
\newblock


\bibitem[Gilchrist et~al\mbox{.}(2005)]%
        {GLN05}
\bibfield{author}{\bibinfo{person}{Alexei Gilchrist}, \bibinfo{person}{Nathan~K. Langford}, {and} \bibinfo{person}{Michael~A Nielsen}.} \bibinfo{year}{2005}\natexlab{}.
\newblock \showarticletitle{Distance measures to compare real and ideal quantum processes}.
\newblock \bibinfo{journal}{\emph{Physical Review A}} \bibinfo{volume}{71}, \bibinfo{number}{6} (\bibinfo{year}{2005}), \bibinfo{pages}{062310}.
\newblock


\bibitem[Goguen and Meseguer(1982)]%
        {GM82}
\bibfield{author}{\bibinfo{person}{Joseph~A. Goguen} {and} \bibinfo{person}{Jos{\'e} Meseguer}.} \bibinfo{year}{1982}\natexlab{}.
\newblock \showarticletitle{Security policies and security models}. In \bibinfo{booktitle}{\emph{1982 IEEE Symposium on Security and Privacy}}. \bibinfo{pages}{11--20}.
\newblock


\bibitem[Goldwasser and Micali(1984)]%
        {GM82b}
\bibfield{author}{\bibinfo{person}{Shafi Goldwasser} {and} \bibinfo{person}{Silvio Micali}.} \bibinfo{year}{1984}\natexlab{}.
\newblock \showarticletitle{Probabilistic encryption}.
\newblock \bibinfo{journal}{\emph{J. Comput. System Sci.}} \bibinfo{volume}{28}, \bibinfo{number}{2} (\bibinfo{year}{1984}), \bibinfo{pages}{270--299}.
\newblock
\href{https://doi.org/10.1016/0022-0000(84)90070-9}{doi:\nolinkurl{10.1016/0022-0000(84)90070-9}}


\bibitem[{G}ottesman(1997)]%
        {Gottesman97}
\bibfield{author}{\bibinfo{person}{Daniel {G}ottesman}.} \bibinfo{year}{1997}\natexlab{}.
\newblock \emph{\bibinfo{title}{Stabilizer codes and quantum error correction}}.
\newblock \bibinfo{thesistype}{Ph.\,D. Dissertation}. \bibinfo{school}{California Institute of Technology}.
\newblock


\bibitem[Gregersen et~al\mbox{.}(2021)]%
        {GBTB21}
\bibfield{author}{\bibinfo{person}{Simon~Oddershede Gregersen}, \bibinfo{person}{Johan Bay}, \bibinfo{person}{Amin Timany}, {and} \bibinfo{person}{Lars Birkedal}.} \bibinfo{year}{2021}\natexlab{}.
\newblock \showarticletitle{Mechanized logical relations for termination-insensitive noninterference}.
\newblock \bibinfo{journal}{\emph{Proceedings of the ACM on Programming Languages}} \bibinfo{volume}{5}, \bibinfo{number}{POPL} (\bibinfo{year}{2021}), \bibinfo{pages}{1--29}.
\newblock


\bibitem[Hausladen and Wootters(1994)]%
        {HW94}
\bibfield{author}{\bibinfo{person}{Paul Hausladen} {and} \bibinfo{person}{William~K. Wootters}.} \bibinfo{year}{1994}\natexlab{}.
\newblock \showarticletitle{A ``pretty good'' measurement for distinguishing quantum states}.
\newblock \bibinfo{journal}{\emph{Journal of Modern Optics}} \bibinfo{volume}{41}, \bibinfo{number}{12} (\bibinfo{year}{1994}), \bibinfo{pages}{2385--2390}.
\newblock


\bibitem[Heintze and Riecke(1998)]%
        {HR98}
\bibfield{author}{\bibinfo{person}{Nevin Heintze} {and} \bibinfo{person}{Jon~G. Riecke}.} \bibinfo{year}{1998}\natexlab{}.
\newblock \showarticletitle{The {SLam} calculus: programming with secrecy and integrity}. In \bibinfo{booktitle}{\emph{Proceedings of the 25th ACM SIGPLAN-SIGACT symposium on Principles of programming languages}}. \bibinfo{pages}{365--377}.
\newblock


\bibitem[Hietala et~al\mbox{.}(2021)]%
        {Rand21}
\bibfield{author}{\bibinfo{person}{Kesha Hietala}, \bibinfo{person}{Robert Rand}, \bibinfo{person}{Shih{-}Han Hung}, \bibinfo{person}{Xiaodi Wu}, {and} \bibinfo{person}{Michael Hicks}.} \bibinfo{year}{2021}\natexlab{}.
\newblock \showarticletitle{A Verified Optimizer for Quantum Circuits}.
\newblock \bibinfo{journal}{\emph{Proceedings of the ACM on Programming Languages}} \bibinfo{volume}{5}, \bibinfo{number}{{POPL}}, Article \bibinfo{articleno}{37} (\bibinfo{year}{2021}), \bibinfo{numpages}{29}~pages.
\newblock
\href{https://doi.org/10.1145/3434318}{doi:\nolinkurl{10.1145/3434318}}


\bibitem[Hoare(1969)]%
        {Hoare69}
\bibfield{author}{\bibinfo{person}{Charles Antony~Richard Hoare}.} \bibinfo{year}{1969}\natexlab{}.
\newblock \showarticletitle{An axiomatic basis for computer programming}.
\newblock \bibinfo{journal}{\emph{Commun. ACM}} \bibinfo{volume}{12}, \bibinfo{number}{10} (\bibinfo{year}{1969}), \bibinfo{pages}{576--580}.
\newblock


\bibitem[Holevo(1979)]%
        {Holevo79}
\bibfield{author}{\bibinfo{person}{Alexander~S. Holevo}.} \bibinfo{year}{1979}\natexlab{}.
\newblock \showarticletitle{On asymptotically optimal hypothesis testing in quantum statistics}.
\newblock \bibinfo{journal}{\emph{Theory of Probability \& Its Applications}} \bibinfo{volume}{23}, \bibinfo{number}{2} (\bibinfo{year}{1979}), \bibinfo{pages}{411--415}.
\newblock


\bibitem[Itzhaky et~al\mbox{.}(2024)]%
        {ISV24}
\bibfield{author}{\bibinfo{person}{Shachar Itzhaky}, \bibinfo{person}{Sharon Shoham}, {and} \bibinfo{person}{Yakir Vizel}.} \bibinfo{year}{2024}\natexlab{}.
\newblock \showarticletitle{Hyperproperty verification as CHC satisfiability}. In \bibinfo{booktitle}{\emph{European Symposium on Programming}}. \bibinfo{pages}{212--241}.
\newblock


\bibitem[Jex et~al\mbox{.}(2004)]%
        {JAC04}
\bibfield{author}{\bibinfo{person}{Igor Jex}, \bibinfo{person}{Erika Andersson}, {and} \bibinfo{person}{Anthony Chefles}.} \bibinfo{year}{2004}\natexlab{}.
\newblock \showarticletitle{Comparing the states of many quantum systems}.
\newblock \bibinfo{journal}{\emph{Journal of Modern Optics}} \bibinfo{volume}{51}, \bibinfo{number}{4} (\bibinfo{year}{2004}), \bibinfo{pages}{505--523}.
\newblock


\bibitem[Joshi and Leino(2000)]%
        {JL00}
\bibfield{author}{\bibinfo{person}{Rajeev Joshi} {and} \bibinfo{person}{K.~Rustan~M. Leino}.} \bibinfo{year}{2000}\natexlab{}.
\newblock \showarticletitle{A semantic approach to secure information flow}.
\newblock \bibinfo{journal}{\emph{Science of Computer Programming}} \bibinfo{volume}{37}, \bibinfo{number}{1-3} (\bibinfo{year}{2000}), \bibinfo{pages}{113--138}.
\newblock


\bibitem[Kissinger and van~de Wetering(2020)]%
        {KW19}
\bibfield{author}{\bibinfo{person}{Aleks Kissinger} {and} \bibinfo{person}{John van~de Wetering}.} \bibinfo{year}{2020}\natexlab{}.
\newblock \showarticletitle{{PyZX}: Large Scale Automated Diagrammatic Reasoning}. In \bibinfo{booktitle}{\emph{Proceedings of the 16th International Conference on Quantum Physics and Logic ({QPL})}} \emph{(\bibinfo{series}{{EPTCS}}, Vol.~\bibinfo{volume}{318})}. \bibinfo{pages}{229--241}.
\newblock
\href{https://doi.org/10.4204/EPTCS.318.14}{doi:\nolinkurl{10.4204/EPTCS.318.14}}


\bibitem[Kissinger and Zamdzhiev(2015)]%
        {KZ15}
\bibfield{author}{\bibinfo{person}{Aleks Kissinger} {and} \bibinfo{person}{Vladimir Zamdzhiev}.} \bibinfo{year}{2015}\natexlab{}.
\newblock \showarticletitle{Quantomatic: A Proof Assistant for Diagrammatic Reasoning}. In \bibinfo{booktitle}{\emph{Proceedings of the 25th International Conference on Automated Deduction ({CADE})}}, Vol.~\bibinfo{volume}{9195}. \bibinfo{pages}{326--336}.
\newblock
\href{https://doi.org/10.1007/978-3-319-21401-6_22}{doi:\nolinkurl{10.1007/978-3-319-21401-6_22}}


\bibitem[Kitaev(1997)]%
        {Kitaev97}
\bibfield{author}{\bibinfo{person}{Alexei~Yu. Kitaev}.} \bibinfo{year}{1997}\natexlab{}.
\newblock \showarticletitle{Quantum computations: algorithms and error correction}.
\newblock \bibinfo{journal}{\emph{Russian Mathematical Surveys}} \bibinfo{volume}{52}, \bibinfo{number}{6} (\bibinfo{year}{1997}), \bibinfo{pages}{1191--1249}.
\newblock


\bibitem[Landauer(1961)]%
        {Landauer61}
\bibfield{author}{\bibinfo{person}{Rolf Landauer}.} \bibinfo{year}{1961}\natexlab{}.
\newblock \showarticletitle{Irreversibility and heat generation in the computing process}.
\newblock \bibinfo{journal}{\emph{IBM journal of research and development}} \bibinfo{volume}{5}, \bibinfo{number}{3} (\bibinfo{year}{1961}), \bibinfo{pages}{183--191}.
\newblock
\href{https://doi.org/10.1147/rd.53.0183}{doi:\nolinkurl{10.1147/rd.53.0183}}


\bibitem[Li and Unruh(2021)]%
        {LU21}
\bibfield{author}{\bibinfo{person}{Yangjia Li} {and} \bibinfo{person}{Dominique Unruh}.} \bibinfo{year}{2021}\natexlab{}.
\newblock \showarticletitle{{Quantum Relational Hoare Logic with Expectations}}. In \bibinfo{booktitle}{\emph{48th International Colloquium on Automata, Languages, and Programming (ICALP 2021)}} \emph{(\bibinfo{series}{Leibniz International Proceedings in Informatics (LIPIcs)}, Vol.~\bibinfo{volume}{198})}. \bibinfo{pages}{136:1--136:20}.
\newblock
\href{https://doi.org/10.4230/LIPIcs.ICALP.2021.136}{doi:\nolinkurl{10.4230/LIPIcs.ICALP.2021.136}}


\bibitem[Liu et~al\mbox{.}(2019)]%
        {Liu19}
\bibfield{author}{\bibinfo{person}{Junyi Liu}, \bibinfo{person}{Bohua Zhan}, \bibinfo{person}{Shuling Wang}, \bibinfo{person}{Shenggang Ying}, \bibinfo{person}{Tao Liu}, \bibinfo{person}{Yangjia Li}, \bibinfo{person}{Mingsheng Ying}, {and} \bibinfo{person}{Naijun Zhan}.} \bibinfo{year}{2019}\natexlab{}.
\newblock \showarticletitle{Formal Verification of Quantum Algorithms Using Quantum Hoare Logic}. In \bibinfo{booktitle}{\emph{Proceedings of the 31st International Conference on Computer Aided Verification ({CAV})}}, Vol.~\bibinfo{volume}{11562}. \bibinfo{pages}{187--207}.
\newblock
\href{https://doi.org/10.1007/978-3-030-25543-5_12}{doi:\nolinkurl{10.1007/978-3-030-25543-5_12}}


\bibitem[Lloyd(1997)]%
        {Lloyd97}
\bibfield{author}{\bibinfo{person}{Seth Lloyd}.} \bibinfo{year}{1997}\natexlab{}.
\newblock \showarticletitle{Capacity of the noisy quantum channel}.
\newblock \bibinfo{journal}{\emph{Physical Review A}} \bibinfo{volume}{55}, \bibinfo{number}{3} (\bibinfo{year}{1997}), \bibinfo{pages}{1613}.
\newblock


\bibitem[Malacaria and Hankin(1999)]%
        {MH99}
\bibfield{author}{\bibinfo{person}{Pasquale Malacaria} {and} \bibinfo{person}{Chris Hankin}.} \bibinfo{year}{1999}\natexlab{}.
\newblock \showarticletitle{Non-deterministic games and program analysis: An application to security}. In \bibinfo{booktitle}{\emph{Proceedings. 14th Symposium on Logic in Computer Science (Cat. No. PR00158)}}. \bibinfo{pages}{443--452}.
\newblock


\bibitem[Mestel(2022)]%
        {Mestel22}
\bibfield{author}{\bibinfo{person}{David Mestel}.} \bibinfo{year}{2022}\natexlab{}.
\newblock \showarticletitle{Beware of Greeks bearing entanglement? Quantum covert channels, information flow and non-local games}. In \bibinfo{booktitle}{\emph{2022 IEEE 35th Computer Security Foundations Symposium (CSF)}}. \bibinfo{pages}{276--288}.
\newblock


\bibitem[Micali et~al\mbox{.}(1986)]%
        {MRS86}
\bibfield{author}{\bibinfo{person}{Silvio Micali}, \bibinfo{person}{Charles Rackoff}, {and} \bibinfo{person}{Bob Sloan}.} \bibinfo{year}{1986}\natexlab{}.
\newblock \showarticletitle{The notion of security for probabilistic cryptosystems}. In \bibinfo{booktitle}{\emph{Conference on the Theory and Application of Cryptographic Techniques}}. Springer, \bibinfo{pages}{381--392}.
\newblock


\bibitem[Montanaro(2019)]%
        {Montanaro19}
\bibfield{author}{\bibinfo{person}{Ashley Montanaro}.} \bibinfo{year}{2019}\natexlab{}.
\newblock \bibinfo{title}{Pretty simple bounds on quantum state discrimination}.
\newblock
\showeprint[arxiv]{1908.08312}~[quant-ph]


\bibitem[Mutolo et~al\mbox{.}(2025)]%
        {MCM+25}
\bibfield{author}{\bibinfo{person}{Vincent Mutolo}, \bibinfo{person}{Devon Campbell}, \bibinfo{person}{Quinn Manning}, \bibinfo{person}{Henri~Witold Dubourg}, \bibinfo{person}{Ruibin Lyu}, \bibinfo{person}{Simha Sethumadhavan}, \bibinfo{person}{Daniel Rubenstein}, {and} \bibinfo{person}{Salvatore Stolfo}.} \bibinfo{year}{2025}\natexlab{}.
\newblock \bibinfo{title}{Quantum computer fingerprinting using error syndromes}.
\newblock
\showeprint[arxiv]{2506.16614}~[quant-ph]


\bibitem[Myers(1999)]%
        {Myers99}
\bibfield{author}{\bibinfo{person}{Andrew~C. Myers}.} \bibinfo{year}{1999}\natexlab{}.
\newblock \showarticletitle{{JFlow}: Practical mostly-static information flow control}. In \bibinfo{booktitle}{\emph{Proceedings of the 26th ACM SIGPLAN-SIGACT symposium on Principles of programming languages}}. \bibinfo{pages}{228--241}.
\newblock


\bibitem[{N}ation and {J}ohnson(2021)]%
        {NJ21}
\bibfield{author}{\bibinfo{person}{Paul {N}ation} {and} \bibinfo{person}{Blake {J}ohnson}.} \bibinfo{year}{2021}\natexlab{}.
\newblock \bibinfo{title}{How to measure and reset a qubit in the middle of a circuit execution}.
\newblock
\urldef\tempurl%
\url{https://www.ibm.com/blogs/research/2021/02/quantum-mid-circuit-measurement/}
\showURL{%
\tempurl}


\bibitem[{\O}rb{\ae}k and Palsberg(1997)]%
        {OP97}
\bibfield{author}{\bibinfo{person}{Peter {\O}rb{\ae}k} {and} \bibinfo{person}{Jens Palsberg}.} \bibinfo{year}{1997}\natexlab{}.
\newblock \showarticletitle{Trust in the $\lambda$-calculus}.
\newblock \bibinfo{journal}{\emph{Journal of Functional Programming}} \bibinfo{volume}{7}, \bibinfo{number}{6} (\bibinfo{year}{1997}), \bibinfo{pages}{557–591}.
\newblock
\href{https://doi.org/10.1017/S0956796897002906}{doi:\nolinkurl{10.1017/S0956796897002906}}


\bibitem[Paykin et~al\mbox{.}(2017)]%
        {Paykin17}
\bibfield{author}{\bibinfo{person}{Jennifer Paykin}, \bibinfo{person}{Robert Rand}, {and} \bibinfo{person}{Steve Zdancewic}.} \bibinfo{year}{2017}\natexlab{}.
\newblock \showarticletitle{{QWIRE}: A Core Language for Quantum Circuits}. In \bibinfo{booktitle}{\emph{Proceedings of the 44th {ACM} {SIGPLAN} Symposium on Principles of Programming Languages ({POPL})}}. \bibinfo{pages}{846--858}.
\newblock
\href{https://doi.org/10.1145/3009837.3009894}{doi:\nolinkurl{10.1145/3009837.3009894}}


\bibitem[Pottier and Simonet(2002)]%
        {PS02}
\bibfield{author}{\bibinfo{person}{Fran{\c{c}}ois Pottier} {and} \bibinfo{person}{Vincent Simonet}.} \bibinfo{year}{2002}\natexlab{}.
\newblock \showarticletitle{Information flow inference for {ML}}. In \bibinfo{booktitle}{\emph{Proceedings of the 29th ACM SIGPLAN-SIGACT symposium on Principles of programming languages}}. \bibinfo{pages}{319--330}.
\newblock


\bibitem[Ren et~al\mbox{.}(2026)]%
        {Ren26}
\bibfield{author}{\bibinfo{person}{Yuanjie Ren}, \bibinfo{person}{Zhicheng Zhang}, \bibinfo{person}{Junyi Liu}, \bibinfo{person}{Jinzheng Li}, \bibinfo{person}{Yidi Qi}, {and} \bibinfo{person}{Mingsheng Ying}.} \bibinfo{year}{2026}\natexlab{}.
\newblock \bibinfo{title}{{QEDIA}: Quantum Program Verification in {L}ean}.
\newblock
\showeprint[arxiv]{2610.17708}~[quant-ph]


\bibitem[Sabelfeld and Myers(2003)]%
        {SM03}
\bibfield{author}{\bibinfo{person}{Andrei Sabelfeld} {and} \bibinfo{person}{Andrew~C. Myers}.} \bibinfo{year}{2003}\natexlab{}.
\newblock \showarticletitle{Language-based information-flow security}.
\newblock \bibinfo{journal}{\emph{IEEE Journal on selected areas in communications}} \bibinfo{volume}{21}, \bibinfo{number}{1} (\bibinfo{year}{2003}), \bibinfo{pages}{5--19}.
\newblock


\bibitem[Sabry et~al\mbox{.}(2018)]%
        {SVV18}
\bibfield{author}{\bibinfo{person}{Amr Sabry}, \bibinfo{person}{Beno{\^i}t Valiron}, {and} \bibinfo{person}{Juliana~Kaizer Vizzotto}.} \bibinfo{year}{2018}\natexlab{}.
\newblock \showarticletitle{From symmetric pattern-matching to quantum control}. In \bibinfo{booktitle}{\emph{Foundations of Software Science and Computation Structures: 21st International Conference, FOSSACS 2018}}. \bibinfo{pages}{348--364}.
\newblock
\href{https://doi.org/10.1007/978-3-319-89366-2_19}{doi:\nolinkurl{10.1007/978-3-319-89366-2_19}}


\bibitem[Schumacher(1996)]%
        {Schumacher96}
\bibfield{author}{\bibinfo{person}{Benjamin Schumacher}.} \bibinfo{year}{1996}\natexlab{}.
\newblock \showarticletitle{Sending entanglement through noisy quantum channels}.
\newblock \bibinfo{journal}{\emph{Physical Review A}} \bibinfo{volume}{54}, \bibinfo{number}{4} (\bibinfo{year}{1996}), \bibinfo{pages}{2614}.
\newblock


\bibitem[Schumacher and Nielsen(1996)]%
        {SN96}
\bibfield{author}{\bibinfo{person}{Benjamin Schumacher} {and} \bibinfo{person}{Michael~A. Nielsen}.} \bibinfo{year}{1996}\natexlab{}.
\newblock \showarticletitle{Quantum data processing and error correction}.
\newblock \bibinfo{journal}{\emph{Physical Review A}} \bibinfo{volume}{54}, \bibinfo{number}{4} (\bibinfo{year}{1996}), \bibinfo{pages}{2629}.
\newblock


\bibitem[Selinger(2004)]%
        {Selinger04}
\bibfield{author}{\bibinfo{person}{Peter Selinger}.} \bibinfo{year}{2004}\natexlab{}.
\newblock \showarticletitle{Towards a quantum programming language}.
\newblock \bibinfo{journal}{\emph{Mathematical Structures in Computer Science}} \bibinfo{volume}{14}, \bibinfo{number}{4} (\bibinfo{year}{2004}), \bibinfo{pages}{527--586}.
\newblock
\href{https://doi.org/10.1017/S0960129504004256}{doi:\nolinkurl{10.1017/S0960129504004256}}


\bibitem[Shannon(1949)]%
        {Shannon49}
\bibfield{author}{\bibinfo{person}{Claude~E. Shannon}.} \bibinfo{year}{1949}\natexlab{}.
\newblock \showarticletitle{Communication theory of secrecy systems}.
\newblock \bibinfo{journal}{\emph{The Bell system technical journal}} \bibinfo{volume}{28}, \bibinfo{number}{4} (\bibinfo{year}{1949}), \bibinfo{pages}{656--715}.
\newblock


\bibitem[Smith and Volpano(1998)]%
        {SV98}
\bibfield{author}{\bibinfo{person}{Geoffrey Smith} {and} \bibinfo{person}{Dennis Volpano}.} \bibinfo{year}{1998}\natexlab{}.
\newblock \showarticletitle{Secure information flow in a multi-threaded imperative language}. In \bibinfo{booktitle}{\emph{Proceedings of the 25th ACM SIGPLAN-SIGACT symposium on Principles of programming languages}}. \bibinfo{pages}{355--364}.
\newblock


\bibitem[Tao et~al\mbox{.}(2021)]%
        {Tao21}
\bibfield{author}{\bibinfo{person}{Runzhou Tao}, \bibinfo{person}{Yunong Shi}, \bibinfo{person}{Jianan Yao}, \bibinfo{person}{John Hui}, \bibinfo{person}{Frederic~T. Chong}, {and} \bibinfo{person}{Ronghui Gu}.} \bibinfo{year}{2021}\natexlab{}.
\newblock \showarticletitle{Gleipnir: Toward Practical Error Analysis for Quantum Programs}. In \bibinfo{booktitle}{\emph{Proceedings of the 42nd {ACM} {SIGPLAN} International Conference on Programming Language Design and Implementation ({PLDI})}}. \bibinfo{pages}{48--64}.
\newblock
\href{https://doi.org/10.1145/3453483.3454029}{doi:\nolinkurl{10.1145/3453483.3454029}}


\bibitem[Trochatos et~al\mbox{.}(2026)]%
        {TKW+25}
\bibfield{author}{\bibinfo{person}{Theodoros Trochatos}, \bibinfo{person}{Christopher Kang}, \bibinfo{person}{Andrew Wang}, \bibinfo{person}{Frederic~T Chong}, {and} \bibinfo{person}{Jakub Szefer}.} \bibinfo{year}{2026}\natexlab{}.
\newblock \showarticletitle{{TraceQ}: Trace-based reconstruction of quantum circuit dataflow in surface-code fault-tolerant quantum computing}. In \bibinfo{booktitle}{\emph{2026 IEEE International Symposium on High Performance Computer Architecture (HPCA)}}. IEEE, \bibinfo{pages}{1--14}.
\newblock


\bibitem[Unruh(2019)]%
        {Unruh19}
\bibfield{author}{\bibinfo{person}{Dominique Unruh}.} \bibinfo{year}{2019}\natexlab{}.
\newblock \showarticletitle{Quantum relational {Hoare} logic}.
\newblock \bibinfo{journal}{\emph{Proceedings of the ACM on Programming Languages}} \bibinfo{volume}{3}, \bibinfo{number}{POPL}, Article \bibinfo{articleno}{33} (\bibinfo{year}{2019}), \bibinfo{numpages}{31}~pages.
\newblock
\href{https://doi.org/10.1145/3290346}{doi:\nolinkurl{10.1145/3290346}}


\bibitem[Unruh(2020)]%
        {Unruh20}
\bibfield{author}{\bibinfo{person}{Dominique Unruh}.} \bibinfo{year}{2020}\natexlab{}.
\newblock \bibinfo{title}{Local variables and quantum relational {Hoare} logic}.
\newblock
\showeprint[arxiv]{2007.14155}~[cs.LO]


\bibitem[Voichick et~al\mbox{.}(2023)]%
        {VLRH23}
\bibfield{author}{\bibinfo{person}{Finn Voichick}, \bibinfo{person}{Liyi Li}, \bibinfo{person}{Robert Rand}, {and} \bibinfo{person}{Michael Hicks}.} \bibinfo{year}{2023}\natexlab{}.
\newblock \showarticletitle{Qunity: A unified language for quantum and classical computing}.
\newblock \bibinfo{journal}{\emph{Proceedings of the ACM on Programming Languages}} \bibinfo{volume}{7}, \bibinfo{number}{POPL} (\bibinfo{year}{2023}), \bibinfo{pages}{921--951}.
\newblock
\href{https://doi.org/10.1145/3571225}{doi:\nolinkurl{10.1145/3571225}}


\bibitem[Volpano et~al\mbox{.}(1996)]%
        {VIS96}
\bibfield{author}{\bibinfo{person}{Dennis Volpano}, \bibinfo{person}{Cynthia Irvine}, {and} \bibinfo{person}{Geoffrey Smith}.} \bibinfo{year}{1996}\natexlab{}.
\newblock \showarticletitle{A sound type system for secure flow analysis}.
\newblock \bibinfo{journal}{\emph{Journal of Computer Security}} \bibinfo{volume}{4}, \bibinfo{number}{2-3} (\bibinfo{year}{1996}), \bibinfo{pages}{167--187}.
\newblock


\bibitem[Watrous(2018)]%
        {Watrous18}
\bibfield{author}{\bibinfo{person}{John Watrous}.} \bibinfo{year}{2018}\natexlab{}.
\newblock \bibinfo{booktitle}{\emph{The theory of quantum information}}.
\newblock \bibinfo{publisher}{Cambridge university press}.
\newblock


\bibitem[{W}ilde(2013)]%
        {Wilde13}
\bibfield{author}{\bibinfo{person}{Mark~M. {W}ilde}.} \bibinfo{year}{2013}\natexlab{}.
\newblock \bibinfo{booktitle}{\emph{Quantum Information Theory}}.
\newblock \bibinfo{publisher}{Cambridge University Press}.
\newblock


\bibitem[Wootters and Zurek(1982)]%
        {WZ82}
\bibfield{author}{\bibinfo{person}{William~K. Wootters} {and} \bibinfo{person}{Wojciech~H. Zurek}.} \bibinfo{year}{1982}\natexlab{}.
\newblock \showarticletitle{A single quantum cannot be cloned}.
\newblock \bibinfo{journal}{\emph{Nature}} \bibinfo{volume}{299}, \bibinfo{number}{5886} (\bibinfo{year}{1982}), \bibinfo{pages}{802--803}.
\newblock


\bibitem[Yan et~al\mbox{.}(2024)]%
        {YJY24}
\bibfield{author}{\bibinfo{person}{Peng Yan}, \bibinfo{person}{Hanru Jiang}, {and} \bibinfo{person}{Nengkun Yu}.} \bibinfo{year}{2024}\natexlab{}.
\newblock \showarticletitle{Approximate relational reasoning for quantum programs}. In \bibinfo{booktitle}{\emph{International Conference on Computer Aided Verification}}. \bibinfo{pages}{495--519}.
\newblock


\bibitem[{Y}ing(2011)]%
        {Ying11}
\bibfield{author}{\bibinfo{person}{Mingsheng {Y}ing}.} \bibinfo{year}{2011}\natexlab{}.
\newblock \showarticletitle{Floyd{-H}oare logic for quantum programs}.
\newblock \bibinfo{journal}{\emph{ACM Transactions on Programming Languages and Systems}} \bibinfo{volume}{33}, \bibinfo{number}{6} (\bibinfo{year}{2011}), \bibinfo{pages}{1--49}.
\newblock


\bibitem[Ying et~al\mbox{.}(2013)]%
        {YFY13}
\bibfield{author}{\bibinfo{person}{Mingsheng Ying}, \bibinfo{person}{Yuan Feng}, {and} \bibinfo{person}{Nengkun Yu}.} \bibinfo{year}{2013}\natexlab{}.
\newblock \showarticletitle{Quantum information-flow security: Noninterference and access control}. In \bibinfo{booktitle}{\emph{2013 IEEE 26th Computer Security Foundations Symposium}}. \bibinfo{pages}{130--144}.
\newblock


\bibitem[Ying et~al\mbox{.}(2012)]%
        {YYF12}
\bibfield{author}{\bibinfo{person}{Mingsheng Ying}, \bibinfo{person}{Nengkun Yu}, {and} \bibinfo{person}{Yuan Feng}.} \bibinfo{year}{2012}\natexlab{}.
\newblock \bibinfo{title}{Defining quantum control flow}.
\newblock
\showeprint[arxiv]{1209.4379}~[quant-ph]


\bibitem[Ying and Zhang(2026)]%
        {YZ24}
\bibfield{author}{\bibinfo{person}{Mingsheng Ying} {and} \bibinfo{person}{Zhicheng Zhang}.} \bibinfo{year}{2026}\natexlab{}.
\newblock \showarticletitle{Verification of recursively defined quantum circuits}.
\newblock \bibinfo{journal}{\emph{Proceedings of the ACM on Programming Languages}} \bibinfo{volume}{10}, \bibinfo{number}{PLDI} (\bibinfo{year}{2026}), \bibinfo{pages}{656--680}.
\newblock


\bibitem[Yuan and Carbin(2022)]%
        {YC22}
\bibfield{author}{\bibinfo{person}{Charles Yuan} {and} \bibinfo{person}{Michael Carbin}.} \bibinfo{year}{2022}\natexlab{}.
\newblock \showarticletitle{Tower: data structures in quantum superposition}.
\newblock \bibinfo{journal}{\emph{Proceedings of the ACM on Programming Languages}} \bibinfo{volume}{6}, \bibinfo{number}{OOPSLA2} (\bibinfo{year}{2022}), \bibinfo{pages}{259--288}.
\newblock
\href{https://doi.org/10.1145/3563297}{doi:\nolinkurl{10.1145/3563297}}


\bibitem[Yuan et~al\mbox{.}(2024)]%
        {YVC24}
\bibfield{author}{\bibinfo{person}{Charles Yuan}, \bibinfo{person}{Agnes Villanyi}, {and} \bibinfo{person}{Michael Carbin}.} \bibinfo{year}{2024}\natexlab{}.
\newblock \showarticletitle{Quantum control machine: The limits of control flow in quantum programming}.
\newblock \bibinfo{journal}{\emph{Proceedings of the ACM on Programming Languages}} \bibinfo{volume}{8}, \bibinfo{number}{OOPSLA1} (\bibinfo{year}{2024}), \bibinfo{pages}{1--28}.
\newblock
\href{https://doi.org/10.1145/3649811}{doi:\nolinkurl{10.1145/3649811}}


\bibitem[Zanotti(2002)]%
        {Zanotti02}
\bibfield{author}{\bibinfo{person}{Mirko Zanotti}.} \bibinfo{year}{2002}\natexlab{}.
\newblock \showarticletitle{Security typings by abstract interpretation}. In \bibinfo{booktitle}{\emph{International Static Analysis Symposium}}. \bibinfo{pages}{360--375}.
\newblock


\bibitem[Zhang and Ying(2025a)]%
        {QAC}
\bibfield{author}{\bibinfo{person}{Zhicheng Zhang} {and} \bibinfo{person}{Mingsheng Ying}.} \bibinfo{year}{2025}\natexlab{a}.
\newblock \bibinfo{title}{Access control threatened by quantum entanglement}.
\newblock
\showeprint[arxiv]{2507.02622}~[quant-ph]


\bibitem[Zhang and Ying(2025b)]%
        {QRM}
\bibfield{author}{\bibinfo{person}{Zhicheng Zhang} {and} \bibinfo{person}{Mingsheng Ying}.} \bibinfo{year}{2025}\natexlab{b}.
\newblock \showarticletitle{Quantum register machine: Efficient implementation of quantum recursive programs}.
\newblock \bibinfo{journal}{\emph{Proceedings of the ACM on Programming Languages}} \bibinfo{volume}{9}, \bibinfo{number}{PLDI} (\bibinfo{year}{2025}), \bibinfo{pages}{822--847}.
\newblock


\bibitem[Zhou et~al\mbox{.}(2023)]%
        {Zhou23}
\bibfield{author}{\bibinfo{person}{Li Zhou}, \bibinfo{person}{Gilles Barthe}, \bibinfo{person}{Pierre{-}Yves Strub}, \bibinfo{person}{Junyi Liu}, {and} \bibinfo{person}{Mingsheng Ying}.} \bibinfo{year}{2023}\natexlab{}.
\newblock \showarticletitle{{CoqQ}: Foundational Verification of Quantum Programs}.
\newblock \bibinfo{journal}{\emph{Proceedings of the ACM on Programming Languages}} \bibinfo{volume}{7}, \bibinfo{number}{{POPL}}, Article \bibinfo{articleno}{29} (\bibinfo{year}{2023}), \bibinfo{numpages}{833--865}~pages.
\newblock
\href{https://doi.org/10.1145/3571222}{doi:\nolinkurl{10.1145/3571222}}


\bibitem[Zhou et~al\mbox{.}(2019)]%
        {ZYY19}
\bibfield{author}{\bibinfo{person}{Li Zhou}, \bibinfo{person}{Nengkun Yu}, {and} \bibinfo{person}{Mingsheng Ying}.} \bibinfo{year}{2019}\natexlab{}.
\newblock \showarticletitle{An applied quantum Hoare logic}. In \bibinfo{booktitle}{\emph{Proceedings of the 40th ACM SIGPLAN Conference on Programming Language Design and Implementation}}. \bibinfo{pages}{1149--1162}.
\newblock


\end{thebibliography}

\newpage

\appendix
\crefalias{section}{appendix}
\crefalias{subsection}{appendix}

\section{More Examples}
\label{sec:more_examples}

In this appendix, we present the verification of secure information flow in \Cref{exp:qec,exp:aeg} using our framework in \Cref{sec:verification}.

\begin{example}[Security verification of quantum error correction]
    \label{exp:veri-qec}
    Consider the program
    \begin{equation*}
        P\equiv \mathit{Enc};\mathit{Err};\mathit{Syn};\mathit{Rec}
    \end{equation*}
    in \Cref{exp:qec}.
    Since $m_1$ and $m_2$ are bit variables, we identify the guards ``$m_1=1$'' and ``$m_2=1$'' with $m_1$ and $m_2$, respectively.
    We first construct the corresponding program $\mathit{Test}$ for $P$.
    Applying \Cref{alg:dilation,def:cdil} yields:
    \begin{equation*}
        \begin{aligned}
            \cdil(P)\equiv\ 
            & \mathit{CX}[m_1,m_1^\circ];\ \mathit{CX}[m_2,m_2^\circ];\\
            &\mathit{SWAP}[q_2,q_2'];\ \mathit{SWAP}[q_3,q_3'];\\
            & \mathit{CNOT}[q_1,q_2];\ \mathit{CNOT}[q_1,q_3];\ \mathit{X}[q_2];\\
            & \mathit{SWAP}[r_1,r_1'];\ \mathit{SWAP}[r_2,r_2'];\\
            & \mathit{CNOT}[q_1,r_1];\ \mathit{CNOT}[q_2,r_1];\
            \mathit{CNOT}[q_2,r_2];\ \mathit{CNOT}[q_3,r_2];\\
            & \mathit{SWAP}[m_1,m_1'];\ \mathit{CX}[r_1,r_1''];\ \mathit{CX}[r_1,m_1];\\
            & \mathit{SWAP}[m_2,m_2'];\ \mathit{CX}[r_2,r_2''];\ \mathit{CX}[r_2,m_2];\\
            & \mathbf{qif}[m_1](\ket{0}\rightarrow\mathbf{qif}[m_2](\ket{0}\rightarrow\mathbf{skip})\square(\ket{1}\rightarrow\mathit{X}[q_3]))\\
            & \qquad\square\ \ (\ket{1}\rightarrow\mathbf{qif}[m_2](\ket{0}\rightarrow\mathit{X}[q_1])\square(\ket{1}\rightarrow\mathit{X}[q_2])).
        \end{aligned}
    \end{equation*}
    where $q_2',q_3',r_1',r_2',r_1'',r_2'',m_1',m_2',m_1^\circ,m_2^\circ$ are fresh environmental variables.

    Further applying \Cref{alg:reverse} gives:
    \begin{equation*}
        \begin{aligned}
            \rev\parens*{\cdil(P)}\equiv\ &
            \mathbf{qif}[m_1](\ket{0}\rightarrow\mathbf{qif}[m_2](\ket{0}\rightarrow\mathbf{skip})\square(\ket{1}\rightarrow\mathit{X}[q_3]))\\
            & \qquad\square\ \ (\ket{1}\rightarrow\mathbf{qif}[m_2](\ket{0}\rightarrow\mathit{X}[q_1])\square(\ket{1}\rightarrow\mathit{X}[q_2]));\\
            & \mathit{CX}[r_2,m_2];\ \mathit{CX}[r_2,r_2''];\ \mathit{SWAP}[m_2,m_2'];\\
            & \mathit{CX}[r_1,m_1];\ \mathit{CX}[r_1,r_1''];\ \mathit{SWAP}[m_1,m_1'];\\
            & \mathit{CNOT}[q_3,r_2];\ \mathit{CNOT}[q_2,r_2];\
            \mathit{CNOT}[q_2,r_1];\ \mathit{CNOT}[q_1,r_1];\\
            & \mathit{SWAP}[r_2,r_2'];\ \mathit{SWAP}[r_1,r_1'];\\
            & \mathit{X}[q_2];\ \mathit{CNOT}[q_1,q_3];\ \mathit{CNOT}[q_1,q_2];\\
            & \mathit{SWAP}[q_3,q_3'];\ \mathit{SWAP}[q_2,q_2'];\\
            & \mathit{CX}[m_2,m_2^\circ];\ \mathit{CX}[m_1,m_1^\circ].
        \end{aligned}
    \end{equation*}

    According to \Cref{eq:test}, the corresponding $\mathit{Test}$ circuit for program $P$ is
    \begin{equation}
        \label{eq:qec-test}
        \mathit{Test}\equiv
        \cdil(P);\mathit{SWAP}[m_1,m_1^\star];\mathit{SWAP}[m_2,m_2^\star];\rev\parens*{\cdil(P)}.
    \end{equation}
    where $m_1^\star,m_2^\star$ are fresh variables with the same types as $m_1,m_2$.

    To show that $P$ satisfies \textbf{NI}, our goal is to establish the judgment in \Cref{eq:test-triple}, which can be achieved by applying the proof rules introduced in \Cref{sub:qhl}, as shown in \Cref{fig:qec_proof}.
    As usual, the proof is annotated by intermediate assertions, connected by applying the (SC-P) rule. 
    Whenever needed, the (Frame) rule is applied implicitly to enable local reasoning.

    \begin{figure}
        \centering
        \includegraphics[width=0.96\textwidth]{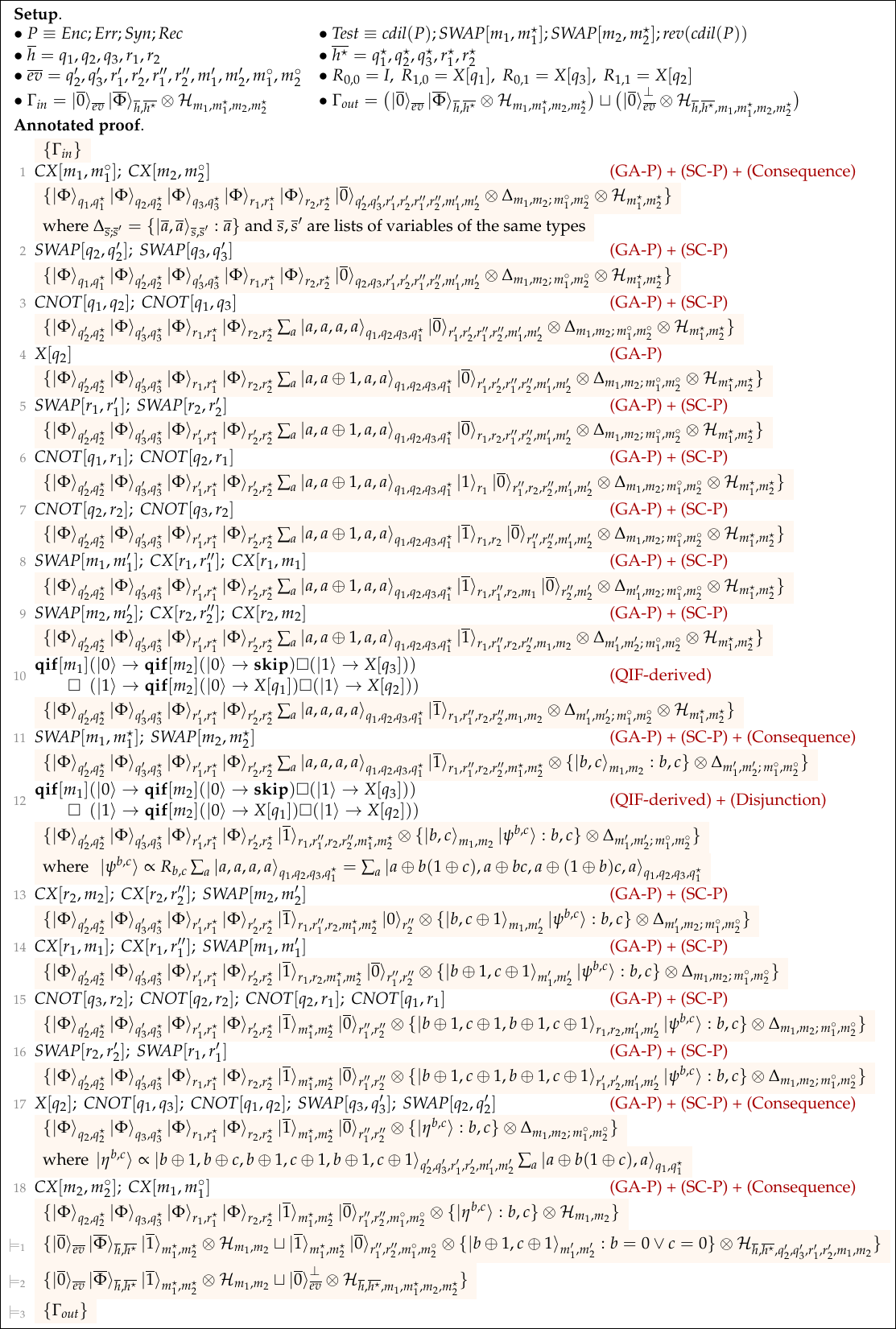}
        \caption{Annotated proof of information-flow security for logical data in quantum error correction.}
        \label{fig:qec_proof}
    \end{figure}

    Let us further explain the proof in \Cref{fig:qec_proof}.
    \begin{itemize}
        \item
            \textbf{Assertions}:
            \begin{enumerate}
                \item 
                All indices $a,b,c$ range over $\braces*{0,1}$, and $\oplus$ denotes binary XOR operation.
                \item
                As in \Cref{exp:veri-qotp}, $\ket{\Phi}$ denotes the maximally entangled state $\sum_j \ket{j,j}$ between variables of the same type, and normalizing factors are omitted in assertions.
                \item
                For convenience, we use $\Delta_{\overline{s};\overline{s'}}$ to denote the set $\braces*{\ket{\overline{a},\overline{a}}_{\overline{s},\overline{s'}}:\overline{a}}$, where $\overline{s}$ and $\overline{s'}$ are lists of variables with the same types.
            \end{enumerate}
        \item
            \textbf{Proof rules}:
            In Steps 10 and 12, the (QIF-derived) denotes the same proof pattern as in \Cref{exp:veri-qotp}:
            \begin{enumerate}
                \item
                    Apply (QIF-P) to handle the $\mathbf{qif}$ statement with branches handled by (SK-P) or (GA-P);
                \item 
                    Use (Frame) for untouched variables; and
                \item 
                    Use (Consequence) to rewrite the preconditions/postconditions.
            \end{enumerate}
            What is slightly more complicated in \Cref{exp:veri-qec} is that the $\mathit{Test}$ program involves nested $\mathbf{qif}$ statements,
            but the proof pattern is essentially the same.
        \item
            \textbf{Further details}:
            \begin{enumerate}
                \item
                    In Step 1, we first obtain
                    \begin{equation*}
                        \calH_{m_1,m_2}\models \braces*{\ket{a_1,a_2}_{m_1,m_2}:a_1,a_2},
                    \end{equation*}
                    where $a_1,a_2\in\braces*{0,1}$.
                    By (Frame), (GA-P), (SC-P), and (Consequence), this further implies
                    \begin{equation*}
                        \vdash\braces*{\calH_{m_1,m_2}\otimes\ket{\overline{0}}_{m_1^\circ,m_2^\circ}}
                        \ \mathit{CX}[m_1,m_1^\circ];\ \mathit{CX}[m_2,m_2^\circ]\ 
                        \braces*{\Delta_{m_1,m_2;m_1^\circ,m_2^\circ}},
                    \end{equation*}
                    which yields the full Hoare triple in Step 1 by (Frame).
                \item
                    In Step 11, we first apply (GA-P) and (SC-P) to derive
                    \begin{equation*}
                        \vdash\braces*{\ket{\overline{1}}_{m_1,m_2}\otimes \calH_{m_1^\star,m_2^\star}}\ \mathit{SWAP}[m_1,m_1^\star];\ \mathit{SWAP}[m_2,m_2^\star]\ 
                        \braces*{\ket{\overline{1}}_{m_1^\star,m_2^\star}\otimes \calH_{m_1,m_2}}.
                    \end{equation*}
                    Next, we can apply (Consequence) due to
                    \begin{equation*}
                        \calH_{m_1,m_2}\models\braces*{\ket{b,c}_{m_1,m_2}:b,c}.
                    \end{equation*}
                    Finally, applying (Frame) gives the full Hoare triple in Step 11. 
                \item
                    In Step 12, analogous to Step 5 in \Cref{exp:veri-qotp}, we first apply the (QIF-derived) proof pattern to obtain for $b,c\in \braces*{0,1}$:
                    \begin{equation*}
                        \begin{split}
                        \vdash&\braces*{\ket{b,c}_{m_1,m_2}\sum\nolimits_{a}\ket{a,a,a,a}_{q_1,q_2,q_3,q_1^\star}}\\
                        &\qquad\mathbf{qif}[m_1](\ket{0}\rightarrow\mathbf{qif}[m_2](\ket{0}\rightarrow\mathbf{skip})\square(\ket{1}\rightarrow\mathit{X}[q_3]))\\ 
                        &\qquad\qquad\square\ \ (\ket{1}\rightarrow\mathbf{qif}[m_2](\ket{0}\rightarrow\mathit{X}[q_1])\square(\ket{1}\rightarrow\mathit{X}[q_2]))\\
                        &\qquad\qquad\qquad\braces*{\ket{\psi^{b,c}}\ket{b,c}_{m_1,m_2}},
                        \end{split}
                    \end{equation*}
                    where we define
                    \begin{equation*}
                        \ket{\psi^{b,c}}\propto R_{b,c}\sum_a\ket{a,a,a,a}_{q_1,q_2,q_3,q_1^\star} =\sum_a\ket{a\oplus b(1\oplus c),a\oplus bc,a\oplus(1\oplus b)c,a}_{q_1,q_2,q_3,q_1^\star}
                    \end{equation*}
                    and $R_{b,c}$ is defined in \Cref{fig:qec_proof}.
                    By applying (Disjunction) and (Frame), one can obtain the full Hoare triple in Step 12.
                \item
                    In Step 17, we exploit the equalities $(a\oplus bc\oplus 1)\oplus (a\oplus b(1\oplus c))=b\oplus 1$ and $(a\oplus b(1\oplus c))\oplus(a\oplus (1\oplus b)c) =b\oplus c$.
                \item
                    Step 18 is analogous to Step 1.
                    The three $\models$ after Step 18 exploit (Frame) and (Consequence).
                    In particular, the first $\models_1$ uses 
                    \begin{equation*}
                        \begin{split}
                            &\braces*{\ket{\eta^{b,c}}:b,c}\models \braces*{\ket{\Phi}_{q_1,q_1^\star}\ket{\overline{0}}_{q_2',q_3',r_1',r_2',m_1',m_2'}}\sqcup \braces*{\ket{\eta^{b,c}}:b=0\vee c=0}\\
                            &\braces*{\ket{\Phi}_{q_2,q_2^\star} \ket{\Phi}_{q_3,q_3^\star}\ket{\Phi}_{r_1,r_1^\star}\ket{\Phi}_{r_2,r_2^\star}}\otimes\braces*{\ket{\eta^{b,c}}:b=0\vee c=0}\\
                            &\qquad\qquad\quad\models \braces*{\ket{b\oplus 1,c\oplus 1}_{m_1',m_2'}:b=0\vee c=0}\otimes \calH_{\overline{h},\overline{h^\star},q_2',q_3',r_1',r_2'}.
                        \end{split}
                    \end{equation*}
                    It groups the states in the postcondition according to the component $\ket{b\oplus 1,c\oplus 1}_{m_1',m_2'}$, so it becomes $\sqcup$ of two assertions corresponding to $b=1\wedge c=1$ and $b=0\vee c=0$.
                    The second $\models_2$ uses
                    \begin{equation*}
                        \braces*{\ket{\overline{0}}_{r_1'',r_2'',m_1^\circ,m_2^\circ}}\otimes\braces*{\ket{b\oplus 1, c\oplus 1}_{m_1',m_2'}:b=0\vee c=0}\otimes \calH_{q_2',q_3',r_1',r_2'}
                        \models \ket{\overline{0}}_{\overline{ev}}^\bot.
                    \end{equation*}
                    The third $\models_3$ uses $\braces*{\ket{\overline{1}}_{m_1^\star,m_2^\star}}\models \calH_{m_1^\star,m_2^\star}$.
            \end{enumerate}
    \end{itemize}
\end{example}

\begin{example}[Security verification of anonymous entanglement generation]
    \label{exp:veri-aeg}
    Consider the program
    \begin{equation*}
        P\equiv\mathit{Tran};\mathit{Rec}
    \end{equation*}
    in \Cref{exp:aeg}.
    By the compositionality of \textbf{NI} in \Cref{thm:compose}, we will verify $\mathit{Tran}$ and $\mathit{Rec}$ independently.

    \paragraph{Verification of $\mathit{Tran}$}
    We first construct the corresponding $\mathit{Test}$ for $\mathit{Tran}$.
    Applying \Cref{alg:dilation,def:cdil} to $\mathit{Tran}$ yields:
    \begin{equation*}
        \begin{split}
            \cdil(\mathit{Tran})\equiv\ 
            &\mathit{CX}[\mathit{id},\mathit{id}^\circ];\ \mathit{CX}[\overline{u},\overline{u^\circ}];\ \mathit{CX}[\overline{v},\overline{v^\circ}];\\
            &\mathit{Rbit}_1;\mathit{Rbit}_2;\mathit{Rbit}_3;\mathit{Rbit}_4;\\
            &\mathit{SWAP}[\overline{q},\overline{q'}];\\
            &\mathit{H}[q_1];\
            \mathit{CNOT}[q_1,q_2];\
            \mathit{CNOT}[q_1,q_3];\
            \mathit{CNOT}[q_1,q_4];\\
            &\mathit{Broad}_1;\mathit{Broad}_2;\mathit{Broad}_3;\mathit{Broad}_4,
        \end{split}
    \end{equation*}
    where quantum circuits $\mathit{Rbit}_j$ and $\mathit{Broad}_j$ are defined as follows:
    \begin{equation*}
        \begin{split}
            \mathit{Rbit}_j\equiv\ &
            \mathit{SWAP}[w_j,w_j'];\ \mathit{H}[w_j];\
            \mathit{SWAP}[u_j,u_j'];\ \mathit{CX}[w_j,w_j''];\ \mathit{CX}[w_j,u_j],\\
            \mathit{Broad}_j\equiv\ &
            \mathit{SWAP}[\mathit{at}_j,\mathit{at}_j'];\
            \mathit{C}_{s(\mathit{id})=j\vee r(\mathit{id})=j}
            [\mathit{at}_j,\mathit{id}];\\
            &
            \mathbf{qif}[\mathit{at}_j]
            (\ket{0}\rightarrow \mathit{H}[q_j];\ \mathit{SWAP}[v_j,v_j''];\
            \mathit{CX}[q_j,q_j''];\ \mathit{CX}[q_j,v_j])\\
            &\qquad\square\ \ \,(\ket{1}\rightarrow
            \mathit{SWAP}[v_j,v_j'];\ \mathit{C}_{u_j}[v_j,u_j])
        \end{split}
    \end{equation*}
    for $j=1$ to $4$, and we use the convention
    \begin{equation*}
        \mathit{SWAP}[\overline{q},\overline{q'}]\equiv \mathit{SWAP}[q_1,q_1'];\ \mathit{SWAP}[q_2,q_2'];\ \mathit{SWAP}[q_3,q_3'];\ \mathit{SWAP}[q_4,q_4'].
    \end{equation*}

    Here, each $\mathit{Rbit}_j$ corresponds to the probabilistic assignment
    $u_j:=\braces*{(1/2,0),(1/2,1)}$, and we expand the syntactic sugar with $w_j$ being the implicit qubit variable, before applying \Cref{alg:dilation}.
    Each $\mathit{Broad}_j$ corresponds to the $\mathbf{if}$ statement in $\mathit{Tran}$. The reversible circuit $\mathit{C}_{s(\mathit{id})=j\vee r(\mathit{id})=j}[\mathit{at}_j,\mathit{id}]$ computes the value of the Boolean expression $s(\mathit{id})=j\vee r(\mathit{id})=j$ into $\mathit{at}_j$ (see the explanation after \Cref{alg:dilation}).
    The fresh environmental variables introduced by \Cref{alg:dilation,def:cdil} include $w_j',w_j'',u_j',q_j',q_j'',\mathit{at}_j,\mathit{at}_j',v_j',v_j'',\mathit{id}^\circ,\overline{u^\circ},\overline{v^\circ}$.

    Next, applying \Cref{alg:reverse} gives:
    \begin{equation*}
        \begin{split}
            \rev\parens*{\cdil(\mathit{Tran})}\equiv\ &
            \mathit{Broad}_4^\dagger;\mathit{Broad}_3^\dagger;
            \mathit{Broad}_2^\dagger;\mathit{Broad}_1^\dagger;\\
            &\mathit{CNOT}[q_1,q_4];\
            \mathit{CNOT}[q_1,q_3];\
            \mathit{CNOT}[q_1,q_2];\
            \mathit{H}[q_1];\\
            &\mathit{SWAP}[\overline{q},\overline{q'}];\\
            &\mathit{Rbit}_4^\dagger;\mathit{Rbit}_3^\dagger;
            \mathit{Rbit}_2^\dagger;\mathit{Rbit}_1^\dagger;\\
            &\mathit{CX}[\overline{v},\overline{v^\circ}];\ \mathit{CX}[\overline{u},\overline{u^\circ}];\ \mathit{CX}[\mathit{id},\mathit{id}^\circ]
        \end{split}
    \end{equation*}
    where quantum circuits $\mathit{Rbit}_j^\dagger$ and $\mathit{Broad}_j^\dagger$ are defined as follows:
    \begin{equation*}
        \begin{split}
            \mathit{Rbit}_j^\dagger\equiv\ &
            \mathit{CX}[w_j,u_j];\ \mathit{CX}[w_j,w_j''];\
            \mathit{SWAP}[u_j,u_j'];\ \mathit{H}[w_j];\ \mathit{SWAP}[w_j,w_j'],\\
            \mathit{Broad}_j^\dagger\equiv\ &
            \mathbf{qif}[\mathit{at}_j]
            (\ket{0}\rightarrow\mathit{CX}[q_j,v_j];\ \mathit{CX}[q_j,q_j''];\
            \mathit{SWAP}[v_j,v_j''];\ \mathit{H}[q_j])\\
            &\qquad\square\ \ \,(\ket{1}\rightarrow
            \mathit{C}_{u_j}[v_j,u_j];\ \mathit{SWAP}[v_j,v_j']);\\
            &\mathit{C}_{s(\mathit{id})=j\vee r(\mathit{id})=j}
            [\mathit{at}_j,\mathit{id}];\
            \mathit{SWAP}[\mathit{at}_j,\mathit{at}_j'].
        \end{split}
    \end{equation*}
    According to \Cref{eq:test}, the corresponding test circuit for $\mathit{Tran}$ is
    \begin{equation}
        \label{eq:aeg-test-tran}
        \mathit{Test}\equiv
        \cdil(\mathit{Tran});
        \mathit{SWAP}[\overline{v},\overline{v^\star}];
        \rev\parens*{\cdil(\mathit{Tran})},
    \end{equation}
    where $\mathit{SWAP}[\overline{v},\overline{v^\star}]$ is defined similarly to $\mathit{SWAP}[\overline{q},\overline{q'}]$.%

    To show that $\mathit{Tran}$ satisfies \textbf{NI}, our goal is to establish the judgment in \Cref{eq:test-triple}, which can be achieved by applying the proof rules introduced in \Cref{sub:qhl}, as shown in \Cref{fig:aeg_tran_proof}.
    As usual, the proof is annotated by intermediate assertions, connected by applying the (SC-P) rule. 
    Whenever needed, the (Frame) rule is applied implicitly to enable local reasoning.

    \begin{figure}
        \centering
        \includegraphics[width=\textwidth]{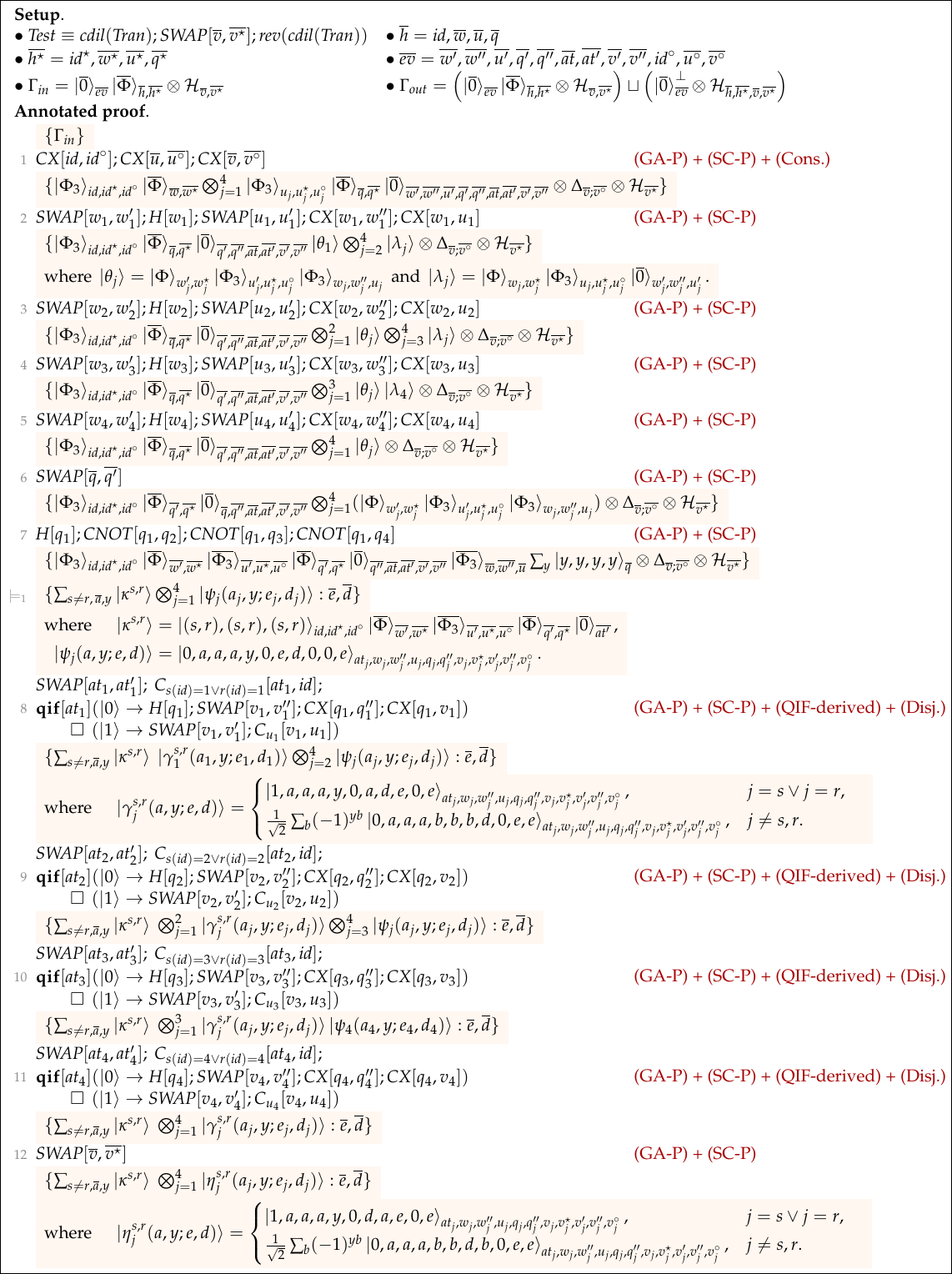}
        \caption{Annotated proof of information-flow security for anonymous entanglement generation: transmission stage.}
        \label{fig:aeg_tran_proof}
    \end{figure}

    \begin{figure}
        \ContinuedFloat
        \centering
        \includegraphics[width=\textwidth]{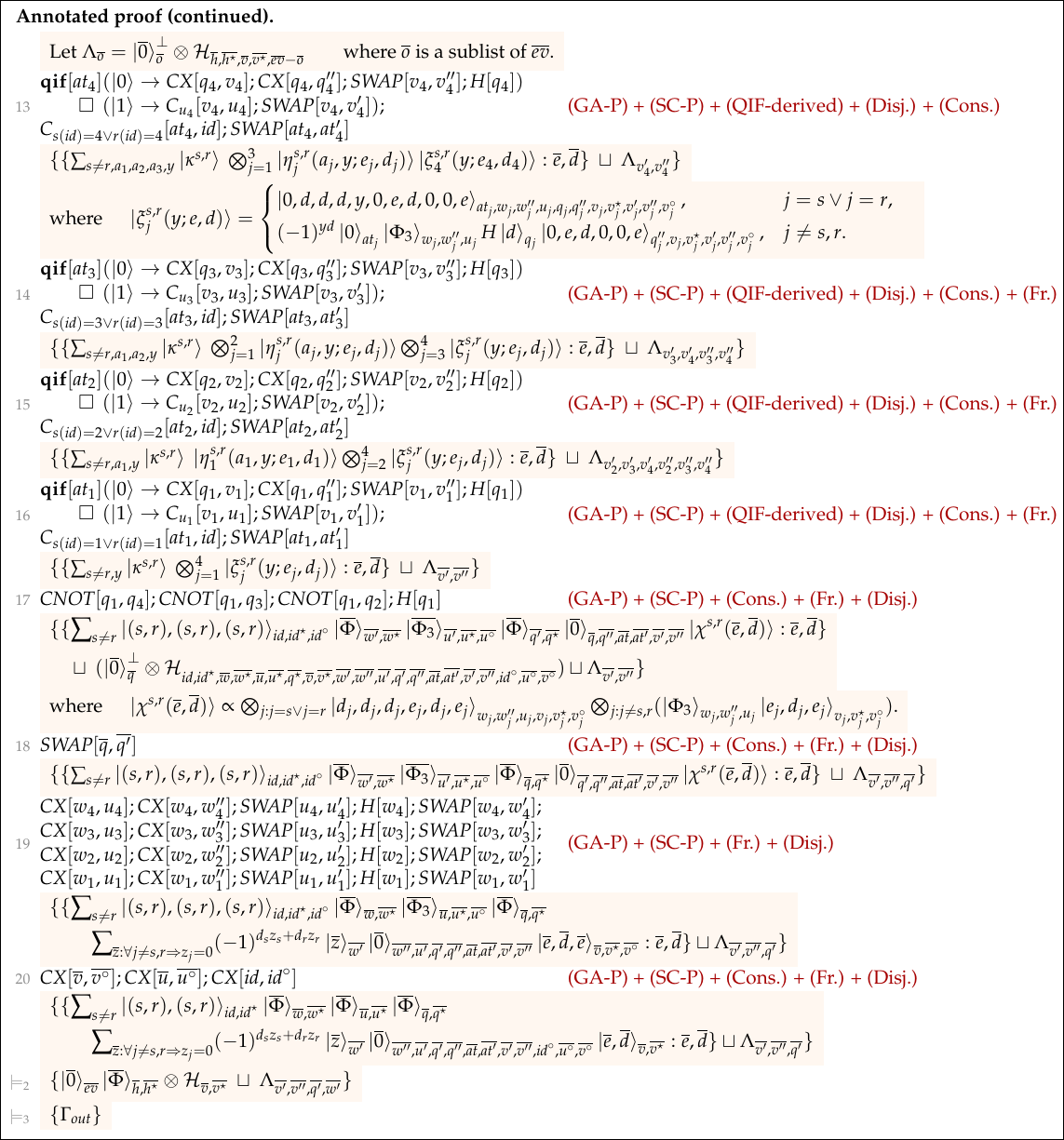}
        \caption{Annotated proof of information-flow security for anonymous entanglement generation: transmission stage.}
    \end{figure}

    Let us further explain the proof in \Cref{fig:aeg_tran_proof}.
    \begin{itemize}
        \item
            \textbf{Assertions}:
            \begin{enumerate}
                \item
                All indices $s,r,j$ range over $\braces*{1,2,3,4}$, where $s\neq r$. 
                All indices $y,b,a_j,e_j,d_j,z_j$ range over $\braces*{0,1}$.
                \item
                As in \Cref{exp:veri-qotp,exp:veri-qec}, $\ket{\Phi}$ denotes $\sum_j\ket{j,j}$ and $\ket{\Phi_3}$ denotes $\sum_j \ket{j,j,j}$.
                Here, the range of $j$ is implicit; e.g., in $\ket{\Phi_3}_{\mathit{id},\mathit{id}^\star,\mathit{id}^\circ}$ (see Step 1), $j$ ranges over all $(s,r)$ with $s\neq r$.
                As usual, normalizing factors are omitted in assertions whenever needed.
                \item
                Similar to $\ket{\overline{\Phi}}_{\overline{q},\overline{q'}}$ (see \Cref{sub:verification_by_self_conjugation}) that denotes $\bigotimes_{j=1}^n\ket{\Phi}_{q_j,q_j'}$ for $\overline{q}=q_1,\ldots,q_n$ and $\overline{q'}=q_1',\ldots,q_n'$,
                we use $\ket{\overline{\Phi_3}}_{\overline{q},\overline{q'},\overline{q''}}$ to denote $\bigotimes_{j=1}^n \ket{\Phi_3}_{q_j,q_j',q_j''}$.
                \item
                As in \Cref{exp:veri-qec}, we use $\Delta_{\overline{q};\overline{q'}}$ to denote the set $\braces*{\ket{\overline{a},\overline{a}}_{\overline{q},\overline{q'}}:\overline{a}}$.
            \end{enumerate}
        \item
            \textbf{Proof rules}:
            In Steps 8--11 and 13--16, the (QIF-derived) denotes the same proof pattern as in \Cref{exp:veri-qotp,exp:veri-qec}.
            On the RHS, (Disj.), (Fr.), and (Cons.) abbreviate (Disjunction), (Frame), and (Consequence), respectively.
        \item
            \textbf{Further details}:
            \begin{enumerate}
                \item
                Step 1 follows the same reasoning as Steps 1 in \Cref{exp:veri-qotp,exp:veri-qec}.
                \item
                In Step 2, we first apply (GA-P) and (SC-P) to derive
                \begin{equation*}
                    \begin{split}
                        \vdash&\braces*{\ket{\Phi}_{w_1,w_1^\star}\ket{\Phi_3}_{u_1,u_1^\star,u_1^\circ}
                        \ket{0,0,0}_{w_1',w_1'',u_1'}}\\
                        &\qquad
                        \mathit{SWAP}[w_1,w_1'];\mathit{H}[w_1];
                        \mathit{SWAP}[u_1,u_1'];\mathit{CX}[w_1,w_1''];\mathit{CX}[w_1,u_1]\\
                        &\qquad\qquad
                        \braces*{\ket{\Phi}_{w_1',w_1^\star}\ket{\Phi_3}_{u_1',u_1^\star,u_1^\circ}
                        \ket{\Phi_3}_{w_1,w_1'',u_1}}.
                    \end{split}
                \end{equation*}
                Applying (Frame) immediately gives the full Hoare triple in Step 2.
                Steps 3--5 are analogous.

                \item
                In Step 7, we first apply (GA-P) and (SC-P) to derive
                \begin{equation*}
                    \vdash\braces*{\ket{\overline{0}}_{\overline q}}\ 
                    \mathit{H}[q_1];\mathit{CNOT}[q_1,q_2];\mathit{CNOT}[q_1,q_3];\mathit{CNOT}[q_1,q_4]\
                    \braces*{\sum\nolimits_y\ket{y,y,y,y}_{\overline{q}}},
                \end{equation*}
                which leads to the full Hoare triple in Step 7 after applying (Frame).
                Then, the $\models_1$ exploits
                \begin{equation*}
                    \Delta_{\overline{v};\overline{v^\circ}}\otimes\calH_{\overline{v^\star}}
                    \models
                    \braces*{\ket{\overline e,\overline d,\overline {e}}_{\overline v,\overline{v^\star},\overline{v^\circ}}:\overline e,\overline d}
                \end{equation*}
                and the definitions of $\ket{\kappa^{s,r}}$ and $\ket{\psi_j(a_j,y;e_j,d_j)}$.
                Note that $\ket{(s,r)}$ is a computational basis state of quantum variable $\mathit{id}$,
                which has type in \Cref{eq:id-type} and therefore requires $s\neq r$.

                \item
                In Steps 8--11, we first apply (GA-P) and (SC-P) to obtain, for $j\in \braces*{1,2,3,4}$:
                \begin{equation*}
                    \begin{split}
                    \vdash &\braces*{\sum\nolimits_{s\neq r}\ket{(s,r)}_{\mathit{id}}\ket{0,0}_{\mathit{at}_j,\mathit{at}_j'}}\\
                    &\qquad\mathit{SWAP}[\mathit{at}_j,\mathit{at}_j'];\
                    C_{s(\mathit{id})=j\vee r(\mathit{id})=j}[\mathit{at}_j,\mathit{id}]\\
                    &\qquad\qquad\braces{\sum\nolimits_{s\neq r}\ket{(s,r)}_{\mathit{id}}\ket{j=s\vee j=r,0}_{\mathit{at}_j,\mathit{at}_j'}}.
                    \end{split}
                \end{equation*}
                Then, we apply the (QIF-derived) proof pattern similar to Step 5 in \Cref{exp:veri-qotp}.
                In particular, for $j\in \braces*{1,2,3,4}$ and $e_j\in \braces*{0,1}$, the two premises of (QIF-P) are
                \begin{equation*}
                    \begin{split}
                    \vdash&\braces*{\sum\nolimits_y\ket{0,y,0,e_j,0}_{\mathit{at}_j,q_j,q_j'',v_j,v_j''}}\\
                    &\qquad\mathit H[q_j];\mathit{SWAP}[v_j,v_j''];
                    \mathit{CX}[q_j,q_j''];\mathit{CX}[q_j,v_j]\\
                    &\qquad\qquad
                    \braces*{
                    \sum\nolimits_{y,b}(-1)^{yb}
                    \ket{0,b,b,b,e_j}_{\mathit{at}_j,q_j,q_j'',v_j,v_j''}}
                    \end{split}
                \end{equation*}
                and
                \begin{equation*}
                    \vdash\braces*{\sum\nolimits_{a_j}\ket{1,a_j,e_j,0}_{\mathit{at}_j,u_j,v_j,v_j'}}\
                    \mathit{SWAP}[v_j,v_j'];\mathit C_{u_j}[v_j,u_j]\
                    \braces*{\sum\nolimits_{a_j}\ket{1,a_j,a_j,e_j}_{\mathit{at}_j,u_j,v_j,v_j'}}.
                \end{equation*}
                By the definitions of $\ket{\psi_j(a_j,y;e_j,d_j)}$ and $\ket{\gamma_j^{s,r}(a_j,y;e_j,d_j)}$, applying (QIF-P), (Frame), and (Consequence), followed by (Disjunction) over $\overline{e},\overline{d}$, gives the full Hoare triples in Steps 8--11.
                \item
                In Steps 13--16, we first apply the (QIF-derived) proof pattern similar to Step 5 in \Cref{exp:veri-qotp}.
                In particular, for $j\in\braces*{1,2,3,4}$ and $e_j,d_j\in\braces*{0,1}$, the two premises of (QIF-P) are
                \begin{equation*}
                    \begin{split}
                    \vdash&\braces*{
                    \sum\nolimits_{y,b}(-1)^{yb}
                    \ket{0,b,b,d_j,e_j}_{\mathit{at}_j,q_j,q_j'',v_j,v_j''}}\\
                    &\ \  
                    \mathit{CX}[q_j,v_j];\mathit{CX}[q_j,q_j''];
                    \mathit{SWAP}[v_j,v_j''];\mathit H[q_j]\\
                    &\quad
                    \braces*{\sum\nolimits_y
                    \parens*{
                    (-1)^{yd_j}
                    \ket0_{\mathit{at}_j}\mathit H\ket{d_j}_{q_j}
                    \ket{0,e_j,0}_{q_j'',v_j,v_j''}
                    +
                    (-1)^{y (d_j \oplus 1)}
                    \ket0_{\mathit{at}_j}\mathit H\ket {d_j\oplus 1}_{q_j}
                    \ket{0,e_j,1}_{q_j'',v_j,v_j''}}},
                    \end{split}
                \end{equation*}
                and
                \begin{equation*}
                    \begin{split}
                    \vdash&\braces*{\sum\nolimits_{a_j}\ket{1,a_j,d_j,e_j}_{\mathit{at}_j,u_j,v_j,v_j'}}\\
                    &\qquad\mathit C_{u_j}[v_j,u_j];\mathit{SWAP}[v_j,v_j']\\
                    &\qquad\qquad
                    \braces*{
                    \ket{1,d_j,e_j,0}_{\mathit{at}_j,u_j,v_j,v_j'}
                    +
                    \ket{1,d_j\oplus 1,e_j,1}_{\mathit{at}_j,u_j,v_j,v_j'}}.
                    \end{split}
                \end{equation*}
                Moreover, applying (GA-P) and (SC-P) gives
                \begin{equation*}
                    \begin{split}
                    \vdash&\braces*{
                    \sum\nolimits_{s\ne r}\ket{(s,r)}_{\mathit{id}}
                    \ket{j=s\vee j=r,0}_{\mathit{at}_j,\mathit{at}_j'}}\\
                    &\qquad
                    C_{s(\mathit{id})=j\vee r(\mathit{id})=j}[\mathit{at}_j,\mathit{id}];\
                    \mathit{SWAP}[\mathit{at}_j,\mathit{at}_j']\\
                    &\qquad\qquad
                    \braces*{
                    \sum\nolimits_{s\ne r}\ket{(s,r)}_{\mathit{id}}
                    \ket{0,0}_{\mathit{at}_j,\mathit{at}_j'}}.
                    \end{split}
                \end{equation*}
                By the definitions of $\ket{\xi_j^{s,r}(y;e_j,d_j)}$ and $\Lambda_{\overline{o}}$, applying (QIF-P), (Frame), and (Consequence), together with (Disjunction) over $\overline e,\overline d$, gives the full Hoare triples in Steps 13--16.
                Note that the terms with $b= d_j\oplus 1$ or $a_j= d_j\oplus 1$ are weakened to the corresponding $\Lambda$ by (Consequence).

                \item
                In Step 17, we first apply (GA-P) and (SC-P), followed by (Consequence), to obtain, for fixed $\overline e,\overline d$:
                \begin{equation*}
                    \begin{split}
                    \vdash&\braces*{
                    \sum\nolimits_{s\ne r}\sum\nolimits_y
                    \ket{(s,r),(s,r),(s,r)}_{\mathit{id},\mathit{id}^\star,\mathit{id}^\circ}
                    \bigotimes\nolimits_{j:j=s\vee j=r}\ket y_{q_j}
                    \bigotimes\nolimits_{j:j\ne s,r}(-1)^{yd_j}\mathit H\ket{d_j}_{q_j}
                    \ket{\chi^{s,r}(\overline e,\overline d)}}\\
                    &\qquad
                    \mathit{CNOT}[q_1,q_4];\mathit{CNOT}[q_1,q_3];
                    \mathit{CNOT}[q_1,q_2];\mathit H[q_1]\\
                    &\qquad\qquad
                    \braces*{
                    \braces*{
                    \sum\nolimits_{s\ne r}
                    \ket{(s,r),(s,r),(s,r)}_{\mathit{id},\mathit{id}^\star,\mathit{id}^\circ}
                    \ket{\overline{0}}_{\overline q}
                    \ket{\chi^{s,r}(\overline e,\overline d)}}
                    \sqcup
                    \parens*{
                    \ket{\overline{0}}_{\overline q}^{\bot}
                    \otimes
                    \calH_{\mathit{id},\mathit{id}^\star,\mathit{id}^\circ,
                    \overline w,\overline{w''},\overline u,
                    \overline v,\overline{v^\star},\overline{v^\circ}}}},
                    \end{split}
                \end{equation*}
                where for obtaining the first component in the postcondition, we use $\bra{y}H\ket{d_j}=\frac{1}{\sqrt{2}}(-1)^{yd_j}$.
                Further applying (Frame) and (Disjunction) gives the full Hoare triple in Step 17.

                \item
                In Step 19, we first apply (GA-P) and (SC-P) to obtain, for $j\in\braces*{1,2,3,4}$:
                \begin{equation*}
                    \begin{split}
                    \vdash&\braces*{
                    \ket{\Phi}_{w_j',w_j^\star}
                    \ket{\Phi_3}_{u_j',u_j^\star,u_j^\circ}
                    \ket{d_j,d_j,d_j}_{w_j,w_j'',u_j}}\\
                    &\qquad
                    \mathit{CX}[w_j,u_j];\mathit{CX}[w_j,w_j''];
                    \mathit{SWAP}[u_j,u_j'];\mathit H[w_j];\mathit{SWAP}[w_j,w_j']\\
                    &\qquad\qquad
                    \braces*{
                    \ket{\Phi}_{w_j,w_j^\star}
                    \ket{\Phi_3}_{u_j,u_j^\star,u_j^\circ}
                    \sum\nolimits_{z_j}(-1)^{d_jz_j}
                    \ket{z_j,0,0}_{w_j',w_j'',u_j'}}.
                    \end{split}
                \end{equation*}
                for $j=s\vee j=r$, and
                \begin{equation*}
                    \begin{split}
                    \vdash&\braces*{
                    \ket{\Phi}_{w_j',w_j^\star}
                    \ket{\Phi_3}_{u_j',u_j^\star,u_j^\circ}
                    \sum\nolimits_{a_j}\ket{a_j,a_j,a_j}_{w_j,w_j'',u_j}}\\
                    &\qquad
                    \mathit{CX}[w_j,u_j];\mathit{CX}[w_j,w_j''];
                    \mathit{SWAP}[u_j,u_j'];\mathit H[w_j];\mathit{SWAP}[w_j,w_j']\\
                    &\qquad\qquad
                    \braces*{
                    \ket{\Phi}_{w_j,w_j^\star}
                    \ket{\Phi_3}_{u_j,u_j^\star,u_j^\circ}
                    \ket{0,0,0}_{w_j',w_j'',u_j'}}.
                    \end{split}
                \end{equation*}
                for $j\ne s,r$.
                Applying (Frame) and (SC-P) to the above for $j=1,2,3,4$ gives the first component in the assertion after Step 19.
                Further applying (Frame) and (Disjunction) preserves the existing $\Lambda_{\overline{v'},\overline{v''},\overline{q'}}$ component and gives the full Hoare triple in Step 19.
            \item
                Step 20 is analogous to Step 1 with additional (Frame) and (Disjunction).
                The $\models_2$ groups the states in the first component according to the state $\ket{\overline{z}}_{\overline{w'}}$, yielding the disjunction $\sqcup$ of two assertions corresponding to $\overline z=\overline{0}$ and $\overline z\ne\overline{0}$ by further applying (Consequence).

                The $\models_3$ follows from the definition of $\Lambda_{\overline{o}}$ and $\Gamma_{\mathit{out}}$.
            \end{enumerate}
    \end{itemize}

    \paragraph{Verification of $\mathit{Rec}$}
    We first construct the corresponding program $\mathit{Test}$ for $\mathit{Rec}$.
    For readability, we define the Boolean expression $\mathit{pt}_j=v_1\oplus v_2\oplus v_3\oplus v_4\oplus v_j$.
    Applying \Cref{alg:dilation,def:cdil} to $\mathit{Rec}$ yields:
    \begin{equation*}
        \begin{split}
            \cdil(\mathit{Rec})\equiv\ 
            &\mathit{CX}[\mathit{id},\mathit{id}^\circ];\ \mathit{CX}[\overline{u},\overline{u^\circ}];\ \mathit{CX}[\overline{v},\overline{v^\circ}];\\
            &\mathit{Send}_1;\mathit{Send}_2;\mathit{Send}_3;\mathit{Send}_4;\\
            &\mathit{Recv}_1;\mathit{Recv}_2;\mathit{Recv}_3;\mathit{Recv}_4
        \end{split}
    \end{equation*}
    where quantum circuits $\mathit{Send}_j$ and $\mathit{Recv}_j$ are defined as follows:
    \begin{equation*}
        \begin{split}
            \mathit{Send}_j\equiv\ &
            \mathit{SWAP}[\mathit{sd}_j,\mathit{sd}_j'];\
            \mathit{C}_{s(\mathit{id})=j\wedge u_j=1}
            [\mathit{sd}_j,\mathit{id},u_j];\\
            &\mathbf{qif}[\mathit{sd}_j](\ket{0}\rightarrow \mathbf{skip})
            \square(\ket{1}\rightarrow \mathit{Z}[q_j])\\
            \mathit{Recv}_j\equiv\ &
            \mathit{SWAP}[\mathit{rc}_j,\mathit{rc}_j'];\
            \mathit{C}_{r(\mathit{id})=j\wedge\mathit{pt}_j=1}
            [\mathit{rc}_j,\mathit{id},\overline{v}];\\
            &\mathbf{qif}[\mathit{rc}_j](\ket{0}\rightarrow \mathbf{skip})
            \square(\ket{1}\rightarrow \mathit{Z}[q_j])
        \end{split}
    \end{equation*}
    for $j=1$ to $4$.

    Here, $\mathit{Send}_j$ corresponds to the first four $\mathbf{if}$ statements in $\mathit{Rec}$ and $\mathit{Recv}_j$ to the last four.
    The reversible circuits $C_{s(\mathit{id})=j\wedge u_j=1}$ and $C_{r(\mathit{id})=j\wedge \mathit{pt}_j=1}$ calculate the Boolean guards of these $\mathbf{if}$ statements. The fresh environmental variables introduced by \Cref{alg:dilation,def:cdil} include $\mathit{sd}_j,\mathit{sd}_j',\mathit{rc}_j,\mathit{rc}_j',\mathit{id}^\circ,\overline{u^\circ},\overline{v^\circ}$.

    Next, applying \Cref{alg:reverse} gives:
    \begin{equation*}
        \begin{aligned}
            \rev\parens*{\cdil(\mathit{Rec})}\equiv\ &
            \mathit{Recv}_4^\dagger;\mathit{Recv}_3^\dagger;
            \mathit{Recv}_2^\dagger;\mathit{Recv}_1^\dagger;\\
            &\mathit{Send}_4^\dagger;\mathit{Send}_3^\dagger;
            \mathit{Send}_2^\dagger;\mathit{Send}_1^\dagger;\\
            &\mathit{CX}[\overline{v},\overline{v^\circ}];\ \mathit{CX}[\overline{u},\overline{u^\circ}];\ \mathit{CX}[\mathit{id},\mathit{id}^\circ]
        \end{aligned}
    \end{equation*}
    where quantum circuits $\mathit{Send}_j^\dagger$ and $\mathit{Recv}_j^\dagger$ are defined as follows:
    \begin{equation*}
        \begin{aligned}
            \mathit{Send}_j^\dagger\equiv\ &
            \mathbf{qif}[\mathit{sd}_j](\ket{0}\rightarrow \mathbf{skip})
            \square(\ket{1}\rightarrow \mathit{Z}[q_j]);\\
            &\mathit{C}_{s(\mathit{id})=j\wedge u_j=1}
            [\mathit{sd}_j,\mathit{id},u_j];\
            \mathit{SWAP}[\mathit{sd}_j,\mathit{sd}_j'],\\
            \mathit{Recv}_j^\dagger\equiv\ &
            \mathbf{qif}[\mathit{rc}_j](\ket{0}\rightarrow \mathbf{skip})
            \square(\ket{1}\rightarrow \mathit{Z}[q_j]);\\
            &\mathit{C}_{r(\mathit{id})=j\wedge\mathit{pt}_j=1}
            [\mathit{rc}_j,\mathit{id},\overline{v}];\
            \mathit{SWAP}[\mathit{rc}_j,\mathit{rc}_j'].
        \end{aligned}
    \end{equation*}
    According to \Cref{eq:test}, 
    the corresponding test circuit for $\mathit{Rec}$ is
    \begin{equation}
        \label{eq:aeg-test-rec}
        \mathit{Test}\equiv
        \cdil(\mathit{Rec});
        \mathit{SWAP}[\overline{v},\overline{v^\star}];
        \rev\parens*{\cdil(\mathit{Rec})}.
    \end{equation}
    To show that $\mathit{Rec}$ satisfies \textbf{NI}, our goal is to establish the judgment in \Cref{eq:test-triple}, which can be achieved by applying the proof rules introduced in \Cref{sub:qhl}, as shown in \Cref{fig:aeg_rec_proof}.
    As usual, the proof is annotated by intermediate assertions, connected by applying the (SC-P) rule. 
    The (Frame) rule is applied implicitly to enable local reasoning.

    \begin{figure}
        \centering
        \includegraphics[width=\textwidth]{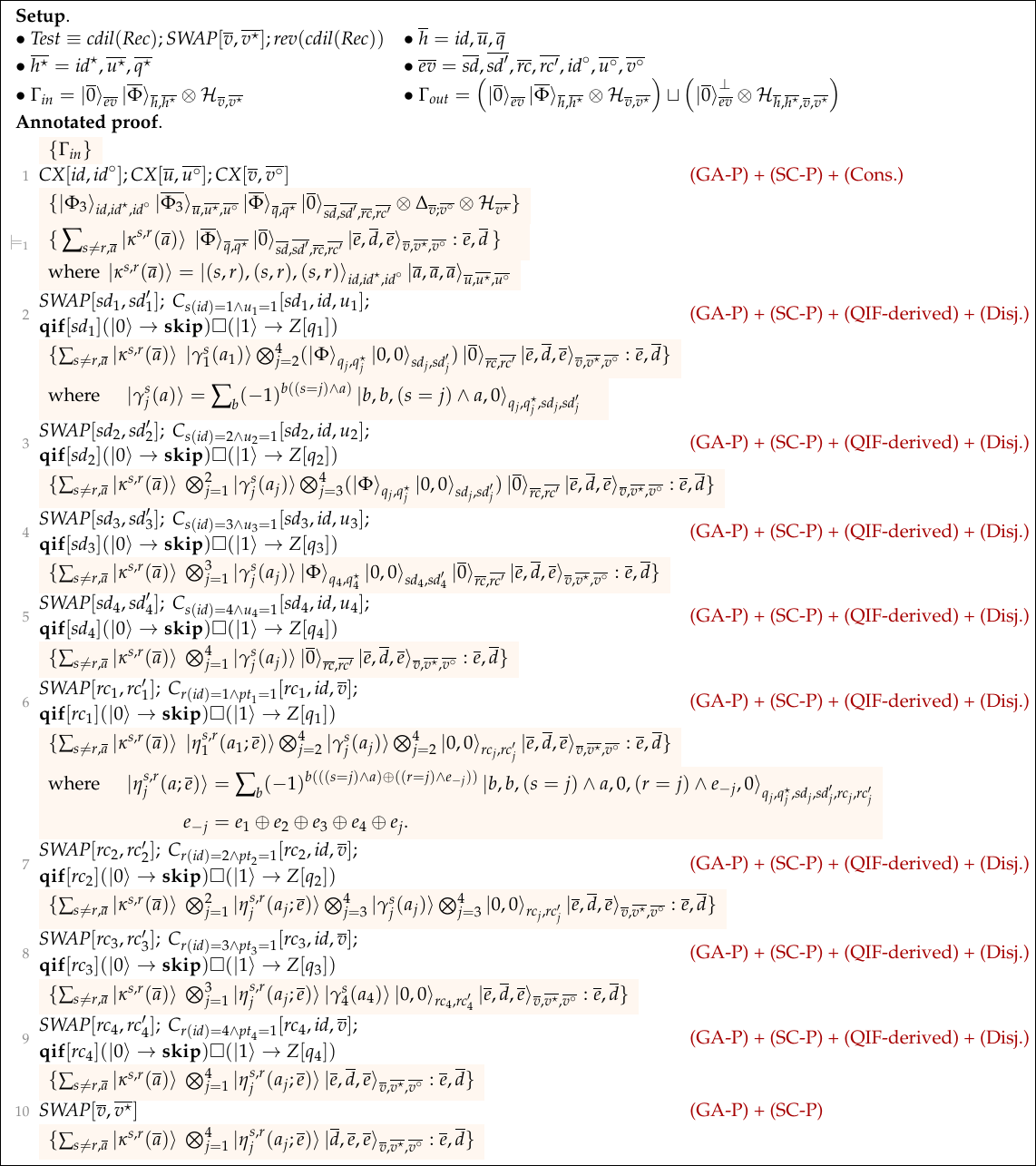}
        \caption{Annotated proof of information-flow security for anonymous entanglement generation: recovery stage.}
        \label{fig:aeg_rec_proof}
    \end{figure}

    \begin{figure}
        \ContinuedFloat
        \centering
        \includegraphics[width=\textwidth]{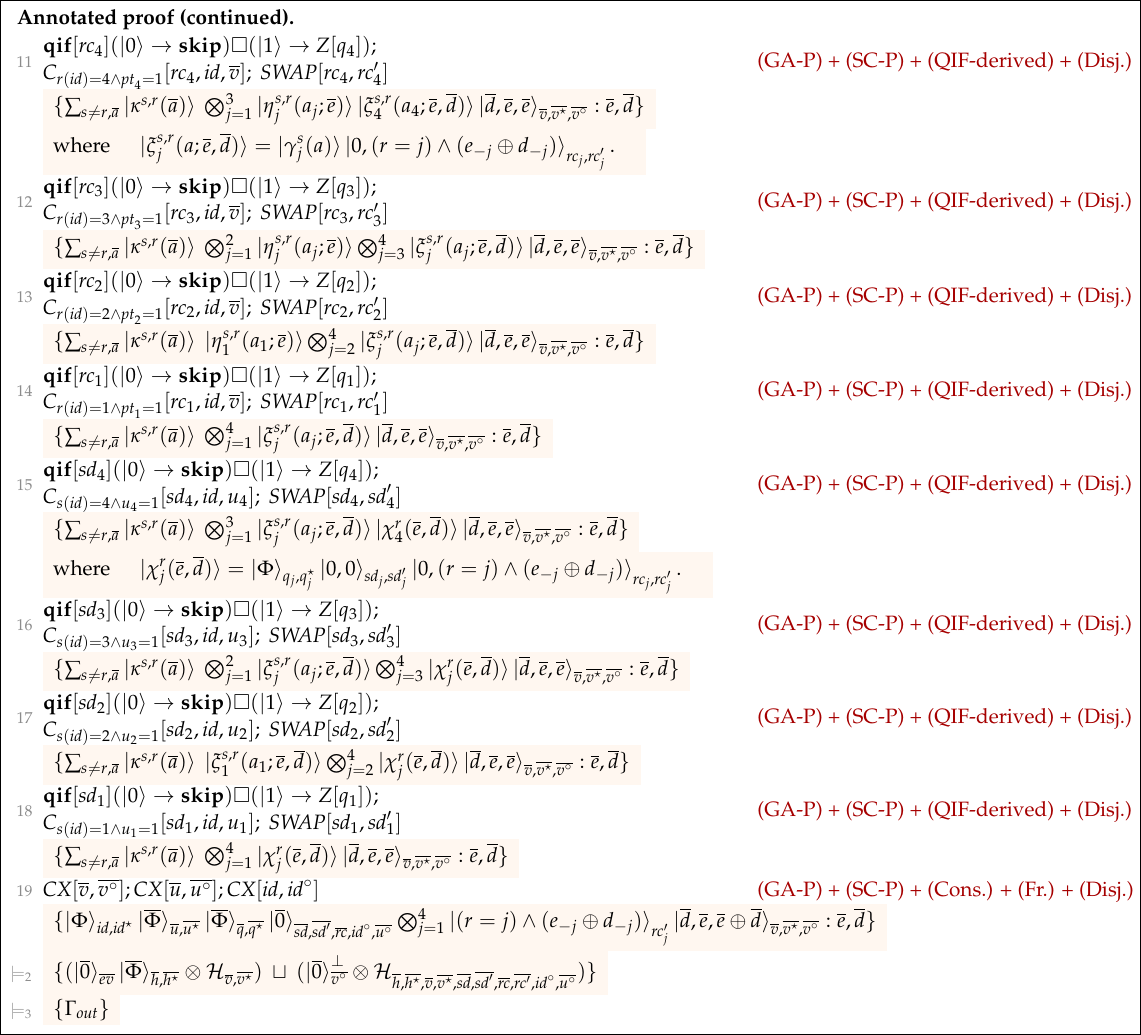}
        \caption{Annotated proof of information-flow security for anonymous entanglement generation: recovery stage.}
    \end{figure}

    Let us further explain the proof in \Cref{fig:aeg_rec_proof}.
    \begin{itemize}
        \item
            \textbf{Assertions}:
            \begin{enumerate}
                \item
                All indices $s,r,j$ range over $\braces*{1,2,3,4}$, where $s\neq r$.
                All indices $a_j,e_j,d_j,b$ range over $\braces*{0,1}$, and $\oplus$ denotes binary XOR operation.
                For $\overline x=x_1,x_2,x_3,x_4$, we use
                \begin{equation*}
                    x_{-j}=x_1\oplus x_2\oplus x_3\oplus x_4\oplus x_j.
                \end{equation*}
                \item
                We use the same notation $\ket{\Phi}$, $\ket{\Phi_3}$,
                $\ket{\overline{\Phi_3}}$, and $\Delta_{\overline q;\overline{q'}}$
                as in the verification of $\mathit{Tran}$.
                As usual, normalizing factors are omitted in assertions whenever needed.
            \end{enumerate}
        \item
            \textbf{Proof rules}:
            In Steps 2--9 and 11--18, the (QIF-derived) denotes the same proof pattern as in \Cref{exp:veri-qotp,exp:veri-qec}.
            On the RHS, (Disj.), (Fr.), and (Cons.) abbreviate (Disjunction), (Frame), and (Consequence), respectively.
        \item
            \textbf{Further details}:
            \begin{enumerate}
                \item
                Step 1 follows the same reasoning as Step 1 in the verification of $\mathit{Tran}$.
                The $\models_1$ after Step 1 exploits
                \begin{equation*}
                    \Delta_{\overline v;\overline{v^\circ}}\otimes
                    \calH_{\overline{v^\star}}
                    \models
                    \braces*{
                    \ket{\overline e,\overline d,\overline e}_{
                    \overline v,\overline{v^\star},\overline{v^\circ}}
                    :\overline e,\overline d}.
                \end{equation*}
                \item
                In Steps 2--5, we first apply (GA-P) and (SC-P) to obtain, for $j\in\braces*{1,2,3,4}$:
                \begin{equation*}
                    \begin{split}
                    \vdash&\braces*{
                    \sum\nolimits_{s\ne r}\ket{(s,r),(s,r),(s,r)}_{\mathit{id},\mathit{id}^\star,\mathit{id}^\circ}
                    \ket{\Phi_3}_{u_j,u_j^\star,u_j^\circ}\ket{0,0}_{\mathit{sd}_j,\mathit{sd}_j'}}\\
                    &\qquad
                    \mathit{SWAP}[\mathit{sd}_j,\mathit{sd}_j'];\
                    \mathit C_{s(\mathit{id})=j\wedge u_j=1}[\mathit{sd}_j,\mathit{id},u_j]\\
                    &\qquad\qquad
                    \braces*{
                    \sum\nolimits_{s\ne r} \sum\nolimits_{a_j}\ket{(s,r),(s,r),(s,r)}_{\mathit{id},\mathit{id}^\star,\mathit{id}^\circ}
                    \ket{a_j,a_j,a_j}_{u_j,u_j^\star,u_j^\circ}
                    \ket{(s=j)\wedge a_j,0}_{\mathit{sd}_j,\mathit{sd}_j'}}.
                    \end{split}
                \end{equation*}
                Then, the two premises of (QIF-P) are
                \begin{equation*}
                    \begin{split}
                    &\vdash\braces*{\ket0_{\mathit{sd}_j}\ket{\Phi}_{q_j,q_j^\star}}\
                    \mathbf{skip}\
                    \braces*{\ket0_{\mathit{sd}_j}\ket{\Phi}_{q_j,q_j^\star}},\\
                    &\vdash\braces*{\ket1_{\mathit{sd}_j}\ket{\Phi}_{q_j,q_j^\star}}\
                    \mathit Z[q_j]\
                    \braces*{\ket1_{\mathit{sd}_j}\sum_b (-1)^{b}\ket{b,b}_{q_j,q_j^\star}}.
                    \end{split}
                \end{equation*}
                By the definition of $\ket{\gamma_j^s(a_j)}$, applying (QIF-P), (Frame), and (Consequence), followed by (Disjunction) over $\overline e,\overline d$, gives the full Hoare triples in Steps 2--5.

                \item
                In Steps 6--9, we first apply (GA-P) and (SC-P) to obtain, for $j\in\braces*{1,2,3,4}$ and $\overline e$:
                \begin{equation*}
                    \begin{split}
                    \vdash&\braces*{
                    \sum\nolimits_{s\ne r}\ket{(s,r),(s,r),(s,r)}_{\mathit{id},\mathit{id}^\star,\mathit{id}^\circ}
                    \ket{\overline e}_{\overline v}\ket{0,0}_{\mathit{rc}_j,\mathit{rc}_j'}}\\
                    &\qquad
                    \mathit{SWAP}[\mathit{rc}_j,\mathit{rc}_j'];\
                    \mathit C_{r(\mathit{id})=j\wedge\mathit{pt}_j=1}
                    [\mathit{rc}_j,\mathit{id},\overline v]\\
                    &\qquad\qquad
                    \braces*{
                    \sum\nolimits_{s\ne r}\ket{(s,r),(s,r),(s,r)}_{\mathit{id},\mathit{id}^\star,\mathit{id}^\circ}
                    \ket{\overline e}_{\overline v}
                    \ket{(r=j)\wedge e_{-j},0}_{\mathit{rc}_j,\mathit{rc}_j'}}.
                    \end{split}
                \end{equation*}
                Then, the two premises of (QIF-P) are
                \begin{equation*}
                    \begin{split}
                    &\vdash\bigg\{
                    \ket0_{\mathit{rc}_j}
                    \sum_{\substack{s\ne r,a_j\\\mathclap{(r=j)\wedge e_{-j}=0}}}
                    \ket{(s,r),(s,r),(s,r)}_{\mathit{id},\mathit{id}^\star,\mathit{id}^\circ}
                    \ket{a_j,a_j,a_j}_{u_j,u_j^\star,u_j^\circ}
                    \ket{\gamma_j^s(a_j)}\bigg\}\ 
                    \mathbf{skip}\\
                    &\quad\qquad \bigg\{
                    \ket0_{\mathit{rc}_j}
                    \sum_{\substack{s\ne r,a_j\\\mathclap{(r=j)\wedge e_{-j}=0}}}
                    \ket{(s,r),(s,r),(s,r)}_{\mathit{id},\mathit{id}^\star,\mathit{id}^\circ}
                    \ket{a_j,a_j,a_j}_{u_j,u_j^\star,u_j^\circ}
                    \ket{\gamma_j^s(a_j)}\bigg\},
                    \end{split}
                \end{equation*}
                and
                \begin{equation*}
                    \begin{split}
                    &\vdash\bigg\{
                    \ket1_{\mathit{rc}_j}
                    \sum_{\substack{s\ne r,a_j\\\mathclap{(r=j)\wedge e_{-j}=1}}}
                    \ket{(s,r),(s,r),(s,r)}_{\mathit{id},\mathit{id}^\star,\mathit{id}^\circ}
                    \ket{a_j,a_j,a_j}_{u_j,u_j^\star,u_j^\circ}
                    \ket{\gamma_j^s(a_j)}\bigg\}\
                    \mathit Z[q_j]\\
                    &\quad\qquad\bigg\{
                    \ket1_{\mathit{rc}_j}
                    \sum_{\substack{s\ne r,a_j,b\\\mathclap{(r=j)\wedge e_{-j}=1}}}
                    (-1)^{b(((s=j)\wedge a_j)\oplus 1)}
                    \ket{(s,r),(s,r),(s,r)}_{\mathit{id},\mathit{id}^\star,\mathit{id}^\circ}\\
                    &\qquad\qquad\ket{a_j,a_j,a_j,b,b,(s=j)\wedge a_j,0}_{u_j,u_j^\star,u_j^\circ,q_j,q_j^\star,\mathit{sd}_j,\mathit{sd}_j'}\bigg\}.
                    \end{split}
                \end{equation*}
                By the definition of $\ket{\eta_j^{s,r}(a_j;\overline e)}$, applying (QIF-P), (Frame), and (Consequence), followed by (Disjunction) over $\overline e,\overline d$, gives the full Hoare triples in Steps 6--9.

                \item
                In Steps 11--14, we first apply the (QIF-derived) proof pattern.
                In particular, for $j\in\braces*{1,2,3,4}$, the two premises of (QIF-P) are
                \begin{equation*}
                    \begin{split}
                    &\vdash\bigg\{
                    \ket{0,0}_{\mathit{rc}_j,\mathit{rc}_j'}
                    \sum_{\substack{s\ne r,a_j\\\mathclap{(r=j)\wedge e_{-j}=0}}}
                    \ket{(s,r),(s,r),(s,r)}_{\mathit{id},\mathit{id}^\star,\mathit{id}^\circ}
                    \ket{a_j,a_j,a_j}_{u_j,u_j^\star,u_j^\circ}
                    \ket{\gamma_j^s(a_j)}\bigg\}\
                    \mathbf{skip}\\
                    &\quad\qquad\bigg\{
                    \ket{0,0}_{\mathit{rc}_j,\mathit{rc}_j'}
                    \sum_{\substack{s\ne r,a_j\\\mathclap{(r=j)\wedge e_{-j}=0}}}
                    \ket{(s,r),(s,r),(s,r)}_{\mathit{id},\mathit{id}^\star,\mathit{id}^\circ}
                    \ket{a_j,a_j,a_j}_{u_j,u_j^\star,u_j^\circ}
                    \ket{\gamma_j^s(a_j)}\bigg\},
                    \end{split}
                \end{equation*}
                and
                \begin{equation*}
                    \begin{split}
                    &\vdash\bigg\{\ket{1,0}_{\mathit{rc}_j,\mathit{rc}_j'}
                    \sum_{\substack{s\ne r,a_j,b\\\mathclap{(r=j)\wedge e_{-j}=1}}}
                    \ket{(s,r),(s,r),(s,r)}_{\mathit{id},\mathit{id}^\star,\mathit{id}^\circ}\\
                    &\quad\qquad(-1)^{b(((s=j)\wedge a_j)\oplus1)}
                    \ket{a_j,a_j,a_j,b,b,(s=j)\wedge a_j,0}_{u_j,u_j^\star,u_j^\circ,q_j,q_j^\star,\mathit{sd}_j,\mathit{sd}_j'}\bigg\}\ \mathit Z[q_j]\\
                    &\qquad\qquad
                    \bigg\{\ket{1,0}_{\mathit{rc}_j,\mathit{rc}_j'}
                    \sum_{\substack{s\ne r,a_j\\\mathclap{(r=j)\wedge e_{-j}=1}}}
                    \ket{(s,r),(s,r),(s,r)}_{\mathit{id},\mathit{id}^\star,\mathit{id}^\circ}
                    \ket{a_j,a_j,a_j}_{u_j,u_j^\star,u_j^\circ}\ket{\gamma_j^s(a_j)}\bigg\}.
                    \end{split}
                \end{equation*}
                By the definition of $\ket{\eta_j^{s,r}(a_j;\overline e)}$, applying (QIF-P), (Frame), and (Consequence) gives
                \begin{equation*}
                    \begin{split}
                    &\vdash\braces*{
                    \sum\nolimits_{s\ne r}\sum\nolimits_{a_j}
                    \ket{(s,r),(s,r),(s,r)}_{\mathit{id},\mathit{id}^\star,\mathit{id}^\circ}
                    \ket{a_j,a_j,a_j}_{u_j,u_j^\star,u_j^\circ}\ket{\eta_j^{s,r}(a_j;\overline e)}}\\
                    &\qquad\mathbf{qif}[\mathit{rc}_j](\ket0\rightarrow\mathbf{skip})
                    \square(\ket1\rightarrow\mathit Z[q_j])\\
                    &\qquad\quad\braces*{
                    \sum\nolimits_{s\ne r}\sum\nolimits_{a_j}
                    \ket{(s,r),(s,r),(s,r)}_{\mathit{id},\mathit{id}^\star,\mathit{id}^\circ}
                    \ket{a_j,a_j,a_j}_{u_j,u_j^\star,u_j^\circ}\ket{\gamma_j^s(a_j)}
                    \ket{(r=j)\wedge e_{-j},0}_{\mathit{rc}_j,\mathit{rc}_j'}}.
                    \end{split}
                \end{equation*}
                Moreover, applying (GA-P) and (SC-P) gives, for $j\in\braces*{1,2,3,4}$ and $\overline e,\overline d$:
                \begin{equation*}
                    \begin{split}
                    \vdash&\braces*{
                    \sum\nolimits_{s\ne r}\ket{(s,r),(s,r),(s,r)}_{\mathit{id},\mathit{id}^\star,\mathit{id}^\circ}
                    \ket{\overline d}_{\overline v}
                    \ket{(r=j)\wedge e_{-j},0}_{\mathit{rc}_j,\mathit{rc}_j'}}\\
                    &\qquad
                    \mathit C_{r(\mathit{id})=j\wedge\mathit{pt}_j=1}
                    [\mathit{rc}_j,\mathit{id},\overline v];\
                    \mathit{SWAP}[\mathit{rc}_j,\mathit{rc}_j']\\
                    &\qquad\qquad
                    \braces*{
                    \sum\nolimits_{s\ne r}\ket{(s,r),(s,r),(s,r)}_{\mathit{id},\mathit{id}^\star,\mathit{id}^\circ}
                    \ket{\overline d}_{\overline v}
                    \ket{0,(r=j)\wedge(e_{-j}\oplus d_{-j})}_{\mathit{rc}_j,\mathit{rc}_j'}}.
                    \end{split}
                \end{equation*}
                By the definition of $\ket{\xi_j^{s,r}(a_j;\overline e,\overline d)}$, applying (Frame), (Consequence), and (Disjunction) over $\overline e,\overline d$ gives the full Hoare triples in Steps 11--14, respectively.

                \item
                In Steps 15--18, we first apply (GA-P), (SC-P), and the (QIF-derived) proof pattern to obtain, for $j\in\braces*{1,2,3,4}$:
                \begin{equation*}
                    \begin{split}
                    \vdash&\braces*{
                    \sum\nolimits_{s\ne r}\sum\nolimits_{a_j}
                    \ket{(s,r),(s,r),(s,r)}_{\mathit{id},\mathit{id}^\star,\mathit{id}^\circ}
                    \ket{a_j,a_j,a_j}_{u_j,u_j^\star,u_j^\circ}
                    \ket{\gamma_j^s(a_j)}
                    \ket{0,(r=j)\wedge(e_{-j}\oplus d_{-j})}_{\mathit{rc}_j,\mathit{rc}_j'}}\\
                    &\qquad
                    \mathbf{qif}[\mathit{sd}_j](\ket0\rightarrow\mathbf{skip})
                    \square(\ket1\rightarrow\mathit Z[q_j]);\\
                    &\qquad
                    \mathit C_{s(\mathit{id})=j\wedge u_j=1}[\mathit{sd}_j,\mathit{id},u_j];\
                    \mathit{SWAP}[\mathit{sd}_j,\mathit{sd}_j']\\
                    &\qquad\qquad
                    \bigg\{
                    \sum\nolimits_{s\ne r}\sum\nolimits_{a_j}
                    \ket{(s,r),(s,r),(s,r)}_{\mathit{id},\mathit{id}^\star,\mathit{id}^\circ}
                    \ket{a_j,a_j,a_j}_{u_j,u_j^\star,u_j^\circ}\\
                    &\qquad\qquad\qquad\ket{\Phi}_{q_j,q_j^\star}\ket{0,0}_{\mathit{sd}_j,\mathit{sd}_j'}
                    \ket{0,(r=j)\wedge(e_{-j}\oplus d_{-j})}_{\mathit{rc}_j,\mathit{rc}_j'}\bigg\}.
                    \end{split}
                \end{equation*}
                By the definition of $\ket{\chi_j^r(\overline e,\overline d)}$, applying (Frame), (Consequence), and (Disjunction) over $\overline e,\overline d$ gives the full Hoare triples in Steps 15--18.

                \item
                Step 19 is analogous to Step 1 with additional applications of (Frame) and (Disjunction).
                The $\models_2$ groups the states according to the component
                $\ket{\overline e\oplus\overline d}_{\overline{v^\circ}}$, yielding the disjunction $\sqcup$ of two assertions corresponding to $\overline e=\overline d$ and $\overline e\ne\overline d$ by further applying (Consequence).
                The $\models_3$ follows from the definition of $\Gamma_{\mathit{out}}$.
            \end{enumerate}
    \end{itemize}
\end{example}

\section{Proof Details}

\subsection{Leakage Probability as Worst-Case Recovery Probability}
\label{sub:justification}

The following lemma explains why $\Pr_{\textup{leak}}$ in \Cref{def:leaking-pr} is a worst-case recovery probability.
It shows that the infimum in \Cref{def:leaking-pr} may be restricted to input states where $\overline{x}$ has a fixed value. For such states, the fidelity in \Cref{eq:leak-pr} is exactly the success probability of recovering that value in $\overline{y}$.

\begin{lemma}
    \label{lmm:operational-leakage-probability}
    Consider a quantum program $P$ and classical variables $\overline{x},\overline{y}$ of the same types.
    Then,
    \begin{equation}
        \label{eq:leak-operational}
        \Pr_{\textup{leak}}\bracks*{P,\overline{x}\rightarrow \overline{y}}
        = \inf_{\overline{a},t} f\parens*{ \ket{\overline{a}}\!\bra{\overline{a}}, \tr_{\overline{ry}}\circ \Bracks*{P} \parens*{ \ket{\overline{a}}\!\bra{\overline{a}}_{\overline{x}}\otimes t } },
    \end{equation}
    where $\overline{a}$ ranges over classical states of $\overline{x}$, and $t$ ranges over all (normalized) states with
    $\var(P)\cup \overline{x}\cup\overline{y}\subseteq \overline{x}\cup\var\parens*{t}$ and $\overline{x}\cap \var(t)=\emptyset$.
\end{lemma}

\begin{proof}
    Let $f^*$ denote the RHS in \Cref{eq:leak-operational}.
    Consider any state $s$ in \Cref{def:leaking-pr}.
    Since $\overline{x}$ is classical,
    \begin{equation*}
        s=\sum_{\overline{a}}
        p_{\overline{a}}
        \ket{\overline{a}}\!\bra{\overline{a}}_{\overline{x}}
        \otimes t_{\overline{a}},
    \end{equation*}
    where $p_{\overline{a}}\geq 0$, $\sum_{\overline{a}}p_{\overline{a}}=1$, and $t_{\overline{a}}$ are normalized.
    Similarly, since $\overline{y}$ is classical,
    \begin{equation*}
        \tr_{\overline{ry}}\circ \Bracks*{P} \parens*{s}
        = \sum_{\overline{a}} q_{\overline{a}}\ket{\overline{a}}\!\bra{\overline{a}}_{\overline{y}}.
    \end{equation*}
    By linearity of $\Bracks*{P}$ and the definition of fidelity, we have
    \begin{equation*}
        q_{\overline{a}} \geq p_{\overline{a}} \bra{\overline{a}} \tr_{\overline{ry}}\circ\Bracks*{P} \parens*{ \ket{\overline{a}}\!\bra{\overline{a}}_{\overline{x}} \otimes t_{\overline{a}} } \ket{\overline{a}} \geq p_{\overline{a}}f^*.
    \end{equation*}
    We can calculate
    \begin{equation*}
        f\parens*{\tr_{\overline{rx}}(s),\tr_{\overline{ry}}\circ \Bracks*{P}(s)}
        = \bigg(\sum_{\overline{a}}\sqrt{p_{\overline{a}}q_{\overline{a}}}\bigg)^2 \notag
        \geq \bigg(\sum_{\overline{a}}\sqrt{p_{\overline{a}}(p_{\overline{a}}f^*)}\bigg)^2
        =f^*.
    \end{equation*}
    Taking the infimum over all state $s$ gives
    \begin{equation}
        \label{eq:left-f-star}
        \Pr_{\textup{leak}}\bracks*{P,\overline{x}\rightarrow \overline{y}}
        \geq f^*.
    \end{equation}

    Conversely, any state $\ket{\overline{a}}\!\bra{\overline{a}}_{\overline{x}}\otimes t$ is a valid state $s$ in \Cref{def:leaking-pr}.
    Hence
    \begin{equation}
        \label{eq:right-f-star}
        \Pr_{\textup{leak}}\bracks*{P,\overline{x}\rightarrow \overline{y}}\leq f^*
    \end{equation}
    Combining \Cref{eq:left-f-star,eq:right-f-star} leads to the conclusion.
\end{proof}

\subsection{Proofs for Properties of Information-Flow Security}
\label{sub:proofs_properties}

\begin{proof}[Proof of \Cref{lmm:res-suffice}]
    The ``only if'' direction is immediate from \Cref{def:noninterference}.
    For the ``if'' direction, suppose that \Cref{eq:low-eq-var-P} holds for any two input states $s_1,s_2$ with $\var(s_1)=\var(s_2)=\var(P)$.
    Let $\overline{h}=\hvar(P)$ and $\overline{l}=\lvar(P)$.
    Define a linear operator
    \begin{equation*}
        \calE(\cdot)=\tr_{\overline{h}}\circ \Bracks*{P}\parens*{\ket{\overline{0}}\!\bra{\overline{0}}_{\overline{h}}\otimes (\cdot)}.
    \end{equation*}
    Then, for any state $s$ with $\var(s)=\var(P)$, we have
    $s =_{\Low} \ket{\overline{0}}\!\bra{\overline{0}}_{\overline{h}}\otimes \tr_{\overline{h}}(s)$,
    which by \Cref{eq:low-eq-var-P} implies 
    \begin{equation}
        \label{eq:reexpress-P-tr}
        \tr_{\overline{h}}\circ \Bracks*{P}(s)=\calE\circ\tr_{\overline{h}}(s).
    \end{equation}

    Now let $\overline{v}$ be any list of variables with $\var(P)\subseteq \overline{v}$, and let $\overline{h'},\overline{l'}$ be the lists of $\High$ and $\Low$ variables in $\overline{v}-\var(P)$, respectively.
    Applying \Cref{lmm:channel-extension} to the two linear operators
    $\tr_{\overline{h}}\circ \Bracks*{P}$ and $\calE\circ \tr_{\overline{h}}$,
    we obtain for any state $s$ with $\var(s)=\overline{v}$:
    \begin{equation*}
        \tr_{\overline{h}}\circ \Bracks*{P}(s)=
        \parens*{\calE\otimes \Id_{\overline{h'},\overline{l'}}}\circ\tr_{\overline{h}}(s),
    \end{equation*}
    which further implies $\tr_{\overline{h},\overline{h'}}\circ \Bracks*{P}(s)=\calE\circ\tr_{\overline{h},\overline{h'}}(s)$ due to \Cref{eq:reexpress-P-tr}.
    It is then easy to verify that for any two input states $s_1,s_2$ to $P$ with $\var(s_1)=\var(s_2)=\overline{v}$, \Cref{eq:low-eq-var-P} also holds. Hence, by \Cref{def:noninterference}, $P$ satisfies \textbf{NI}.
\end{proof}

\begin{proof}[Proof of \Cref{thm:compose}]
    Suppose programs $P_1$ and $P_2$ both satisfy \textbf{NI}.
    Using the \textbf{NI} of $P_1$, we have for any two input states $s_1$ and $s_2$ to $P_1;P_2$:
    \begin{equation}
        \label{eq:P_1-NI}
        s_1 =_{\Low} s_2 \Rightarrow \Bracks*{P_1}(s_1)=_{\Low} \Bracks*{P_1}(s_2).
    \end{equation}
    By the \textbf{NI} of $P_2$ and the semantics in \Cref{sec:quantum_programs}, we obtain
    \begin{equation}
        \label{eq:P_2-NI}
        \Bracks*{P_1}(s_1)=_{\Low} \Bracks*{P_1}(s_2)
        \Rightarrow
        \Bracks*{P_1;P_2}(s_1)=_{\Low} \Bracks*{P_1;P_2}(s_2).
    \end{equation}
    Combining \Cref{eq:P_1-NI,eq:P_2-NI} immediately yields that the sequential composition $P_1;P_2$ also satisfies \textbf{NI}.
\end{proof}

\subsection{Proof of \Cref{lmm:obstruction}}
\label{sub:proof_no_quantum_equal}

\begin{proof}[Proof of \Cref{lmm:obstruction}]
    It suffices to consider qubit variables $q$ and $q'$.
    Suppose for contradiction that such a quantum observable $\Lambda$ exists.
    From $0\leq \Lambda\leq \Id$, 
    taking $\rho=\ket{0}\!\bra{0}$ and $\rho=\ket{1}\!\bra{1}$ in \Cref{eq:quantum-equality-accept} implies
    \begin{equation*}
        \Lambda\ket{0,0}=\ket{0,0},\qquad \Lambda\ket{1,1}=\ket{1,1}.
    \end{equation*}
    Taking $\rho= \ket{0}\!\bra{0},\sigma=\ket{1}\!\bra{1}$ and $\rho=\ket{1}\!\bra{1},\sigma=\ket{0}\!\bra{0}$ in \Cref{eq:quantum-equality-reject} implies
    \begin{equation*}
        \Lambda\ket{0,1}=0,\qquad \Lambda\ket{1,0}=0.
    \end{equation*}
    Let $\ket{+}=\frac{1}{\sqrt{2}}(\ket{0}+\ket{1})$.
    By \Cref{eq:quantum-equality-accept}, we should have
    $\bra{+,+}\Lambda\ket{+,+}=1$.
    On the other hand,
    \begin{equation*}
        \Lambda\ket{+,+}
        =\frac{1}{2}\Lambda\parens*{\ket{0,0}+\ket{0,1}+\ket{1,0}+\ket{1,1}}
        =\frac{1}{2}\parens*{\ket{0,0}+\ket{1,1}},
    \end{equation*}
    and therefore $\bra{+,+}\Lambda\ket{+,+}=\frac{1}{2}$,
    which leads to a contradiction.
\end{proof}

\subsection{Soundness of the Proof Rules}
\label{sub:soundness}

We prove the soundness of the proof rules in \Cref{fig:proof_system}.
Here, $\vdash$ denotes derivability by the proof rules in \Cref{fig:proof_system}.

\begin{theorem}[Soundness]
    \label{thm:qhl-soundness}
    For any quantum circuit $C$ and quantum assertions $\Gamma,\Gamma'$,
    if $\vdash \braces*{\Gamma}\ C\ \braces*{\Gamma'}$
    then $\models \braces*{\Gamma}\ C\ \braces*{\Gamma'}$.
\end{theorem}

\begin{proof}
    We prove the theorem by induction on the derivation of
    $\vdash \braces*{\Gamma}\ C\ \braces*{\Gamma'}$.
    By \Cref{def:qht-semantics} and linearity, it is easy to verify that if
    $\models \braces*{\Gamma}\ C\ \braces*{\Gamma'}$,
    then for any $\ket{\psi}\in\Span(\Gamma)$, the transition $(C,\ket{\psi})\rightarrow^*{(\downarrow,\ket{\psi'})}$ (see \Cref{def:operational_semantics_circ}) implies $\ket{\psi'}\in\Span(\Gamma')$.
    Consider the last rule used in the derivation:
    \begin{itemize}
        \item \textup{(SK-P).}
        For any $\ket{\psi}$ with $\ket{\psi}\models\Gamma$,
        (SK) gives $(\mathbf{skip},\ket{\psi})\rightarrow(\downarrow,\ket{\psi})$,
        which implies $\models \braces*{\Gamma}\ \mathbf{skip}\ \braces*{\Gamma}$.

        \item \textup{(GA-P).}
        We need to prove
        \begin{equation}
            \label{eq:ga-soundness}
            \models \braces*{\Gamma}\ U[\overline{q}]\ \braces*{U_{\overline{q}}(\Gamma)}.
        \end{equation}
        If $\Gamma=\braces*{\ket{\phi}}$, then any $\ket{\psi}\models\Gamma$ satisfies $\ket{\psi}=\ket{\phi}$.
        From (GA) we have $(U[\overline{q}],\ket{\psi})\rightarrow (\downarrow, U_{\overline{q}}\ket{\psi})$,
        which, combined with $U_{\overline{q}}(\Gamma)=\braces*{U_{\overline{q}}\ket{\phi}}$, yields \Cref{eq:ga-soundness}.
        Otherwise, $\ket{\psi}\models\Gamma$ means $\ket{\psi}\in\Span(\Gamma)$.
        By linearity,
        $U_{\overline{q}}\ket{\psi}\in U_{\overline{q}}(\Span(\Gamma))=\Span(U_{\overline{q}}(\Gamma))$, which also leads to \Cref{eq:ga-soundness}.

        \item \textup{(SC-P).}
        Suppose the last rule is
        \begin{equation*}
            \frac{\vdash\braces*{\Gamma}\ C_1\ \braces*{\Gamma'}\qquad
            \vdash\braces*{\Gamma'}\ C_2\ \braces*{\Gamma''}}
            {\vdash\braces*{\Gamma}\ C_1;C_2\ \braces*{\Gamma''}}.
        \end{equation*}
        By the induction hypothesis, we have $\models \braces*{\Gamma}\ C_1\ \braces*{\Gamma'}$
        and $\models \braces*{\Gamma'}\ C_2\ \braces*{\Gamma''}$.
        For any $\ket{\psi}\models\Gamma$, if
        $(C_1,\ket{\psi})\rightarrow^*(\downarrow,\ket{\psi_1})$
        and
        $(C_2,\ket{\psi_1})\rightarrow^*(\downarrow,\ket{\psi_2})$,
        then $\ket{\psi_1}\models\Gamma'$ and $\ket{\psi_2}\models\Gamma''$.
        As $(C_1;C_2,\ket{\psi})\rightarrow^*(\downarrow, \ket{\psi_2})$ from (SC), we obtain $\models \braces*{\Gamma}\ C_1;C_2\ \braces*{\Gamma''}$.

        \item \textup{(QIF-P).}
        Suppose the last rule is
        \begin{equation*}
            \frac{
            \vdash\braces{\ket{\psi_j}}\ C_j\ \braces{\ket{\psi_j'}}\quad (j=0,1)
            \qquad \abs*{\alpha}^2+\abs*{\beta}^2=1}
            {\vdash\braces*{\alpha\ket{0}_q\ket{\psi_0}+\beta\ket{1}_q\ket{\psi_1}}\
            \mathbf{qif}[q](\ket{0}\rightarrow C_0)\square(\ket{1}\rightarrow C_1)\
            \braces*{\alpha\ket{0}_q\ket{\psi_0'}+\beta\ket{1}_q\ket{\psi_1'}}}.
        \end{equation*}
        By the induction hypothesis, for $j=0,1$,
        we have $\models\braces{\ket{\psi_j}}\ C_j\ \braces{\ket{\psi_j'}}$,
        which means $(C_j,\ket{\psi_j})\rightarrow^*(\downarrow,\ket{\psi_j'})$
        as the execution of quantum circuit $C_j$ is deterministic and terminating.
        The (QIF) rule gives
        \begin{equation*}
            \parens*{\mathbf{qif}[q](\ket{0}\rightarrow C_0)\square(\ket{1}\rightarrow C_1),
            \alpha\ket{0}_q\ket{\psi_0}+\beta\ket{1}_q\ket{\psi_1}}
            \rightarrow^*
            \parens*{\downarrow,
            \alpha\ket{0}_q\ket{\psi_0'}+\beta\ket{1}_q\ket{\psi_1'}},
        \end{equation*}
        which immediately implies
        \begin{equation*}
            \models
            \braces*{\alpha\ket{0}_q\ket{\psi_0}+\beta\ket{1}_q\ket{\psi_1}}
            \ \mathbf{qif}[q](\ket{0}\rightarrow C_0)\square(\ket{1}\rightarrow C_1)\
            \braces*{\alpha\ket{0}_q\ket{\psi_0'}+\beta\ket{1}_q\ket{\psi_1'}}.
        \end{equation*}

        \item \textup{(Consequence).}
        Suppose the last rule is
        \begin{equation*}
            \frac{\Lambda\models\Gamma\qquad
            \vdash\braces*{\Gamma}\ C\ \braces*{\Gamma'}\qquad
            \Gamma'\models\Lambda'}
            {\vdash\braces*{\Lambda}\ C\ \braces*{\Lambda'}}.
        \end{equation*}
        By the induction hypothesis,
        $\models \braces*{\Gamma}\ C\ \braces*{\Gamma'}$.
        For $\ket{\psi}\models\Lambda$,
        since $\Lambda\models\Gamma$, it is easy to see $\ket{\psi}\models\Gamma$ by the definition of entailment $\models$ (see discussion below \Cref{fig:proof_system}).
        If $(C,\ket{\psi})\rightarrow^*(\downarrow,\ket{\psi'})$,
        then $\ket{\psi'}\models\Gamma'$, which further implies $\ket{\psi'}\models\Lambda'$
        because $\Gamma'\models\Lambda'$.
        Consequently, $\models \braces*{\Lambda}\ C\ \braces*{\Lambda'}$.

        \item \textup{(Conjunction).}
        Suppose the last rule is
        \begin{equation*}
            \frac{\vdash\braces{\Gamma_j}\ C\ \braces{\Gamma_j'}\quad (j=0,1)}
            {\vdash\braces*{\Gamma_0\sqcap \Gamma_1}\ C\ \braces*{\Gamma_0'\sqcap \Gamma_1'}}.
        \end{equation*}
        By the induction hypothesis,
        $\models\braces{\Gamma_j}\ C\ \braces{\Gamma_j'}$ for $j=0,1$.
        Suppose $\ket{\psi}\models\Gamma_0\sqcap\Gamma_1$ and
        $(C,\ket{\psi})\rightarrow^*(\downarrow,\ket{\psi'})$.
        As $\Gamma_0\sqcap\Gamma_1=\Span(\Gamma_0)\cap\Span(\Gamma_1)$,
        we have $\ket{\psi}\in\Span(\Gamma_j)$,
        which further implies $\ket{\psi'}\in\Span(\Gamma_j')$ for $j=0,1$.
        Therefore, $\ket{\psi'}\in\Span(\Gamma_0')\cap\Span(\Gamma_1')=\Gamma_0'\sqcap\Gamma_1'$,
        which means $\ket{\psi'}\models\Gamma_0'\sqcap\Gamma_1'$.
        This proves $\models \braces*{\Gamma_0\sqcap\Gamma_1}\ C\ \braces*{\Gamma_0'\sqcap\Gamma_1'}$.

        \item \textup{(Disjunction).}
        Suppose the last rule is
        \begin{equation*}
            \frac{\vdash\braces{\Gamma_j}\ C\ \braces{\Gamma_j'}\quad (j=0,1)}
            {\vdash\braces*{\Gamma_0\sqcup \Gamma_1}\ C\ \braces*{\Gamma_0'\sqcup \Gamma_1'}}.
        \end{equation*}
        By the induction hypothesis,
        $\models\braces{\Gamma_j}\ C\ \braces{\Gamma_j'}$ for $j=0,1$.
        Suppose $\ket{\psi}\models\Gamma_0\sqcup\Gamma_1$ and
        $(C,\ket{\psi})\rightarrow^*(\downarrow,\ket{\psi'})$.
        As $\Gamma_0\sqcup\Gamma_1=\overline{\operatorname{span}(\Gamma_0\cup\Gamma_1)}$
        and $C$ only involves finitely many variables with finite types,
        $\ket{\psi}$ can be written as a finite linear combination of states in
        $\Gamma_0\cup\Gamma_1$.
        By linearity,
        $\ket{\psi'}\in\overline{\operatorname{span}(\Gamma_0'\cup\Gamma_1')}$,
        which further implies
        $\ket{\psi'}\models\Gamma_0'\sqcup\Gamma_1'$.
        This proves $\models \braces*{\Gamma_0\sqcup\Gamma_1}\ C\ \braces*{\Gamma_0'\sqcup\Gamma_1'}$.

        \item \textup{(Frame).}
        Suppose the last rule is
        \begin{equation*}
            \frac{\vdash\braces*{\Gamma}\ C\ \braces*{\Gamma'}\qquad
            \qvar(\Lambda)\cap\qvar(C)=\emptyset}
            {\vdash\braces*{\Gamma\otimes \Lambda}\ C\ \braces*{\Gamma'\otimes \Lambda}}.
        \end{equation*}
        By the induction hypothesis,
        $\models\braces*{\Gamma}\ C\ \braces*{\Gamma'}$.
        We first consider a product state $\ket{\psi}\ket{\phi}$ with
        $\ket{\psi}\models\Gamma$ and $\ket{\phi}\models\Lambda$.
        Since $\qvar(\Lambda)\cap\qvar(C)=\emptyset$, by the semantics in \Cref{sub:semantics}, $C$ acts trivially on variables in $\qvar(\Lambda)$.
        If $(C,\ket{\psi})\rightarrow^*(\downarrow,\ket{\psi'})$, then
        $(C,\ket{\psi}\ket{\phi})\rightarrow^*(\downarrow,\ket{\psi'}\ket{\phi})$
        and the induction hypothesis gives $\ket{\psi'}\models\Gamma'$.
        Hence, it is easy to see $\ket{\psi'}\ket{\phi}\models\Gamma'\otimes\Lambda$.

        Now consider any $\ket{\theta}\models\Gamma\otimes\Lambda$.
        If $\Gamma\otimes\Lambda$ contains a single state, then $\ket{\theta}$ is a product state covered by the above.
        Otherwise, $\ket{\theta}\in\Span(\Gamma\otimes\Lambda)$,
        which is a finite linear combination of product states
        $\ket{\psi_i}\ket{\phi_i}$ with $\ket{\psi_i}\in\Gamma$ and $\ket{\phi_i}\in\Lambda$ due to \Cref{eq:wedge-vee}.
        By linearity and the previous product case,
        if $(C,\ket{\theta})\rightarrow^* (\downarrow,\ket{\theta'})$, then
        $\ket{\theta'}\in \Span(\Gamma'\otimes\Lambda)$,
        which means $\ket{\theta'}\models\Gamma'\otimes\Lambda$.
        This proves $\models \braces*{\Gamma\otimes\Lambda}\ C\ \braces*{\Gamma'\otimes\Lambda}$.
    \end{itemize}
\end{proof}

\subsection{Proof of \Cref{lmm:dil-is-dilation}}
\label{sub:proof_dilation}

\begin{proof}[Proof of \Cref{lmm:dil-is-dilation}]
    We prove the lemma by induction on the structure of $P$. 
    From \Cref{alg:dilation}, the condition
    \begin{equation*}
        \qvar(\dil(P))=\lift(\cvar(P))\uplus \qvar(P)\uplus \overline{ev}
    \end{equation*}
    is rather easy to verify.
    Thus, we focus on proving
    that for any input state $s$ to $P$ with
    $\var(s)\cap\overline{ev}=\emptyset$,
    \begin{equation}
        \label{eq:induction_hypo}
        \tr_{\overline{ev}}\circ\Bracks*{\dil(P)}
        \parens*{s\otimes\ket{\overline{0}}\!\bra{\overline{0}}_{\overline{ev}}}
        =
        \Bracks*{P}\parens*{s},
    \end{equation}
    which is also our induction hypothesis.
    By \Cref{def:denotational_semantics} and linearity, it suffices to
    consider input states $s=\ket{\sigma}\!\bra{\sigma}\otimes\rho$.
    Consider the form of $P$:
    \begin{enumerate}
        \item 
            $P\equiv\mathbf{skip}\mid C$. This case is trivial as $\dil(P)\equiv P$ and $\overline{ev}=\emptyset$. 
        \item 
            $P\equiv P_1;P_2$. Then, $\dil(P)=\dil(P_1);\dil(P_2)$ and $\overline{ev}=\overline{ev_1},\overline{ev_2}$, where $\overline{ev_j}$ are the environmental variables introduced by $\dil(P_j)$. 
            For any input state $s$ to $P_1;P_2$, by the induction hypothesis and (SC),
            we have
            \begin{align*}
                \tr_{\overline{ev}}
                \circ\Bracks*{\dil(P_1);\dil(P_2)}
                \parens*{ s\otimes \ket{\overline{0}}\!\bra{\overline{0}}_{\overline{ev}} }
                &= \tr_{\overline{ev_2}}\circ\Bracks*{\dil(P_2)} \parens*{ \tr_{\overline{ev_1}}\circ\Bracks*{\dil(P_1)} \parens*{ s\otimes \ket{\overline{0}}\!\bra{\overline{0}}_{\overline{ev_1}} } \otimes \ket{\overline{0}}\!\bra{\overline{0}}_{\overline{ev_2}} }\\ 
                &= \tr_{\overline{ev_2}}\circ\Bracks*{\dil(P_2)} \parens*{ \Bracks*{P_1}(s)\otimes \ket{\overline{0}}\!\bra{\overline{0}}_{\overline{ev_2}} }\\
                &= \Bracks*{P_1;P_2}(s),
            \end{align*}
            where the first equality uses that $\tr_{\overline{ev_1}}$ commutes with $\Bracks*{\dil(P_2)}$.
        \item 
            \label{case:as}
            $P\equiv x:=e$. 
            Then $\dil(P)\equiv \mathit{SWAP}[x,x']; C_{e'}[x,\overline{u}]$ and $\overline{ev}=x'$,
            where $x'$ is a fresh environmental variable, $e'=e[x\leftarrow x']$, $\overline{u}$ is a list of variables in $\var(e')$, and $C_{e'}$ is as defined in \Cref{alg:dilation}.
            For $s=\ket{\sigma}\!\bra{\sigma}\otimes\rho$, by (SC), (GA), and (AS),
            we have
            \begin{equation*}
                \tr_{x'}\circ\Bracks*{\dil(x:=e)} \parens*{ \ket{\sigma}\!\bra{\sigma}\otimes\rho \otimes\ket{0}\!\bra{0}_{x'} }
                = \ket{\sigma[x:=\sigma(e)]}\!\bra{\sigma[x:=\sigma(e)]} \otimes\rho=\Bracks*{x:=e}(s).
            \end{equation*}
        \item 
            \label{case:if}
            $P\equiv\mathbf{if}\ y\ \mathbf{then}\ P_1\ \mathbf{else}\ P_2$, where $y$ is a bit variable and $y\notin\var(P_1)\cup\var(P_2)$. 
            Then,
            \begin{equation}
                \label{eq:if-to-qif}
                \dil(P)\equiv \mathbf{qif}[y](\ket{0}\rightarrow\dil(P_2)) \square(\ket{1}\rightarrow\dil(P_1)),
            \end{equation}
            and $\overline{ev}=\overline{ev_1},\overline{ev_2}$,
            where $\overline{ev_j}$ are the environmental variables introduced by $\dil(P_j)$.
            By the induction hypothesis and freshness of $\overline{ev_j}$, the assumption $y\notin\var(P_j)$ implies
            $y\notin\qvar(\dil(P_1))\cup \qvar(\dil(P_2))$ and thus \Cref{eq:if-to-qif} is well-defined.

            For $j=1,2$ and $s=\ket{\sigma}\!\bra{\sigma}\otimes\rho$ with $\sigma(y)=2-j$,
            using (QIF), (QC), the induction hypothesis for $P_j$, and (IF), we have 
            \begin{align*}
                \tr_{\overline{ev}}\circ\Bracks*{\dil(P)} \parens*{ \ket{\sigma}\!\bra{\sigma}\otimes\rho\otimes \ket{\overline{0}}\!\bra{\overline{0}}_{\overline{ev}} }
                &= \tr_{\overline{ev_j}}\circ\Bracks*{\dil(P_j)} \parens*{ \ket{\sigma}\!\bra{\sigma}\otimes\rho\otimes \ket{\overline{0}}\!\bra{\overline{0}}_{\overline{ev_j}} }\\
                &= \Bracks*{P_j}\parens*{ \ket{\sigma}\!\bra{\sigma}\otimes\rho }\\
                &= \Bracks*{\mathbf{if}\ y\ \mathbf{then}\ P_1\ \mathbf{else}\ P_2} \parens*{ \ket{\sigma}\!\bra{\sigma}\otimes\rho }.
            \end{align*}
        \item 
            $P\equiv\mathbf{if}\ b\ \mathbf{then}\ P_1\ \mathbf{else}\ P_2$.
            Let $y$ be the fresh bit variable introduced by \Cref{alg:dilation}.
            Then,
            \begin{equation*}
                \dil(P)\equiv \dil(y:=b;\mathbf{if}\ y\ \mathbf{then}\ P_1\ \mathbf{else}\ P_2) \equiv \dil(y:=b); \dil(\mathbf{if}\ y\ \mathbf{then}\ P_1\ \mathbf{else}\ P_2).
            \end{equation*}
            Let $y'$ and $\overline{ev_j}$ be the fresh environmental variables introduced by $\dil(y:=b)$ and $\dil(P_j)$, respectively.
            Since the lifted fresh variable $y$ is included in $\overline{ev}$, we have $\overline{ev}= y,y',\overline{ev_1},\overline{ev_2}$.

            For $s=\ket{\sigma}\!\bra{\sigma}\otimes\rho$, %
            since $y$ is fresh and $y\notin\var(b)$, applying \Cref{case:as} to $y:=b$ gives
            \begin{equation*}
                \tr_{y'}\parens*{ \Bracks*{\dil(y:=b)} \parens*{ \ket{\sigma}\!\bra{\sigma}\otimes\rho\otimes \ket{\overline{0}}\!\bra{\overline{0}}_{y,y'} }}
                = \ket{\sigma}\!\bra{\sigma}\otimes\rho\otimes \ket{\sigma(b)}\!\bra{\sigma(b)}_y.
            \end{equation*}
            Further, using (SC) and \Cref{case:if} on $\mathbf{if}\ y\ \mathbf{then}\ P_1\ \mathbf{else}\ P_2$, we obtain
            \begin{align*}
                &\tr_{\overline{ev}} \circ \Bracks*{\dil(P)} \parens*{ \ket{\sigma}\!\bra{\sigma}\otimes\rho \otimes \ket{\overline{0}}\!\bra{\overline{0}}_{y,y',\overline{ev_1},\overline{ev_2}} }\\
                &\quad = \tr_y\circ \Bracks*{\mathbf{if}\ y\ \mathbf{then}\ P_1\ \mathbf{else}\ P_2} \parens*{ \ket{\sigma}\!\bra{\sigma}\otimes\rho\otimes \ket{\sigma(b)}\!\bra{\sigma(b)}_y}.
            \end{align*}
            If $\sigma\models b$, then $\sigma(b)=1$.
            By (IF) and $y\notin\var(P_1)$, we have
            \begin{align*}
                &\tr_y\circ \Bracks*{\mathbf{if}\ y\ \mathbf{then}\ P_1\ \mathbf{else}\ P_2} \parens*{ \ket{\sigma}\!\bra{\sigma}\otimes\rho\otimes \ket{\sigma(b)}\!\bra{\sigma(b)}_y}\\
                &\quad = \tr_y\circ\Bracks*{P_1} \parens*{ \ket{\sigma}\!\bra{\sigma}\otimes\rho\otimes \ket{1}\!\bra{1}_y}\\
                &\quad = \Bracks*{P_1}\parens*{\ket{\sigma}\!\bra{\sigma}\otimes\rho}\\
                &\quad = \Bracks*{\mathbf{if}\ b\ \mathbf{then}\ P_1\ \mathbf{else}\ P_2} \parens*{\ket{\sigma}\!\bra{\sigma}\otimes\rho}.
            \end{align*}
            Similarly, if $\sigma\models\neg b$, then $\sigma(b)=0$.  
            By (IF) and $y\notin\var(P_2)$, we can also obtain
            \begin{align*}
                &\tr_y\circ \Bracks*{\mathbf{if}\ y\ \mathbf{then}\ P_1\ \mathbf{else}\ P_2} \parens*{  \ket{\sigma}\!\bra{\sigma}\otimes\rho\otimes \ket{\sigma(b)}\!\bra{\sigma(b)}_y}\\
                &\quad = \Bracks*{\mathbf{if}\ b\ \mathbf{then}\ P_1\ \mathbf{else}\ P_2} \parens*{\ket{\sigma}\!\bra{\sigma}\otimes\rho}.
            \end{align*}
        \item 
            $P\equiv q:=\ket{e}$.
            Then $\dil(P)\equiv C_e[q',\overline{u}];\mathit{SWAP}[q,q']$ and $\overline{ev}=q'$,
            where $q'$ is a fresh environmental variable, $\overline{u}$ is a list of variables in $\var(e)$,
            and $C_e$ is as defined in \Cref{alg:dilation}.
            For $s=\ket{\sigma}\!\bra{\sigma}\otimes\rho$, we can write
            \begin{equation*}
                \rho=\sum_{j,k}s_{j,k}\otimes\ket{j}\!\bra{k}_{q}.
            \end{equation*}
            By (GA), (SC), and (RS), we have
            \begin{align*}
                \tr_{q'}\circ\Bracks*{\dil(q:=\ket{e})} \parens*{ \ket{\sigma}\!\bra{\sigma}\otimes\rho \otimes\ket{0}\!\bra{0}_{q'} }
                &= \ket{\sigma}\!\bra{\sigma}\otimes \sum_j s_{j,j}\otimes\ket{\sigma(e)}\!\bra{\sigma(e)}_q\\
                &= \ket{\sigma}\!\bra{\sigma}\otimes \tr_q(\rho)\otimes\ket{\sigma(e)}\!\bra{\sigma(e)}_q\\
                &= \Bracks*{q:=\ket{e}}(s).
            \end{align*}
        \item 
            $P\equiv x:=\mathbf{meas}[q]$.
            Then,
            \begin{equation*}
                \dil(P)\equiv
                \mathit{SWAP}[x,x'];\mathit{CX}[q,q'];\mathit{CX}[q,x],
            \end{equation*}
            and $\overline{ev}=x',q'$, where $x'$ and $q'$ are fresh environmental variables.
            For $s=\ket{\sigma}\!\bra{\sigma}\otimes\rho$, we write
            \begin{equation*}
                \rho= \sum_{j,k} s_{j,k}\otimes \ket{j}\!\bra{k}_{q}.
            \end{equation*}
            By (GA), (SC), and (MS), we have
            \begin{align*}
                &\tr_{\overline{ev}}\circ\Bracks*{\dil(x:=\mathbf{meas}[q])} \parens*{ \ket{\sigma}\!\bra{\sigma}\otimes\rho\otimes\ket{\overline{0}}\!\bra{\overline{0}}_{\overline{ev}} }\\
                &\quad = \tr_{q',x'}\sum_{j,k} \ket{\sigma[x:=j]}\!\bra{\sigma[x:=k]} \otimes s_{j,k} \otimes \ket{j}\!\bra{k}_{q} \otimes \ket{j}\!\bra{k}_{q'}\otimes\ket{\sigma(x)}\!\bra{\sigma(x)}_{x'}\\
                &\quad = \tr_{x'}\sum_j \ket{\sigma[x:=j]}\!\bra{\sigma[x:=j]} \otimes s_{j,j} \otimes \ket{j}\!\bra{j}_{q} \otimes\ket{\sigma(x)}\!\bra{\sigma(x)}_{x'}\\
                &\quad = \sum_j \ket{\sigma[x:=j]}\!\bra{\sigma[x:=j]} \otimes s_{j,j}\otimes\ket{j}\!\bra{j}_{q}\\
                &\quad = \sum_j \ket{\sigma[x:=j]}\!\bra{\sigma[x:=j]} \otimes \parens*{\Id\otimes\ket{j}\!\bra{j}_{q}} \rho \parens*{\Id\otimes\ket{j}\!\bra{j}_{q}}^{\dagger}\\
                &\quad = \Bracks*{x:=\mathbf{meas}[q]}(s).
            \end{align*}
    \end{enumerate}
\end{proof}

\subsection{Proof for Self-Conjugation}
\label{sub:proof_self_conjugation}

\begin{proof}[Proof of \Cref{thm:test-NI}]
    According to \Cref{def:qht-semantics}, the quantum Hoare triple in \Cref{eq:test-triple} means:
    for any input state $s=\ket{\psi}\!\bra{\psi}$ to program $\mathit{Test}$ with %
    \begin{equation*}
        \ket{\psi}= \ket{\overline{0}}_{\overline{ev}}\ket{\overline{\Phi}}_{\overline{h},\overline{h'}}\ket{\eta}_{\overline{l},\overline{l'}},
    \end{equation*}
    the output state $\Bracks*{\mathit{Test}}(s)=\ket{\psi'}\!\bra{\psi'}$ must have the form
    \begin{equation}
        \label{eq:no-l-perturbation} 
        \ket{\psi'}=\alpha \ket{\overline{0}}_{\overline{ev}} \ket{\overline{\Phi}}_{\overline{h},\overline{h'}}\ket{\eta'}_{\overline{l},\overline{l'}}+\beta\ket{\bot}
    \end{equation}
    for some $\ket{\eta'}$ and $\alpha,\beta$ with $\abs*{\alpha}^2+\abs*{\beta}^2=1$,
    where %
    $(\ket{\overline{0}}\!\bra{\overline{0}}_{\overline{ev}})\ket{\bot}=0$.

    Since $\mathit{Test}$, $\mathit{SWAP}[\overline{l},\overline{l'}]$, and $\cdil(P)$ are quantum circuits, we can define unitaries $U_T$, $U_S$, and $U_P$ such that
    \begin{equation}
        \label{eq:def-U_P}
        \Bracks*{\mathit{Test}}(\cdot)=U_T(\cdot)U_T^\dagger,\qquad
        \Bracks*{\mathit{SWAP}[\overline{l},\overline{l'}]}(\cdot)=U_S(\cdot)U_S^\dagger,\qquad
        \Bracks*{\cdil(P)}(\cdot)=U_P(\cdot)U_P^\dagger.
    \end{equation}
    We prove \Cref{thm:test-NI} by establishing \textbf{NI} $\Leftrightarrow$ \Cref{eq:no-l-perturbation} through two directions.
    \begin{enumerate}
        \item 
        We first prove \textbf{NI} $\Rightarrow$ \Cref{eq:no-l-perturbation}.
        Suppose program $P$ satisfies \textbf{NI}.
        Let us fix a list of variables $\overline{v}$ with $\overline{h},\overline{l},\overline{l'},\overline{ev}\subseteq \overline{v}$.
        
        As in the proof of \Cref{thm:non-inter-eq-classical-sec}, we construct
        \begin{equation*}
            Q\equiv \overline{hc}:=\overline{0};\overline{hq}:=\ket{\overline{0}}; P,
        \end{equation*}
        where $\overline{hc}$ and $\overline{hq}$ are classical and quantum variables in $\overline{h}$, respectively.
        Let $R\equiv Q(\overline{h}\leftarrow \overline{g})$
        be obtained by replacing $\High$ variables $\overline{h}$
        in $Q$ by $\Low$ variables $\overline{g}$ with the same type and $\overline{g}\cap \overline{v}=\emptyset$.
        Note that the choice of $\overline{g}$ depends on the fixed $\overline{v}$.
        Then, similar to \Cref{eq:RP-id-5}, 
        for any state $s$ with $\overline{h},\overline{l},\overline{g}\subseteq \var(s)$:
        \begin{equation*}
            \tr_{\overline{g}}\circ\Bracks*{R}(s)=_{\Low}
            \tr_{\overline{g}}\circ\Bracks*{P}(s),
        \end{equation*}
        which further leads to
        \begin{equation}
            \label{eq:RP-relation-veri}
            \tr_{\overline{h},\overline{g}}\circ\Bracks*{R}(s)=
            \tr_{\overline{h},\overline{g}}\circ\Bracks*{P}(s),
        \end{equation}
        using \Cref{lmm:channel-extension} and $\var(R)\cup \var(P)=\overline{h},\overline{l},\overline{g}$.

        Next, we examine two programs
        \begin{align*}
            R_1&\equiv P;  \overline{hc}:=\overline{0}; \overline{hq}:=\ket{\overline{0}}\\
            R_2&\equiv R; \overline{hc}:=\overline{0}; \overline{hq}:=\ket{\overline{0}}.
        \end{align*} 
        Note that by the construction of $R_1,R_2$ and the semantics in \Cref{sec:quantum_programs}, for any state $s$ with $\overline{h},\overline{l},\overline{g}\subseteq \var(s)$:
        \begin{equation}
            \label{eq:tr-g-R_1-R_2}
            \begin{split}
                \tr_{\overline{g}}\circ\Bracks*{R_1}(s)&=\ket{\overline{0}}\!\bra{\overline{0}}_{\overline{h}}\otimes \tr_{\overline{h},\overline{g}}\Bracks*{P}(s)\\
                \tr_{\overline{g}}\circ\Bracks*{R_2}(s)&=\ket{\overline{0}}\!\bra{\overline{0}}_{\overline{h}}\otimes \tr_{\overline{h},\overline{g}}\Bracks*{R}(s).
            \end{split}
        \end{equation}
        Then, we have for any state $s$ with $\overline{h}, \overline{l}\subseteq \var(s)$ and $\overline{g}\cap \var(s)=\emptyset$,
        \begin{equation}
            \label{eq:R_2-R_1-rel}
            \Bracks*{R_1}(s)=\tr_{\overline{g}} \circ \Bracks*{R_1}(s\otimes \ket{\overline{0}}\!\bra{\overline{0}}_{\overline{g}})=
            \tr_{\overline{g}}\circ\Bracks*{R_2}(s\otimes \ket{\overline{0}}\!\bra{\overline{0}}_{\overline{g}}),
        \end{equation}
        where the last equality comes from \Cref{eq:RP-relation-veri,eq:tr-g-R_1-R_2}.
        Let $C\equiv \dil(\overline{hc}:=\overline{0}; \overline{hq}:=\ket{\overline{0}})$ with environmental variables $\overline{\mathit{evc}}$
        and suppose $\dil(R)$ has environmental variables $\overline{\mathit{evr}}$.
        Without loss of generality, 
        we can assume $\overline{evc},\overline{evr}\cap \overline{v}=\emptyset$ (by slightly modifying \Cref{alg:dilation}),
        which means $C$ and $\dil(R)$ can depend on $\overline{v}$.
        Then, $\dil(R); C$ is a dilation of $R_2$ %
        and therefore a dilation of $R_1$ with environmental variables $\overline{\mathit{evr}},\overline{\mathit{evc}},\overline{g}$,
        by \Cref{eq:R_2-R_1-rel}.
        
        On the other hand, suppose $\dil(P)$ has environmental variables $\overline{evp}$.
        It is easy to see that $\dil(P);C$ is another dilation of $R_1$ with environmental variables $\overline{evp},\overline{\mathit{evc}}$ and it acts trivially on $\overline{evr},\overline{g}$.

        Let $\overline{c}=\lift(\cvar(P))=\lift(\cvar(R_1))$ and $\overline{c^\circ}$ be fresh quantum variables with the same types as $\overline{c}$.
        Then, $\cdil(P)$ has environmental variables $\overline{ev}=\overline{evp},\overline{c^\circ}$.
        Using \Cref{lmm:unique-dilation} with $Q_1\equiv \dil(R); C$, $Q_2\equiv \dil(P); C$, 
        and $\overline{c_1^\circ}=\overline{c_2^\circ}=\overline{c^\circ}$,
        we have for any state $s$
        with $\var(s)=\overline{v}-\overline{ev}$: %
        \begin{equation*}
            \Bracks*{\mathit{CX}[\overline{c},\overline{c^\circ}]; \dil(R); C}(s\otimes \ket{\overline{0}}\!\bra{\overline{0}}_{\overline{ev},\overline{\mathit{evr}},\overline{\mathit{evc}},\overline{g}})=\calU\circ\Bracks*{\cdil(P); C}(s\otimes \ket{\overline{0}}\!\bra{\overline{0}}_{\overline{ev},\overline{\mathit{evr}},\overline{\mathit{evc}},\overline{g}})
        \end{equation*}
        for some unitary channel $\calU$ that acts trivially on $\overline{v}-\overline{ev}$.
        This further implies
        \begin{equation}
            \label{eq:dilP-decompose}
            \Bracks*{\cdil(P)}(s\otimes \ket{\overline{0}}\!\bra{\overline{0}}_{\overline{ev},\overline{\mathit{evr}},\overline{\mathit{evc}},\overline{g}})
            =\calV\circ \Bracks*{\mathit{CX}[\overline{c},\overline{c^\circ}];\dil(R)}(s\otimes \ket{\overline{0}}\!\bra{\overline{0}}_{\overline{ev},\overline{\mathit{evr}},\overline{\mathit{evc}},\overline{g}}),
        \end{equation}
        where the unitary channel $\calV=\Bracks*{C}^\dagger \circ \calU^\dagger \circ \Bracks*{C}$ acts trivially on variables $\overline{l},\overline{l'}$
        since $\overline{l},\overline{l'}\cap \var(C)=\emptyset$.

        Suppose 
        \begin{equation*}
            \Bracks*{\mathit{CX}[\overline{c},\overline{c^\circ}];\dil(R)}(\cdot)=U_R(\cdot)U_R^\dagger
        \end{equation*}
        for some unitary $U_R$.
        Let projector $\Pi_0=\ket{\overline{0}}\!\bra{\overline{0}}_{\overline{ev},\overline{evr},\overline{evc},\overline{g}}$.
        Consider two linear operators $U_P \Pi_0$ and $U_R\Pi_0$,
        where $U_P$ is defined in \Cref{eq:def-U_P}.
        From \Cref{eq:dilP-decompose} and \Cref{lmm:same-up-to-phase},
        we obtain
        \begin{equation}
            \label{eq:U_P_U_R_V}
            U_P\Pi_0 = V U_R \Pi_0
        \end{equation}
        for some unitary $V$ such that $\calV(\cdot)=V(\cdot)V^\dagger$.

        Now we are ready to prove \Cref{eq:no-l-perturbation}.
        Consider any state $s=\ket{\xi}\!\bra{\xi}$ with $\var(s)=\overline{v},\overline{h'},\overline{evr},\overline{evc},\overline{g}$, where $\overline{h'}$ are freshly introduced by $\mathit{Test}$, and $\ket{\xi}$ has the form
        \begin{equation*}
            \ket{\xi}=\ket{\overline{0}}_{\overline{ev},\overline{evr},\overline{evc},\overline{g}}\ket{\overline{\Phi}}_{\overline{h},\overline{h'}}\ket{\eta}_{\overline{l},\overline{l'}}.
        \end{equation*}
        Consider the output state $\ket{\xi'}=U_T\ket{\xi}$, where $U_T$ is defined in \Cref{eq:def-U_P}. By the denotational semantics in \Cref{sub:semantics}, we have
        \begin{equation*}
            U_T\ket{\xi}=U_P^\dagger U_S U_P\ket{\xi}=\parens*{U_P^\dagger U_S U_P}\Pi_0\ket{\xi}.
        \end{equation*}
        Combined with \Cref{eq:U_P_U_R_V}, this gives
        \begin{equation}
            \label{eq:U_T_in_terms_of_U_R}
            \Pi_0 U_T\ket{\xi}=\Pi_0 U_R^\dagger V^\dagger U_S V U_R \Pi_0\ket{\xi}
            = \Pi_0 U_R^\dagger U_S U_R \Pi_0\ket{\xi},
        \end{equation}
        where the last equality holds because $V$ acts trivially on variables $\overline{l},\overline{l'}$ and therefore commutes with $U_S$.

        Since $\hvar(R)=\emptyset$, the definition of $R$ implies that $U_R$ must have the form $U_R=V_R\otimes W_R$. Here, $V_R$ satisfies $V_R(\cdot)V_R^\dagger=\Bracks*{\mathit{CX}[\overline{hc},\overline{hc^\circ}]}(\cdot)$, where $\overline{hc^\circ}\subseteq \overline{c^\circ}$ has the same type as $\overline{hc}$, and $W_R$ acts trivially on variables $\overline{h},\overline{h'}$.
        Meanwhile, $U_S$ acts nontrivially only on variables $\overline{l},\overline{l'}$ and therefore commutes with $V_R$.
        Consequently, $U_R^\dagger U_S U_R=W_R^\dagger U_S W_R$ acts trivially on $\overline{h},\overline{h'}$.
        Together with \Cref{eq:U_T_in_terms_of_U_R}, this implies
        \begin{equation}
            \label{eq:project-post-state}
            \Pi_0 U_T\ket{\xi}=\alpha\ket{\overline{0}}_{{\overline{ev},\overline{evr},\overline{evc},\overline{g}}}\ket{\overline{\Phi}}_{\overline{h},\overline{h'}}\ket{\eta'}_{\overline{l},\overline{l'}}
        \end{equation}
        for some $\ket{\eta'}$ and $\alpha$ with $\abs*{\alpha}\leq 1$,
        where $\ket{\overline{0}}_{{\overline{ev},\overline{evr},\overline{evc},\overline{g}}}$ is enforced by $\Pi_0$ on the LHS of \Cref{eq:project-post-state}.
        As $U_T$ acts trivially on variables $\overline{evr},\overline{evc},\overline{g}$, 
        \Cref{eq:project-post-state} leads to
        \begin{equation*}
            \ket{\overline{0}}\!\bra{\overline{0}}_{\overline{ev}}U_T\parens*{\ket{\overline{0}}_{\overline{ev}}\ket{\overline{\Phi}}_{\overline{h},\overline{h'}}\ket{\eta}_{\overline{l},\overline{l'}}}=\alpha\ket{\overline{0}}_{\overline{ev}}\ket{\overline{\Phi}}_{\overline{h},\overline{h'}}\ket{\eta'}_{\overline{l},\overline{l'}}.
        \end{equation*}
        It immediately follows that \Cref{eq:no-l-perturbation} holds.

        \item
        Next, we prove that \Cref{eq:no-l-perturbation} $\Rightarrow$
        \textbf{NI}.
        Let us prove the contrapositive.
        Suppose that $P$ does not satisfy \textbf{NI}.
        Then, by \Cref{lmm:res-suffice}, there exist two input states $s_1$ and $s_2$ with $\var(s_1)=\var(s_2)=\var(P)$
        such that $s_1=_{\Low}s_2$ but $s_1'=\Bracks*{P}\parens*{s_1}\neq_{\Low}\Bracks*{P}\parens*{s_2}=s_2'$.
        This means there exists some $\ket{\chi}_{\overline{l}}$ such that
        \begin{equation}
            \label{eq:chi-difference}
            \bra{\chi}\tr_{\overline{h}}(s_1')\ket{\chi}
            \neq \bra{\chi}\tr_{\overline{h}}(s_2')\ket{\chi}.
        \end{equation}

        Let us define a linear operator $K_{\chi}$ acting on quantum variables $\overline{h},\overline{l}$:
        \begin{equation*}
            K_\chi = \bra{\overline{0}}_{\overline{ev}} U_P^\dagger\ket{\chi}\!\bra{\chi}_{\overline{l}}U_P\ket{\overline{0}}_{\overline{ev}},
        \end{equation*}
        where $U_P$ is a unitary corresponding to the canonical dilation program $\cdil(P)$ (which is also a dilation of $P$), as defined in \Cref{eq:def-U_P}. 
        Then, by \Cref{def:cdil,def:dilation}, for any input state $s$ to program $P$,
        \begin{equation*}
            \tr(K_\chi s)=\tr\parens*{\Bracks*{P}(s)\ket{\chi}\!\bra{\chi}_{\overline{l}} }=\bra{\chi}\parens*{\tr_{\overline{h}}\circ\Bracks*{P}(s)}\ket{\chi}.
        \end{equation*}
        From \Cref{eq:chi-difference}, we further obtain
        \begin{equation*}
            \tr(K_\chi s_1)\neq \tr(K_\chi s_2).
        \end{equation*}
        Consequently, this implies that $K_\chi$ acts nontrivially on $\overline{h}$, because $s_1=_{\Low}s_2$.

        Now consider any input state $s=\ket{\psi}\!\bra{\psi}$ to program $\mathit{Test}$ with
        \begin{equation}
            \label{eq:Test-input-state-chi}
            \ket{\psi}=\ket{\overline{0}}_{\overline{ev}}\ket{\overline{\Phi}}_{\overline{h},\overline{h'}}\ket{\eta}_{\overline{l}}\ket{\chi}_{\overline{l'}},
        \end{equation}
        where $\ket{\chi}$ is held by variables $\overline{l'}$. Then,
        \begin{equation}
            \label{eq:post-state-proj-chi}
            \parens*{\ket{\overline{0}}\!\bra{\overline{0}}_{\overline{ev}}\otimes \ket{\chi}\!\bra{\chi}_{\overline{l'}}}U_{T}\ket{\psi}=\parens*{\ket{\overline{0}}\!\bra{\overline{0}}_{\overline{ev}}\otimes \ket{\chi}\!\bra{\chi}_{\overline{l'}}}U_P^\dagger U_S U_P\ket{\psi},
        \end{equation}
        where $U_T,U_S,U_P$ are defined in \Cref{eq:def-U_P}.
        Note that $U_P$ acts trivially on variables $\overline{l'}$.
        Moreover, for any $\ket{\varphi}_{\overline{l}}$, we have
        \begin{equation*}
            \bra{\chi}_{\overline{l'}}U_S\parens*{\ket{\varphi}_{\overline{l}}\ket{\chi}_{\overline{l'}}}
            =\ket{\chi}_{\overline{l}}\bra{\chi}_{\overline{l'}}\ket{\varphi}_{\overline{l'}}=
            \parens*{\ket{\chi}\!\bra{\chi}_{\overline{l}}} \ket{\varphi}_{\overline{l}}.
        \end{equation*}
        Consequently, combined with \Cref{eq:Test-input-state-chi}, we can simplify the RHS of \Cref{eq:post-state-proj-chi} as
        \begin{equation}
            \label{eq:proj-chi-RHS}
            K_{\chi}\ket{\overline{0}}_{\overline{ev}}\ket{\overline{\Phi}}_{\overline{h},\overline{h'}}\ket{\eta}_{\overline{l}}\ket{\chi}_{\overline{l'}},
        \end{equation}
        where $K_\chi$ acts nontrivially on variables $\overline{h},\overline{l}$.

        Suppose for contradiction that \Cref{eq:no-l-perturbation} holds,
        which means the LHS of \Cref{eq:post-state-proj-chi} has the form
        \begin{equation}
            \label{eq:proj-chi-LHS}
            \alpha \ket{\overline{0}}_{\overline{ev}}\ket{\overline{\Phi}}_{\overline{h},\overline{h'}}\ket{\eta'}_{\overline{l}}\ket{\chi}_{\overline{l'}}.
        \end{equation}
        Thus, \Cref{eq:no-l-perturbation} means
        \begin{equation*}
            K_\chi \ket{\overline{\Phi}}_{\overline{h},\overline{h'}}\ket{\eta}_{\overline{l}}=
            \alpha\ket{\overline{\Phi}}_{\overline{h},\overline{h'}}\ket{\eta'}_{\overline{l}}
        \end{equation*}
        for any state $\ket{\eta}_{\overline{l}}$.
        Finally, applying \Cref{lmm:maxent-trivial-high} yields that $K_\chi=\Id_{\overline{h}}\otimes L_{\overline{l}}$ for some $L$, which contradicts that $K_\chi$ acts nontrivially on $\overline{h}$. The conclusion immediately follows.

    \end{enumerate}
\end{proof}

\subsection{Technical Lemmas}

This section supplements several technical lemmas used in prior proofs.

\begin{lemma}[Unique dilation up to environmental transformations]
    \label{lmm:unique-dilation}
    Suppose quantum circuits $Q_1$ and $Q_2$ are two dilations of a given program $P$ with environmental variables $\overline{ev_1}$ and $\overline{ev_2}$, respectively.
    Let $\overline{c}=\lift(\cvar(P))$, and let $\overline{c_1^\circ}$ and $\overline{c_2^\circ}$ be fresh quantum variables with the same types as $\overline{c}$.
    Let $\overline{v}$ be a list of all variables in
    $\overline{ev_1}\cup\overline{c_1^\circ}\cup\overline{ev_2}\cup\overline{c_2^\circ}$.
    Then, there exists a unitary channel $\calU$ acting nontrivially only on $\overline{v}$ such that
    for any quantum state $s$ with $\overline{c},\qvar(P)\subseteq \var(s)$ and $\var(s)\cap \overline{v}=\emptyset$:
    \begin{equation*}
        \Bracks*{\mathit{CX}[\overline{c},\overline{c_1^\circ}]; Q_1}(s\otimes \ket{\overline{0}}\!\bra{\overline{0}}_{\overline{v}})
        =\calU\circ
        \Bracks*{\mathit{CX}[\overline{c},\overline{c_2^\circ}]; Q_2}(s\otimes \ket{\overline{0}}\!\bra{\overline{0}}_{\overline{v}}).
    \end{equation*}
\end{lemma}

\begin{proof}
    Any quantum state $s$ with $\overline{c},\qvar(P)\subseteq \var(s)$ and $\var(s)\cap \overline{v}=\emptyset$ has the form:
    \begin{equation*}
        s=\sum_{j,k}
        \ket{j}\!\bra{k}_{\overline{c}}
        \otimes s_{j,k}.
    \end{equation*}
    By \Cref{def:denotational_semantics}, \Cref{def:dilation}, and linearity, for $b=1,2$:
    \begin{equation*}
        \tr_{\overline{ev_b},\overline{c_b^\circ}}
        \circ\Bracks*{\mathit{CX}[\overline{c},\overline{c_b^\circ}]; Q_b}
        \parens*{s\otimes
        \ket{\overline{0}}\!\bra{\overline{0}}_{\overline{ev_b},\overline{c_b^\circ}}}=
        \sum_{j}
        \Bracks*{P}
        \parens*{\ket{j}\!\bra{j}_{\overline{c}}\otimes s_{j,j}},
    \end{equation*}
    where we use $\tr_{\overline{c_b^\circ}}(\ket{j,j}\!\bra{k,k}_{\overline{c},\overline{c_b^\circ}})=\ket{j}\!\bra{j}_{\overline{c}}$ if $j=k$ and $0$ otherwise.
    This further implies that $\mathit{CX}[\overline{c},\overline{c_b^\circ}]; Q_b$ for $b=1,2$ are two Stinespring dilations of the same quantum channel.
    The conclusion then follows from the uniqueness of Stinespring dilation, a standard result in quantum information theory~\cite[Corollary 2.24]{Watrous18}.
\end{proof}

\begin{lemma}
    \label{lmm:linear-density}
    For any Hilbert space $\calH$, 
    any matrix in $\calL(\calH)$ can be represented as a linear combination of density matrices in $\calD\parens*{\calH}$.
\end{lemma}

\begin{proof}
    Any matrix in $\calL(\calH)$ can be represented as a linear combination of
    the basis matrices $\ket{j}\!\bra{k}$, where $\ket{j}$ and $\ket{k}$ are computational basis states of $\calH$.
    Thus, it suffices to prove that any $\ket{j}\!\bra{k}$ can be represented as a linear combination of density matrices.
    If $j=k$, $\ket{j}\!\bra{j}$ itself is a density matrix.
    If $j\neq k$, then 
    \begin{align*}
        \ket{j}\!\bra{k} = \ket{\psi_0}\!\bra{\psi_0} + i \ket{\psi_1}\!\bra{\psi_1} -\frac{1+i}{2}\ket{j}\!\bra{j} - \frac{1+i}{2}\ket{k}\!\bra{k},
    \end{align*}
    where $\ket{\psi_0}=\frac{1}{\sqrt{2}}\parens*{\ket{j}+\ket{k}}$ and $\ket{\psi_1}=\frac{1}{\sqrt{2}}\parens*{\ket{j}+i\ket{k}}$.
\end{proof}

\begin{lemma}
    \label{lmm:channel-extension}
    Let $\overline{v}$ be a list of variables.
    Suppose two linear transformations $\calN$ and $\calM$ satisfy
    $\calN(s) = \calM(s)$ for any input state $s$ with $\var(s)=\overline{v}$.
    Then, $\calN\otimes \Id(s')=\calM\otimes \Id(s')$ for any input state $s'$ with $\overline{v}\subseteq \var(s')$,
    where $\Id$ is an identity transformation acting on variables $\var(s')-\overline{v}$.
\end{lemma}

\begin{proof}
    By induction on the variables in $\var(s')-\overline{v}$, it suffices to only consider
    the following two cases:
    \begin{itemize}
        \item 
        If $\var(s')=\overline{v}\uplus \braces*{x}$ for some classical variable $x$,
        then we can write
        \begin{equation}
            \label{eq:s-classical-decomp}
            s'=\sum_{j} p_j s_j\otimes \ket{\sigma_j}\!\bra{\sigma_j}_{x},
        \end{equation}
        where $p_j$ are probabilities, $s_j$ are states with $\var(s_j)=\overline{v}$,
        and $\sigma_j$ are classical states of variable $x$.
        Combining \Cref{eq:s-classical-decomp}, $\calN(s_j)=\calM(s_j)$, and the linearity immediately yields the conclusion.
    
        \item
        If $\var(s')=\overline{v}\uplus \braces*{q}$ for some quantum variable $q$, then
        suppose $\overline{c}$ and $\overline{r}$ 
        are lists of classical and quantum variables in $\overline{v}$, respectively.
        We can write
        \begin{equation}
            \label{eq:s-quantum-decomp}
            s'=\sum_{j,k,l} \ket{\sigma_l}\!\bra{\sigma_l}\otimes s_{j,k,l} \otimes \ket{j}\!\bra{k},
        \end{equation}
        where $\sigma_l$ are classical states of variables $\overline{c}$,
        $\ket{j}$ and $\ket{k}$ are computational-basis states of variable $q$,
        and $s_{j,k,l}$ are matrices in the space $\calL(\calH_{\overline{r}})$.
        From \Cref{lmm:linear-density},
        we can further decompose $s_{j,k,l}$ and $\ket{j}\!\bra{k}$ in \Cref{eq:s-quantum-decomp} as linear combinations of quantum states of variables $\overline{r}$ and $q$, respectively. 
        Combined with $\calN(s)=\calM(s)$ and the linearity, we obtain the conclusion.
    \end{itemize}
\end{proof}

\begin{lemma}
    \label{lmm:maxent-trivial-high}
    Suppose $\overline{a},\overline{a'},\overline{b}$ are quantum variables,
    and $\overline{a},\overline{a'}$ are of the same type.
    Consider a linear operator $K$ acting on $\calH_{\overline{a},\overline{b}}$ such that for any $\ket{\eta}_{\overline{b}}$:
    \begin{equation}
        \label{eq:K-operator}
        (K\otimes \Id_{\overline{a'}})\ket{\overline{\Phi}}_{\overline{a},\overline{a'}}\ket{\eta}_{\overline{b}}
        =\alpha_\eta\ket{\overline{\Phi}}_{\overline{a},\overline{a'}}\ket{\eta'}_{\overline{b}},
    \end{equation}
    where $\ket{\overline{\Phi}}_{\overline{a},\overline{a'}}$ is a maximally entangled state.
    Then, $K=\Id_{\overline{a}}\otimes L$ for some $L$ acting on variables $\overline{b}$.
\end{lemma}

\begin{proof}
    Writing $\overline{a}=a_1,\ldots,a_n$ and $\overline{a'}=a_1',\ldots,a_n'$, we have
    \begin{equation*}
        \ket{\overline{\Phi}}_{\overline{a},\overline{a'}}
        =\bigotimes_{k=1}^n\sum_{j_k=0}^{d_k-1}\frac{1}{\sqrt{d_k}}
        \ket{j_k}_{a_k}\ket{j_k}_{a_k'},
    \end{equation*}
    where each $d_k=\dim(\calH_k)$ is the dimension of the Hilbert space of variable $a_k$.
    Let $\overline{j}=j_1,\ldots,j_n$.
    Comparing the coefficients of $\ket{\overline{j}}_{\overline{a'}}$ on both sides of \Cref{eq:K-operator}, we obtain for any $\overline{j}$ and $\ket{\eta}_{\overline{b}}$:
    \begin{equation*}
        K\ket{\overline{j}}_{\overline{a}}\ket{\eta}_{\overline{b}}
        =\alpha_\eta\ket{\overline{j}}_{\overline{a}}\ket{\eta'}_{\overline{b}}.
    \end{equation*}
    Define an operator $L=\bra{\overline{0}}_{\overline{a}}K\ket{\overline{0}}_{\overline{a}}$ acting on variables $\overline{b}$. Then, 
    \begin{equation*}
        K\ket{\overline{j}}_{\overline{a}}\ket{\eta}_{\overline{b}}
        =\ket{\overline{j}}_{\overline{a}}\parens*{L\ket{\eta}_{\overline{b}}}
    \end{equation*}
    for any $\overline{j}$ and $\ket{\eta}_{\overline{b}}$.
    Consequently, $K=\Id_{\overline{a}}\otimes L$.
\end{proof}

\begin{lemma}
    \label{lmm:same-up-to-phase}
    Let $A,B:\calH\rightarrow \calH'$ be linear operators.
    Suppose that for any pure state $\ket{\psi}$, 
    \begin{equation}
        \label{eq:linear-on-any}
        A\ket{\psi}\!\bra{\psi}A^\dagger=B\ket{\psi}\!\bra{\psi}B^\dagger.
    \end{equation}
    Then, there exists $\alpha\in \Co$ with $\abs{\alpha}=1$ such that $A=\alpha B$.
\end{lemma}

\begin{proof}
    If $B=0$, then \Cref{eq:linear-on-any} implies $A\ket{\psi}=0$ for every pure state $\ket{\psi}$, and hence $A=0$. Otherwise, choose a state $\ket{\psi_0}$ with $B\ket{\psi_0}\neq 0$. By \Cref{eq:linear-on-any}, there exists $\alpha\in\Co$ with $\abs{\alpha}=1$ such that $A\ket{\psi_0}=\alpha B\ket{\psi_0}$. By applying \Cref{eq:linear-on-any} to states $\propto\ket{\phi}\pm\ket{\psi}$ and $\propto\ket{\phi}\pm i\ket{\psi}$ and using linearity, we obtain
    \begin{equation*}
        A\ket{\phi}\!\bra{\psi}A^\dagger=B\ket{\phi}\!\bra{\psi}B^\dagger
    \end{equation*}
    for any pure states $\ket{\phi},\ket{\psi}$. Taking $\ket{\psi}=\ket{\psi_0}$ and using $A\ket{\psi_0}=\alpha B\ket{\psi_0}$, we have
    \begin{equation*}
        \overline{\alpha}\,A\ket{\phi}\!\bra{\psi_0}B^\dagger
        =B\ket{\phi}\!\bra{\psi_0}B^\dagger .
    \end{equation*}
    Since $B\ket{\psi_0}\neq 0$, it follows that $A\ket{\phi}=\alpha B\ket{\phi}$ for every pure state $\ket{\phi}$. Therefore, $A=\alpha B$.
\end{proof}

\section{Discussion: What if the Source of Confidential Information is Quantum?}
\label{sub:secure_quantum_information_flow}

In the information-flow security of quantum programs defined in \Cref{sub:information_flow_security}, the source of confidential information is classical.
In particular, for the simulation-based information-flow security (\textbf{SIM}, see~\Cref{def:classical-inf-sec}),
the source of the confidential information is explicitly modeled by a list of classical variables $\overline{x}$;
and for the noninterference-based information-flow security (\textbf{NI}, see~\Cref{def:noninterference}), we can intuitively think of the distinguishability between two states as one bit of confidential information.

In this appendix, we further investigate the notion of information-flow security when the source of confidential information is quantum.
This concerns the case when the attacker aims to learn a piece of quantum (rather than classical) information about the input states of $\High$ variables.
To learn such confidential information,
the attacker is also required to recover any entanglement involved.

Our goal is to understand whether the quantum source of confidential information will lead to a distinct notion of information-flow security,
and more generally, how the rich properties of quantum computing can affect the formalization of information-flow security.
In the following, we start by extending the simulation-based security notion \textbf{SIM} in \Cref{def:classical-inf-sec} to quantum information source.

\subsection{Information-Flow Security with Quantum Information Source}

Just as \textbf{SIM} is defined using leakage probability (\Cref{def:leaking-pr}), we first introduce a notion of quantum leakage probability, which quantifies how well a $\Low$ attacker can recover confidential information (from a quantum source) encoded in the $\High$ part of the input state. 
Here, the confidential information is stored in quantum variables, which can be entangled with other quantum variables.
The $\Low$ attacker aims to recover the confidential information and must preserve the entanglement with other quantum variables.

\begin{definition}[Quantum leakage probability]
    \label{def:qleaking-pr}
    Consider a quantum program $P$ and quantum variables $\overline{q},\overline{o}$ of the same type.
    Let $\overline{q'}$ be fresh quantum variables of the same type as $\overline{q}$ with $\overline{q'}\cap \var(P)=\emptyset$.
    The quantum leakage probability of $P$ from $\overline{q}$ to $\overline{o}$ is defined as
    \begin{equation}
        \label{eq:qleak-pr}
        \Pr_{\textup{qleak}}\bracks*{P,\overline{q}\rightarrow \overline{o}}=\inf_{s}
            f\parens*{\tr_{\overline{ro}}\circ \Bracks*{P}(s),\tr_{\overline{rq}}(s)},
    \end{equation}
    where $f(\cdot,\cdot)$ is the quantum (squared) fidelity function,
    $s$ ranges over all (normalized) states with $\var(P)\cup \overline{q}\cup\overline{o}\cup \overline{q'}\subseteq \var(s)$,
    $\overline{rq}=\var(s)-\overline{q}-\overline{q'}$, and $\overline{ro}=\var(s)-\overline{o}-\overline{q'}$.
\end{definition}

Intuitively, $\Pr_{\textup{qleak}}\bracks*{P,\overline{q}\rightarrow \overline{o}}$ characterizes the worst-case success probability of coherently recovering the input state of $\overline{q}$ from the output state $\overline{o}$ after executing program $P$.
Compared to the previous leakage probability $\Pr_{\textup{leak}}$ (see \Cref{def:leaking-pr}),
the main difference is that \Cref{def:qleaking-pr} considers quantum variables $\overline{q},\overline{o}$ rather than
classical variables $\overline{x},\overline{y}$.
The quantum nature of $\overline{q},\overline{o}$ means that they can be entangled with other quantum variables $\overline{q'}$ untouched by program $P$.
We further explain \Cref{def:qleaking-pr} as follows.
\begin{itemize}
    \item
        Note that $\overline{rq}=\var(s)-\overline{q}-\overline{q'}$ and $\overline{ro}=\var(s)-\overline{o}-\overline{q'}$ in \Cref{def:qleaking-pr}.
        Here, $\tr_{\overline{ro}}\circ\Bracks*{P}(s)$ represents the output state of $\overline{o},\overline{q'}$ after executing program $P$,
        and $\tr_{\overline{rq}}(s)$ represents the input state of $\overline{q},\overline{q'}$.
    \item 
        The above choice of $\overline{rq}$ and $\overline{ro}$ implies that
        \Cref{eq:qleak-pr} measures the worst-case overlap between the output of $\overline{o},\overline{q'}$ and the input of $\overline{q},\overline{q'}$, which can be entangled.

    \item
        Fresh quantum variables $\overline{q'}$ play the role of a reference, which is untouched by program $P$.
        Large $\Pr_{\textup{qleak}}\bracks*{P,\overline{q}\rightarrow \overline{o}}$ implies that executing $P$ can also preserve the entanglement between $\overline{q}$ and $\overline{q'}$ well (rather than destroying it by measurements).
        
        The role of $\overline{q'}$ is in the same spirit as the ``reference quantum system'' in the definitions of entanglement fidelity~\cite{Schumacher96} and coherent information~\cite{SN96}.
        The idea has also been used in defining quantum channel capacity~\cite{SN96,Lloyd97} in quantum Shannon theory (see e.g., \cite{Wilde13}).
    \item
        We restrict $\overline{q'}$ to be of the same type as $\overline{q}$,
        which is sufficient to capture the worst case
        for a similar reason to \cite[Appendix A.1]{GLN05}.
    \item
        \Cref{def:qleaking-pr} remains well-defined even if $\overline{q},\overline{o}\not\subseteq \qvar(P)$.
\end{itemize}

To better understand $\Pr_{\textup{qleak}}$ and how it differs from $\Pr_{\textup{leak}}$, we consider a concrete example.
\begin{example}
    \label{exp:qleak}
    Suppose $q,q',o$ are all qubit variables and $b$ is a bit variable.
    Consider two programs
    \begin{align*}
        P_1&\equiv b:=0;\ \mathit{SWAP}[q,o]\\
        P_2&\equiv b:=\mathbf{meas}[q];\ \mathbf{if}\ b=1\ \mathbf{then}\ X[o].
    \end{align*}
    Then, for either $P_1$ or $P_2$,
    we have $\overline{rq}=b,o$ and $\overline{ro}=b,q$.
    
    Now consider two states
    \begin{align*}
        s_1&=\ket{0}\!\bra{0}_b\otimes \parens*{\frac{1}{2}\ket{00}\!\bra{00}_{q,q'}+\frac{1}{2}\ket{11}\!\bra{11}_{q,q'}}\otimes \ket{0}\!\bra{0}_{o}\\
        s_2&=\ket{0}\!\bra{0}_b\otimes \ket{\Phi}\!\bra{\Phi}_{q,q'}\otimes \ket{0}\!\bra{0}_{o},
    \end{align*}
    where $\ket{\Phi}=\frac{1}{\sqrt{2}}\parens*{\ket{00}_{q,q'}+\ket{11}_{q,q'}}$
    is a maximally entangled state.

    For state $s_1$, we can calculate that 
    \begin{equation*}
    f\parens*{\tr_{\overline{ro}}\circ \Bracks*{P_1}(s_1),\tr_{\overline{rq}}(s_1)}=1\quad\textup{and}\quad f\parens*{\tr_{\overline{ro}}\circ \Bracks*{P_2}(s_1),\tr_{\overline{rq}}(s_1)}=1.
    \end{equation*}
    That is, after executing either $P_1$ or $P_2$, the output state of $o,q'$
    perfectly recovers the input state of $q,q'$, which is 
    $\frac{1}{2}\ket{00}\!\bra{00}_{q,q'}+\frac{1}{2}\ket{11}\!\bra{11}_{q,q'}$.

    For state $s_2$, we can calculate that 
    \begin{equation*}
        f\parens*{\tr_{\overline{ro}}\circ \Bracks*{P_1}(s_2),\tr_{\overline{rq}}(s_2)}=1\quad\textup{and}\quad f\parens*{\tr_{\overline{ro}}\circ \Bracks*{P_2}(s_2),\tr_{\overline{rq}}(s_2)}=\frac{1}{2}.
    \end{equation*} 
    That is, after executing $P_1$, the output state of $o,q'$ perfectly recovers the input state of $q,q'$, which is $\ket{\Phi}\!\bra{\Phi}$, while for $P_2$ this is not the case.

    Indeed, one can verify that
    \begin{equation*}
        \Pr_{\textup{qleak}}\bracks*{P_1,\overline{q}\rightarrow \overline{o}}=1 \quad \textup{and}\quad \Pr_{\textup{qleak}}\bracks*{P_2,\overline{q}\rightarrow \overline{o}}=\frac{1}{2}.
    \end{equation*}
    This is because $P_1$ preserves the entanglement between $q$ and $q'$, while $P_2$ performs a measurement on $q$ and therefore destroys any entanglement between $q$ and $q'$.

    This example highlights the role of entanglement between $\overline{q}$ and $\overline{q'}$ in the quantum leakage probability $\Pr_{\textup{qleak}}$, which distinguishes it from the leakage probability $\Pr_{\textup{leak}}$.
\end{example}

    We can now define the (simulation-based) information-flow security with quantum information source (abbreviated as \textbf{SIMQ}) of quantum programs.
The definition is similar to that of \textbf{SIM} but based on the quantum leakage probability $\Pr_{\textup{qleak}}$.

\begin{definition}[Information-flow security with quantum information source (\textbf{SIMQ})]
    \label{def:quantum-inf-sec}
    A quantum program $P$ satisfies \textbf{SIMQ} iff
    for any $\High$ quantum variables $\overline{q}$ and $\Low$ quantum variables $\overline{o}$ of the same type,
    any program $\mathit{E}$,
    and any $\Low$ program $\mathit{D}$,
    there exists a $\Low$ program $\mathit{D}'$ such that
    \begin{equation}
        \label{eq:qleak-cond}
        \Pr_{\textup{qleak}}\bracks*{\mathit{E}; P; \mathit{D}, \overline{q}\rightarrow \overline{o}} \leq \Pr_{\textup{qleak}}\bracks*{\mathit{E}; \mathit{D}', \overline{q}\rightarrow \overline{o}}.
    \end{equation}
\end{definition}

The difference between \textbf{SIMQ} and \textbf{SIM} is that the variables $\overline{q},\overline{o}$ for storing the confidential information are quantum variables and we use $\Pr_{\textup{qleak}}$ rather than $\Pr_{\textup{leak}}$ in the security condition \Cref{eq:qleak-cond}.

\subsection{Equivalence to Classical Information Source}

At first glance, one might expect that \textbf{SIMQ} (see \Cref{def:quantum-inf-sec}) is a weaker security notion than \textbf{SIM} (see \Cref{def:classical-inf-sec}),
because it seems harder for a $\Low$ attacker to learn the confidential information from a quantum source,
which also requires preserving entanglement. 
Nevertheless, we show that \textbf{SIMQ} and \textbf{SIM} are equivalent notions for information-flow security.

\begin{theorem}[\textbf{SIMQ} $\Leftrightarrow$ \textbf{SIM}]
    \label{thm:sim-implies-simc}
    The \textbf{SIMQ} in \Cref{def:quantum-inf-sec} is equivalent to \textbf{SIM} in \Cref{def:classical-inf-sec}.
\end{theorem}

\begin{proof}
    \quad
    \begin{enumerate}
        \item 
        We first prove \textbf{SIM} $\Rightarrow$ \textbf{SIMQ}.
        By \Cref{thm:non-inter-eq-classical-sec}, it suffices to prove \textbf{NI} $\Rightarrow$ \textbf{SIMQ}.
        The proof is similar to the \textbf{NI} $\Rightarrow$ \textbf{SIM} part in the proof of \Cref{thm:non-inter-eq-classical-sec}.
        We only need to note the following differences:
        \begin{itemize}
            \item 
                Classical variables $\overline{x},\overline{y}$ now become quantum variables $\overline{q},\overline{o}$.
            \item
                The list of variables $\overline{v}$ now includes the reference $\overline{q'}$,
                and therefore the input state $s$ also satisfies $\overline{q'}\subseteq \var(s)$.
            \item
                Since $\Pr_{\textup{qleak}}$ in \Cref{def:qleaking-pr} does not depend on whether variables are $\High$ or $\Low$, without loss of generality, we can set the reference $\overline{q'}$ to be $\Low$ variables.
        \end{itemize}
        \item
        Next, we prove \textbf{SIMQ} $\Rightarrow$ \textbf{SIM}.
        By \Cref{thm:non-inter-eq-classical-sec}, it suffices to prove $\neg$\textbf{NI} $\Rightarrow$ $\neg$\textbf{SIMQ}.
        Consider any quantum program $P$ that does not satisfy \textbf{NI}.
        Let us exploit the construction in the \textbf{SIM} $\Rightarrow$ \textbf{NI} part in the proof of \Cref{thm:non-inter-eq-classical-sec}.
        In particular, let $\mathit{E}_1$ be the program defined in \Cref{eq:E-SIM-NI} and $\mathit{D}_1$ be the program defined in \Cref{eq:D-SIM-NI}.
        From \Cref{eq:E-D-SIM-NI}, we have 
        \begin{equation}
            \label{eq:leak-assume-SIMQ-SIM}
            \Pr_{\textup{leak}}\bracks*{\mathit{E}_1;P;\mathit{D}_1, x\rightarrow y} > \frac{1}{2},
        \end{equation}
        where $x$ and $y$ are $\High$ and $\Low$ bit variables.

        Let $\overline{v}$ be the list of variables in $\var(P)\cup \var(\mathit{E}_1)\cup \var(\mathit{D}_1)$.
        Let us define 
        \begin{align*}
            \mathit{Pre}&\equiv o:=\ket{0};\ \mathit{CNOT}[q,o];\ H[q];\ x:=\mathbf{meas}[q]\\
            \mathit{Post}&\equiv \mathbf{if}\ y=1\ \mathbf{then}\ Z[o]\\
            \mathit{E}_2&\equiv \mathit{Pre};\ \mathit{E}_1\\
            \mathit{D}_2&\equiv \mathit{D}_1;\ \mathit{Post},
        \end{align*}
        where $q$ and $o$ are $\High$ and $\Low$ qubit variables, respectively, with $q,o\notin\overline{v}$.
        Let $q'$ be the fresh reference variable introduced by \Cref{def:qleaking-pr}.

        Consider state $s$ with $\overline{v}\uplus \braces*{q,o,q'}\subseteq \var(s)$.
        Using the semantics in \Cref{def:denotational_semantics}, it is easy to verify 
        \begin{equation}
            \label{eq:perfect-trans}
            \tr_{\overline{ro}}\circ\Bracks*{\mathit{Pre};y:=x;\mathit{Post}}(s) = \tr_{\overline{rq}}(s),
        \end{equation}
        where in the equality we neglect which quantum variables hold the states.
        Indeed, one can obtain \Cref{eq:perfect-trans} by first validating the special case $\tr_{\overline{rq}}(s)=\ket{\psi}\!\bra{\psi}_{q,q'}$ for some pure state $\ket{\psi}_{q,q'}=\alpha \ket{0}_{q}\ket{\phi_0}_{q'}+\beta\ket{1}_q\ket{\phi_1}_{q'}$ and then extending to the general case by linearity.
        
        Now our goal is to lower bound the quantum leakage probability
        \begin{align}
            \label{eq:E2-D2-qleak-1}
            \Pr_{\textup{qleak}}\bracks*{\mathit{E}_2;P;\mathit{D}_2, q\rightarrow o}
            &=\inf_s f\parens*{\tr_{\overline{ro}}\circ\Bracks*{\mathit{E}_2; P;\mathit{D}_2}(s), \tr_{\overline{rq}}(s)}\\
            \label{eq:E2-D2-qleak-2}
            &=\inf_s f\parens*{\tr_{\overline{ro}}\circ\Bracks*{\mathit{Pre};\mathit{E}_1; P;\mathit{D}_1;\mathit{Post}}(s), \tr_{\overline{rq}}(s)}.
        \end{align}
        Let us calculate the RHS of \Cref{eq:E2-D2-qleak-2}.
        Suppose 
        \begin{equation}
            \label{eq:assume-after-pre}
            \Bracks*{\mathit{Pre}}(s)=p\ket{0}\!\bra{0}_x\otimes s_0 +(1-p)\ket{1}\!\bra{1}_x\otimes s_1,
        \end{equation}
        for some $0\leq p\leq 1$ and states $s_0,s_1$.
        Then,
        \begin{equation*}
            \begin{split}
                \Bracks*{\mathit{E}_1; P;\mathit{D}_1}\circ \Bracks*{\mathit{Pre}}(s)=
                p\ket{0}\!\bra{0}_x \otimes \parens*{p_0\ket{0}\!\bra{0}_y\otimes s_{00}+(1-p_0)\ket{1}\!\bra{1}_y\otimes s_{01}}+\\
                (1-p)\ket{1}\!\bra{1}_x \otimes \parens*{(1-p_1)\ket{0}\!\bra{0}_y\otimes s_{10}+p_1\ket{1}\!\bra{1}_y\otimes s_{11}},
            \end{split}
        \end{equation*}
        for some $0\leq p_0,p_1\leq 1$ and states $s_{00},s_{01},s_{10},s_{11}$.
        Moreover, due to \Cref{eq:leak-assume-SIMQ-SIM} and that $\Bracks*{\mathit{E}_1; P;\mathit{D}_1}$ acts trivially on $q,o,q'$, 
        we have
        \begin{equation}
            \label{eq:after-E1-P-D1-cond}
            p_a \geq \Pr_{\textup{leak}}\bracks*{\mathit{E}_1;P;\mathit{D}_1, x\rightarrow y} > \frac{1}{2}\quad\textup{and}\quad\tr_{\overline{u}}(s_{ab})=\tr_{\overline{u}}(s_a) \quad\textup{for}\ a,b\in\braces*{0,1},
        \end{equation}
        where $\overline{u}=\overline{v}-\braces*{x,y}$.

        Next, we calculate
        \begin{equation}
            \label{eq:after-post-E1-P-D1}
            \begin{split}
                \Bracks*{Post}\circ\Bracks*{\mathit{E}_1; P;\mathit{D}_1}\circ \Bracks*{\mathit{Pre}}(s)=
                p\ket{0}\!\bra{0}_x \otimes \parens*{p_0\ket{0}\!\bra{0}_y\otimes s_{00}'+(1-p_0)\ket{1}\!\bra{1}_y\otimes s_{01}'}+\\
                (1-p)\ket{1}\!\bra{1}_x \otimes \parens*{(1-p_1)\ket{0}\!\bra{0}_y\otimes s_{10}'+p_1\ket{1}\!\bra{1}_y\otimes s_{11}'},
            \end{split}
        \end{equation}
        for some states $s_{00}',s_{01}',s_{10}',s_{11}'$.
        Moreover,
        using \Cref{eq:after-E1-P-D1-cond}, $\overline{u}\cap \parens*{\var(Post)\cup \braces*{x,y}}=\emptyset$, and the semantics in \Cref{sec:quantum_programs},  we have for $a\in \braces*{0,1}$:
        \begin{align*}
            \tr_{\overline{ro}}(\ket{a}\!\bra{a}_{x}\otimes \ket{a}\!\bra{a}_y\otimes s_{aa}')
            &=
            \tr_{\overline{ro}}\circ \Bracks*{\mathit{Post}}\circ \tr_{\overline{u}}(\ket{a}\!\bra{a}_{x}\otimes \ket{a}\!\bra{a}_y\otimes s_{aa})\\
            &=\tr_{\overline{ro}}\circ \Bracks*{\mathit{Post}}\circ\Bracks*{y:=x}\circ \tr_{\overline{u}}(\ket{a}\!\bra{a}_{x}\otimes s_{a})\\
            &=\tr_{\overline{ro}}\circ \Bracks*{\mathit{Post}}\circ\Bracks*{y:=x}(\ket{a}\!\bra{a}_{x}\otimes s_{a}).
        \end{align*}
        This, together with \Cref{eq:assume-after-pre,eq:perfect-trans} implies
        \begin{equation}
            \label{eq:just-rqs}
            \begin{split}
                \tr_{\overline{rq}}(s)&=
                \tr_{\overline{ro}}\circ \Bracks*{\mathit{Post}}\circ\Bracks*{y:=x}\circ \Bracks*{Pre}(s)\\
                &=
                \tr_{\overline{ro}}(p\ket{0}\!\bra{0}_{x}\otimes \ket{0}\!\bra{0}_y\otimes s_{00}'
                + (1-p)\ket{1}\!\bra{1}_{x}\otimes \ket{1}\!\bra{1}_{y}\otimes s_{11}').
            \end{split}
        \end{equation}

        Since $p_0,p_1>\frac{1}{2}$, from \Cref{eq:after-post-E1-P-D1,eq:just-rqs} it is easy to calculate
        \begin{equation}
            \label{eq:E2-D2-greater-than-half}
            \inf_s f\parens*{\tr_{\overline{ro}}\circ\Bracks*{\mathit{Pre};\mathit{E}_1; P;\mathit{D}_1;\mathit{Post}}(s), \tr_{\overline{rq}}(s)} > \frac{1}{2}.
        \end{equation}

        Finally, consider any $\Low$ program $\mathit{D}_2'$.
        We can pick $\tilde{s}=\ket{\Phi}\!\bra{\Phi}_{q,q'}\otimes \ket{\overline{0}}\!\bra{\overline{0}}_{\overline{w}}$,
        where $\ket{\Phi}_{q,q'}=\frac{1}{\sqrt{2}}(\ket{00}_{q,q'}+\ket{11}_{q,q'})$
        and $\overline{w}$ represents variables in $\var(\mathit{E}_2;\mathit{D}_2')-\braces*{q,q'}$.
        Then, using the semantics in \Cref{def:denotational_semantics}, 
        we have $\tr_{\overline{ro}}\circ\Bracks*{Pre}(\tilde{s})=\frac{1}{2}(\ket{00}\!\bra{00}_{q',o}+\ket{11}\!\bra{11}_{q',o})$, and the output state $\tr_{\overline{ro}}\circ\Bracks*{\mathit{E}_2;\mathit{D}_2'}(\tilde{s})$ %
        has the form 
        \begin{equation*}
            \rho=\frac{1}{2}\ket{0}\!\bra{0}_{q'}\otimes \rho^0_{o}+\frac{1}{2}\ket{1}\!\bra{1}_{q'}\otimes \rho^1_{o},
        \end{equation*}
        which is separable and therefore has fidelity $\leq \frac{1}{2}$ with $\ket{\Phi}$.
        Consequently,
        \begin{equation}
            \label{eq:D2-prime}
            \Pr_{\textup{qleak}}\bracks*{\mathit{E}_2;\mathit{D}_2', q\rightarrow o}
            =\inf_s f\parens*{\tr_{\overline{ro}}\circ\Bracks*{\mathit{E}_2;\mathit{D}_2'}(s), \tr_{\overline{rq}}(s)}
            \leq f\parens*{\rho,\ket{\Phi}\!\bra{\Phi}}\leq \frac{1}{2}.
        \end{equation}
        Thus, combining with \Cref{eq:E2-D2-qleak-2,eq:E2-D2-greater-than-half,eq:D2-prime} yields
        \begin{equation*}
            \Pr_{\textup{qleak}}\bracks*{\mathit{E}_2;\mathit{D}_2', q\rightarrow o} < \Pr_{\textup{qleak}}\bracks*{\mathit{E}_2;P;\mathit{D}_2, q\rightarrow o},
        \end{equation*}
        which means $P$ does not satisfy \textbf{SIMQ}.
    \end{enumerate}
\end{proof}

\section{Quantum State Discrimination}
\label{sec:quantum_state_discrimination}

It is standard that distinct quantum states can be discriminated by a single measurement strictly better than a random guess, even in the worst case.
For completeness, in this appendix, we provide a proof of how the pretty good measurement~\cite{Belavkin75,Holevo79,HW94,Montanaro19} can achieve this.

\begin{lemma}
    \label{lmm:pretty-good-measurement}
    For two distinct quantum states $\rho_1\neq \rho_2$,
    the pretty good measurement $M=\braces*{M_1,M_2}$ can discriminate these two states with worst-case success probability $>\frac{1}{2}$;
    i.e., $\tr\parens*{M_1\rho_1 M_1^\dagger}>\frac{1}{2}$ and $\tr\parens*{M_2\rho_2 M_2^\dagger}>\frac{1}{2}$.
    Here, 
    \begin{equation*}
        \begin{split}
            M_1&=\parens*{\parens*{\rho_1+\rho_2}^{-1/2} \rho_1 \parens*{\rho_1+\rho_2}^{-1/2}}^{1/2}\\
            M_2&=\parens*{\parens*{\rho_1+\rho_2}^{-1/2} \rho_2 \parens*{\rho_1+\rho_2}^{-1/2}}^{1/2}\\
        \end{split},
    \end{equation*}
    and $(\rho_1+\rho_2)^{-1/2}$ is defined via pseudoinverse on $\operatorname{supp}(\rho_1+\rho_2)$. 
\end{lemma}

\begin{proof}
    Let $E_1=M_1^\dagger M_1$ and $E_2=M_2^\dagger M_2$.
    Our goal is to show $\tr\parens*{E_1\rho_1}>\frac{1}{2}$ and $\tr\parens*{E_2\rho_2}>\frac{1}{2}$.

    Using the linearity and cyclic property of trace, 
    we first note that
    \begin{equation}
        \label{eq:E0}
        \tr\parens*{E_1\rho_1} +\tr\parens*{E_1\rho_2}=\tr\parens*{\parens*{\rho_1+\rho_2}^{-1/2} \rho_1 \parens*{\rho_1+\rho_2}^{-1/2}\parens*{\rho_1+\rho_2}}=1.
    \end{equation}
    Similarly, we have
    \begin{equation}
        \label{eq:E1}
        \tr\parens*{E_2\rho_1}+\tr\parens*{E_2\rho_2}=1.
    \end{equation}
    Next, observe that
    \begin{equation}
        \label{eq:symmetric}
        \begin{split}
            \tr\parens*{E_1\rho_2}&=\tr\parens*{\parens*{\rho_1+\rho_2}^{-1/2} \rho_1 \parens*{\rho_1+\rho_2}^{-1/2}\rho_2}\\
            &=\tr\parens*{\parens*{\rho_1+\rho_2}^{-1/2}\rho_2\parens*{\rho_1+\rho_2}^{-1/2} \rho_1}\\
            &=\tr\parens*{E_2\rho_1}.
        \end{split}
    \end{equation}
    \Cref{eq:E0,eq:E1,eq:symmetric} together imply 
    \begin{equation}
        \label{eq:E0E1}
        \begin{split}
            \tr\parens*{E_1\rho_1}=\frac{1+\tr\parens*{E_1(\rho_1-\rho_2)}}{2}=
            \frac{1+\tr\parens*{E_2(\rho_2-\rho_1)}}{2}=\tr\parens*{E_2\rho_2}.
        \end{split}
    \end{equation}
    Further, we can calculate:
    \begin{equation}
        \label{eq:A2}
        \begin{split}
            &\tr\parens*{E_1\parens*{\rho_1-\rho_2}}+\tr\parens*{E_2\parens*{\rho_2-\rho_1}}\\
            =&\tr\parens*{\parens*{\rho_1+\rho_2}^{-1/2} \parens*{\rho_1-\rho_2} \parens*{\rho_1+\rho_2}^{-1/2}\parens*{\rho_1-\rho_2}}\\
            =&\tr\parens*{A^2},
        \end{split}
    \end{equation}
    where $A=\parens*{\rho_1+\rho_2}^{-1/4} \parens*{\rho_2-\rho_1}\parens*{\rho_1+\rho_2}^{-1/4}$ is a Hermitian operator.
    Since $\rho_1\neq \rho_2$, we have $A\neq 0$. Consequently, $A^2$ is positive semidefinite and nonzero, which implies the quantity in $\textup{\Cref{eq:A2}}$ is positive.
    Finally, combined with \Cref{eq:E0E1}, we obtain $\tr\parens*{E_1\rho_1}=\tr\parens*{E_2\rho_2}>\frac{1}{2}$ as desired.
\end{proof}

\end{document}